\documentclass[12pt]{article}
\usepackage{amsthm,amsmath,latexsym,amssymb,amsfonts,amscd}
\usepackage{graphics,lscape,fancyhdr,array,stmaryrd,euscript,wrapfig}
\usepackage{empheq,wrapfig}
\usepackage{verbatim,slashed}
\numberwithin{equation}{section}
\usepackage{hyperref,setspace}
\usepackage{tikz-cd}
\usepackage{mathrsfs}
\usepackage[numbers,sort&compress]{natbib}
\usepackage[nottoc]{tocbibind}

\newcommand{\pl}{\partial}
\newcommand{\plb}{\bar{\partial}}

\newcommand{\be}{\begin{align}}
\newcommand{\ee}{\end{align}}

\newcommand{\aAt}{{\ensuremath{\mathtt{A}}}}
\newcommand{\aBt}{{\ensuremath{\mathtt{B}}}}

\newcommand{\bry}{{{\bar{y}}}}

\newcommand{\hs}{{\mathfrak{hs}}}

\newcommand{\fud}[2]{{}^{#1}{}_{#2}\,}
\newcommand{\fdu}[2]{{}_{#1}{}^{#2}\,}

\DeclareMathOperator{\sign}{sign}

\newcommand{\hhbar}{{\lambda}}

\newcommand{\besubeqs}{\begin{subequations}}
\newcommand{\esubeqs}{\end{subequations}}

\usepackage{tensor}
\usepackage{todonotes}

\renewcommand{\bar}[1]{\overline{#1}}

\newcommand{\pluk}{{\boldsymbol{p}}}

\newtheorem{theorem}{\fbox{\color{violet}{Theorem}}}[section]

\newtheorem{definition}[theorem]{Definition}

\usepackage[all,cmtip]{xy}

\begin{document}
\pagenumbering{gobble}
\hfill
\vskip 0.01\textheight
\begin{center}
{\Large\bfseries 
More on Chiral Higher Spin Gravity \\ [3mm]and Convex Geometry}

\vspace{0.4cm}

\vskip 0.03\textheight
\renewcommand{\thefootnote}{\fnsymbol{footnote}}
Alexey \textsc{Sharapov}${}^{a}$, 
Evgeny \textsc{Skvortsov}\footnote{Research Associate of the Fund for Scientific Research -- FNRS, Belgium}${}^{b,c}$, Arseny \textsc{Sukhanov}${}^d$ \& Richard \textsc {Van Dongen}${}^{b}$
\renewcommand{\thefootnote}{\arabic{footnote}}
\vskip 0.03\textheight

{\em ${}^{a}$Physics Faculty, Tomsk State University, \\Lenin ave. 36, Tomsk 634050, Russia}\\
\vspace*{5pt}
{\em ${}^{b}$ Service de Physique de l'Univers, Champs et Gravitation, \\ Universit\'e de Mons, 20 place du Parc, 7000 Mons, 
Belgium}\\
\vspace*{5pt}
{\em ${}^{c}$ Lebedev Institute of Physics, \\
Leninsky ave. 53, 119991 Moscow, Russia}\\
\vspace*{5pt}
{\em ${}^{d}$
Moscow Institute of Physics and Technology, \\
Institutskiy per. 7, Dolgoprudnyi, 141700 Moscow region, Russia}

\end{center}

\vskip 0.02\textheight

\begin{abstract}
Recently, a unique class of local Higher Spin Gravities with propagating massless fields in $4d$ -- Chiral Higher Spin Gravity -- was given a covariant formulation both in flat and $(A)dS_4$ spacetimes at the level of equations of motion. We unfold the corresponding homological perturbation theory as to explicitly obtain all interaction vertices. The vertices reveal a remarkable simplicity after an appropriate change of variables. Similarly to formality theorems the $A_\infty/L_\infty$ multi-linear products can be represented as integrals over a configuration space, which in our case is the space of convex polygons. The $A_\infty$-algebra underlying Chiral Theory is of pre-Calabi--Yau type. As a consequence, the equations of motion have the Poisson sigma-model form.
\end{abstract}

\newpage
\tableofcontents
\newpage
\section{Introduction}
\label{sec:}
\pagenumbering{arabic}
\setcounter{page}{2}
Despite decades of efforts \cite{Bekaert:2022poo} there is only one class of Higher Spin Gravities (HiSGRA) with propagating massless fields whose existence does not require relaxing basic field theory concepts such as locality and does not face any open problem \cite{Boulanger:2015ova,Bekaert:2015tva,Maldacena:2015iua,Sleight:2017pcz,Ponomarev:2017qab}. This is Chiral HiSGRA \cite{Metsaev:1991mt,Metsaev:1991nb,Ponomarev:2016lrm,Skvortsov:2018jea,Skvortsov:2020wtf}. Other well-defined examples of HiSGRA with either topological or conformal fields are $3d$ models with (partially-)massless and conformal higher spin fields \cite{Blencowe:1988gj,Bergshoeff:1989ns,Campoleoni:2010zq,Henneaux:2010xg,Pope:1989vj,Fradkin:1989xt,Grigoriev:2019xmp,Grigoriev:2020lzu} and a higher spin extension of $4d$ conformal gravity \cite{Segal:2002gd,Tseytlin:2002gz,Bekaert:2010ky}. Further interesting ideas include \cite{deMelloKoch:2018ivk,Aharony:2020omh} and \cite{Sperling:2017dts,Tran:2021ukl,Steinacker:2022jjv}.

Chiral Theory is an explicit counter-example to many folklore no-go-type statements regarding higher spin theories: (i) it exists in flat space, thus avoiding (or rather obeying in a delicate way) the Weinberg and Coleman--Mandula theorems \cite{Weinberg:1964ew, Coleman:1967ad}; (ii) it smoothly deforms to $(A)dS_4$, whereas the need for the cosmological constant and singularity of the flat limit are usually overstated; (iii) it is perturbatively local,  even though many results indicate that generic HiSGRA have to go beyond the usual modest definitions of locality \cite{Bekaert:2015tva,Maldacena:2015iua,Sleight:2017pcz,Ponomarev:2017qab}; (iv) its holographic $S$-matrix cannot coincide with a free CFT \cite{Skvortsov:2018uru} as it is generally the case for HiSGRA \cite{Sezgin:2002rt,Klebanov:2002ja,Sezgin:2003pt,Leigh:2003gk,Maldacena:2011jn,Boulanger:2013zza,Alba:2013yda,Alba:2015upa}. 

The main price to pay for all these properties is an apparent lack of unitarity even though it gets restored \cite{Skvortsov:2018jea,Skvortsov:2020wtf,Skvortsov:2020gpn} in flat space. In $AdS_4$ Chiral Theory can be used to obtain unitary results \cite{Skvortsov:2018uru} due to the fact that it should be a consistent truncation of the holographic dual of (Chern--Simons) vector models \cite{Skvortsov:2018uru,Sharapov:2022awp}. The latter faces open challenges \cite{Boulanger:2015ova,Bekaert:2015tva,Maldacena:2015iua,Sleight:2017pcz,Ponomarev:2017qab} that prevent one to give a bulk definition of the theory. Being a consistent truncation entails many useful properties: all classical solutions to Chiral Theory are simultaneously solutions to the full theory and the same is true for amplitudes, i.e. the amplitudes of Chiral Theory are subsets of unitary amplitudes. More generally, any nice property, e.g. one-loop finiteness \cite{Skvortsov:2018jea,Skvortsov:2020wtf,Skvortsov:2020gpn}, can be tested first in Chiral Theory and, if true, it has a chance to hold in the complete theory.

Originally, Chiral Theory was found in the light-cone gauge in flat space \cite{Metsaev:1991mt,Metsaev:1991nb,Ponomarev:2016lrm} and conjectured to have a smooth deformation to $(A)dS_4$ \cite{Ponomarev:2016lrm}. The latter was supported \cite{Metsaev:2018xip,Skvortsov:2018uru} by an extension of the light-cone analysis to $AdS_4$ \cite{Metsaev:2018xip}. Nevertheless, one can hardly deny that a covariant form of the theory would be more than useful. While a covariant action of Chiral Theory is an open problem (see \cite{Tran:2022tft} for the recent progress) its classical equations of motion were constructed in \cite{Skvortsov:2022syz,Sharapov:2022faa} for vanishing cosmological constant and in \cite{Sharapov:2022awp} for a nonvanishing one. Significantly, the equations of motion appear to be  perturbatively local.

The covariant form of Chiral Theory was constructed via the standard homological perturbation theory: there is a differential graded Lie algebra that encodes the free theory, whereas its simple deformation leads to a nontrivial $L_\infty$-algebra $\mathbb{L}$ that encodes the interaction vertices. It was shown in \cite{Sharapov:2022faa,Sharapov:2022awp} that the vertices are local, and hence, well-defined; this was also illustrated with examples that go well beyond the state of the art. Nevertheless, concrete applications call for an explicit form of all interaction vertices, which we provide in the present paper, see \cite{Sharapov:2022wpz} for a short summary. 

First of all, the aforementioned $L_\infty$-algebra $\mathbb{L}$ is obtained by symmetrization of a certain $A_\infty$-algebra $\hat{\mathbb{A}}$. It is the latter algebra which  structure maps (or products) we compute.
Another observation is that all nontrivial algebraic structures defining the interaction vertices are effectively low-dimensional. To put it more formally, the $A_\infty$-algebra $\hat{\mathbb{A}}$ is given by a tensor product of a smaller $A_\infty$-algebra $\mathbb{A}$ with some associative algebra $B$. While $B$ enters trivially and can be replaced with any other associative algebra, e.g. $\mathrm{Mat}_N$, the theory based on $\mathbb{A}$ is effectively low-dimensional. The effective dimension can be seen from the functional dimension of the vector space underlying $\mathbb{A}$, which is $2$. By definition, both $\hat{\mathbb{A}}$ and $\mathbb{A}$ have natural pairings that make them into cyclic $A_\infty$-algebras, or more specifically, pre-Calabi--Yau algebras of degree two \cite{kontsevich2021pre}. This cyclic structure appears to be very useful in linking different interaction vertices to each other.

\begin{wrapfigure}{r}{0.3\textwidth}
\begin{tikzpicture}
\draw[-]  (0,4) -- (4,4)--(4,0)--cycle;
\draw[-]  (0,4) -- (4,0);
\draw[thick]  (0,0) -- (0,4)--(0.5,1.5)--(1,0.7)--(2.5,0.15)--(4,0)--cycle;
 \fill[black!5!white] (0,0) -- (0,4)--(0.5,1.5)--(1,0.7)--(2.5,0.15)--(4,0)--cycle;
\filldraw (0,4) circle (1.0 pt);
\filldraw (0.5,1.5) circle (1.5 pt);
\filldraw (1,0.7) circle (1.5 pt);
\filldraw (2.5,0.15) circle (1.5 pt);
\filldraw (0,0) circle (1 pt);
\filldraw (4,0) circle (1 pt);
\node[] at (0.4,0.7) {A};
\node[] at (1.7,1.5) {B};
\end{tikzpicture}
\caption{A convex polygon B and a swallowtail A.}\label{F0}
\end{wrapfigure}
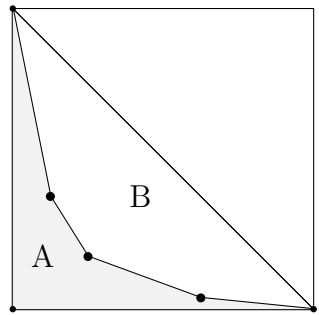
Secondly, the vertices that come out of homological perturbation theory can be considerably simplified by performing a certain change of variables, which, among other things, makes the  cyclic structures of $\hat{\mathbb{A}}$ and $\mathbb{A}$ explicit. Remarkably, the final result is that, pretty much like in Kontsevich \cite{Kontsevich:1997vb} and Shoikhet--Tsygan--Kontsevich Formality \cite{Shoikhet:2000gw}, the structure maps can be written as integrals over a certain configuration space. The configuration space, which will be defined in detail in Section \ref{sec:config}, can be described as the space of concave polygons (region A in Fig. \ref{F0}), which we call swallowtails. Alternatively, it is the space of convex polygons with one edge corresponding to the diagonal of the square, i.e., polygons B inscribed into a protractor triangle ($45^\circ-90^\circ-45^\circ$). The area of region A also plays a role and appears in front of the cosmological constant term in the $A_\infty$-structure maps. The example in Fig. \ref{F0} corresponds to quintic structure maps that are given by an integral over the six-dimensional configuration space, the positions of the three points in between A and B. The compactness of the configuration space implies that the vertices are formally well-defined. Importantly, the vertices also obey an additional property that translates into locality from the field theory vantage point. The vertices we found turn out to be maximally local, which corresponds to a certain distinguished coordinate system from the $A_\infty/L_\infty$-perspective.

The outline of the paper is as follows. In Section \ref{sec:initialdata}, we recall some basic aspects of Chiral Theory \cite{Skvortsov:2022syz,Sharapov:2022faa,Sharapov:2022awp} and of the well-known formalism which dates back to \cite{Vasiliev:1986td,Vasiliev:1988sa} in the HiSGRA context. In Section \ref{sec:vertices}, we first present a few vertices already known explicitly and then proceed to getting a general formula for vertices in all orders. We also identify a change of variables that drastically simplifies the vertices and allows us to represent them via integrals over a simple configuration space. This is discussed in Section \ref{sec:config}. The derivation is supported by a few technical Appendices A, B, and C. We end up with some conclusions in Section \ref{sec:conclusions}.  

\section{Initial data}
\label{sec:initialdata}
It can be shown within the light-front approach that Chiral Theory is a unique class of theories that completes a single nontrivial higher spin self-interaction to a Lorentz invariant local theory. The spectrum of the theory contains fields of all spins including the graviton and scalar field, see \cite{Metsaev:1991mt,Metsaev:1991nb,Ponomarev:2016lrm,Skvortsov:2020wtf} for more detail. Chiral Theory admits two simple contractions \cite{Ponomarev:2017nrr} where fields interact via either Yang--Mills or gravitational interactions (they are defined as one- and two-derivative vertices, respectively). These interactions are not binding enough, do not fix the spectrum uniquely, and the spin-zero field can be dropped. Nevertheless, the contractions are very useful since they have a simple manifestly Lorentz invariant actions both in flat and $(A)dS_4$ spacetimes \cite{Krasnov:2021nsq}. It turned out that the field variables suitable for Chiral Theory are not Fronsdal fields, i.e., symmetric tensors $\Phi_{\mu_1\cdots\mu_s}$, rather they originate from the twistor approach to massless helicity fields \cite{Hughston:1979tq,Eastwood:1981jy,Woodhouse:1985id}. The free action reads \cite{Krasnov:2021nsq}
\begin{align}\label{niceaction}
    S= \int \Psi^{A(2s)}\wedge e_{AB'}\wedge e\fdu{A}{B'}\wedge \nabla \omega_{A(2s-2)}\,,
\end{align}
where $\omega^{A(2s-2)}\equiv \omega^{A(2s-2)}_\mu \,dx^\mu$ is a one-form that is a symmetric rank-$(2s-2)$ spin-tensor\footnote{Following Penrose and Rindler \cite{penroserindler}, $A,B,\ldots =1,2$ and $A',B',\ldots =1,2$ are the indices of the two $2$-dimensional representations of the Lorentz algebra, e.g. of the fundamental and anti-fundamental of $sl(2,\mathbb{C})$ in the case of the Lorentz signature. A group of symmetric (or to be symmetrized) indices $A_1\ldots A_k$ is abbreviated as $A(k)$. } and $\Psi^{A(2s)}$ is a zero-form that is a symmetric  spin-tensor of rank $2s$. In case $s=1$, these two fields  represent the gauge potential $A_\mu$ and the self-dual part $\Psi^{AB}$ of the strength two-form $F_{\mu\nu}$ (treated as an independent field).  For $s=2$ they correspond to the self-dual part of the spin-connection $\omega^{AB}$ and of the Weyl tensor $\Psi^{ABCD}$ (treated as an independent field). The set of one forms $e^{AA'}\equiv e^{AA'}_\mu \, dx^\mu$ defines a vierbein compatible with the spin-connection, $\nabla e^{AA'}=0$. The action enjoys a gauge symmetry of the form
\begin{align}\label{lin-gauge}
    \delta \omega^{A(2s-2)}&= \nabla \xi^{A(2s-2)} +e\fud{A}{C'} \eta^{A(2s-3),C'}\,,& \delta\Psi^{A(2s)}&=0\,,
\end{align}
where $\xi^{A(2s-2)}$ and $\eta^{A(2s-3),C'}$ are zero-forms. The action is gauge invariant on any self-dual background, i.e., where $\nabla^2 \chi^A\equiv0$ for an arbitrary test spinor $\chi^A$; this is more general than Flat or $(A)dS$ spaces the Fronsdal fields can consistently propagate on. The challenge is to complete the free action \eqref{niceaction} with appropriate interaction vertices. A much simpler problem is to construct the associated field equations, which was solved in \cite{Skvortsov:2022syz,Sharapov:2022faa,Sharapov:2022awp}. The equations were sought for in the form of a Free Differential Algebra, the idea that was put forward in the higher spin context long ago \cite{Vasiliev:1988sa}. Doing so requires infinitely many auxiliary fields, which are model independent and determined by the physical degrees of freedom one wants to describe rather than a specific  structure of interactions. Therefore, they are exactly the same as in \cite{Vasiliev:1986td,Vasiliev:1988sa} and can be packaged into generating functions 
\begin{align*}
    \omega(y,\bry)&= \sum_{n+m=\text{even}}\tfrac{1}{n!m!} \omega_{A(n),A'(m)}\, y^A\cdots y^A\, \bry^{A'}\cdots \bry^{A'} 
 \end{align*}  
for one-forms and, likewise, 
\begin{align*}
 C(y,\bry)&= \sum_{n+m=\text{even}}\tfrac{1}{n!m!} C_{A(n),A'(m)}\, y^A\cdots y^A\, \bry^{A'}\cdots \bry^{A'}
\end{align*}
for zero-forms. The dynamical fields that appear in \eqref{niceaction} can be identified with $\Psi(y)= C(y,\bry=0)$ and $\omega(y)=\omega(y,\bry=0)$. That the sum $n+m$ is even means that the fields are bosonic. Super-symmetric extensions can be studied without much effort as well as Yang--Mills gaugings. Naturally, $C$ has room for the scalar field $C(0,0)$, which is necessarily present in Chiral Theory. 

With one-form $\omega$ and zero-form $C$ the most general equations of the Free Differential Algebra form read
\besubeqs\label{eq:chiraltheory}
\begin{align} 
    d\omega&= \mathcal{V}(\omega, \omega) +\mathcal{V}(\omega,\omega,C)+\mathcal{V}(\omega,\omega,C,C)+\ldots\,,\\
    dC&= \mathcal{U}(\omega,C)+ \mathcal{U}(\omega,C,C)+\ldots \,.
\end{align}
\esubeqs
The structure maps $\mathcal{V}$ and $\mathcal{U}$ satisfy the $L_\infty$-relations, which ensure the formal consistency of the equations and lead to a natural gauge symmetry. The equations are more compactly written in the form $d\Phi=Q(\Phi)$, where $\Phi(x)=\{\omega(x),C(x)\}$ are maps from space-time to the target space (a supermanifold) with coordinates $\Phi=\{\omega,C\}$ and the latter space is equipped with an odd nilpotent vector field $Q$ (a homological vector field). As is well-known \cite{Alexandrov:1995kv}, the nilpotency condition $QQ=0$ is equivalent to $L_\infty$-relations in the formal neighbourhood of a stationary point. Since we access the $Q$ of Chiral Theory perturbatively, it is more convenient to use the $L_\infty$-language. 

With the help of the generating function, the equations of motion that follow from \eqref{niceaction} (together with the scalar field and auxiliary fields) can be reformulated as
\besubeqs\label{linearizeddata}
\begin{align}
    \nabla\omega &= e^{BB'}(\hhbar\, \bry_{B'} \pl_{B}+y_{B} \plb_{B'}) \omega +e\fdu{A}{B'}\wedge e^{AB'} \plb_{B'}\plb_{B'}C(y=0,\bry)\,,\\
    \nabla C&= e^{BB'}(\hhbar\, y_{B} \bry_{B'}-\pl_B \plb_{B'}) C\,.
\end{align}
\esubeqs
Here $\nabla$ is the Lorentz covariant derivative and the parameter $\hhbar$ is related to the cosmological constant. The free equations should be compared with the linearization of \eqref{eq:chiraltheory} over a purely gravitational background
\begin{align}\label{adsfour}
    \omega_0&= \tfrac14 \omega^{AB}\, y_A y_B+ \tfrac12 e^{AA'}\, y_A\bry_{A'}+\tfrac14 \omega^{A'B'}\, \bry_{A'} \bry_{B'}\,.
\end{align}
The linearized equations read
\begin{align}\label{linearizeddataA}
    d\omega &= \mathcal{V}(\omega_0, \omega) + \mathcal{V}(\omega, \omega_0)+\mathcal{V}(\omega_0,\omega_0,C)\,,& 
    d C&= \mathcal{U}(\omega_0, C)\,.
\end{align}
Eqs. \eqref{linearizeddataA} vs. \eqref{linearizeddata} set the boundary conditions for $\mathcal{V}(\omega_0, \omega)$, $\mathcal{V}(\omega_0, \omega_0,C)$ and $\mathcal{U}(\omega_0, C)$. Proceeding from these boundary conditions, the whole set of vertices was explicitly constructed in \cite{Skvortsov:2022syz,Sharapov:2022faa,Sharapov:2022awp}. This was achieved via  homological perturbation theory starting with an appropriate multiplicative resolution. Even though it is quite easy to show that the vertices obtained this way are well-defined \cite{Sharapov:2022faa,Sharapov:2022awp}, an explicit form is required for practical applications. This is the problem we address in the present paper. We also show that, after an appropriate change of variables, the vertices reveal a remarkable simplicity and allow us to identify the underlying configuration space in a form that is reminiscent of the Formality theorems.

\section{Vertices}
\label{sec:vertices}
For completeness we begin with explicit examples of  low order vertices. In order to present the results in the most compact way we employ the language of symbols of poly-differential operators. The recipe on how to unfold the homological perturbation theory in order to get the actual vertices is briefly explained in Appendix \ref{app:homo}, but see Appendices in \cite{Sharapov:2022faa,Sharapov:2022awp} for more detail. Below we concentrate on the final form of the vertices. 

\paragraph{Poly-differential operators.} Vertices $\mathcal{V}$ and $\mathcal{U}$ encode certain contractions of indices of their arguments, e.g.
\begin{align}
    \mathcal{V}(\omega,\omega,C,\ldots,C)= \sum y_A\cdots y_A \,\omega\fud{A\cdots}{B\cdots M \cdots}\wedge \omega\fud{A\cdots B\cdots}{N\cdots} C^{A\cdots M\cdots N\cdots} \ldots\,,
\end{align}
where we omitted the $\bry$'s. It is convenient to represent such structures via poly-differential operators
\begin{align}
    \mathcal{V}(f_1,\ldots,f_n)&= \mathcal{V}(y, \pl_1,\ldots,\pl_n)\, f_1(y_1)\cdots f_n(y_n) \Big|_{y_i=0}\,.
    \end{align}
We prefer to work with the corresponding symbols, obtained by replacing  the arguments according to $y^A\equiv p_0^A$, $\pl^{y_i}_{A}\equiv p_{A}^i$. The Lorentz symmetry requires the symbols to depend only on $p_{ij}\equiv p_i \cdot p_j\equiv -\epsilon_{AB}p^A_{i}p_{j}^B=p^A_{i}p_{jA}$. These scalars are defined so that $\exp[p_0\cdot p_i]f(y_i)=f(y_i+y)$ represents the shift operator. We will also use the $q$'s for poly-differential operators in $\bry$'s, e.g. $\bry^{A'}\equiv q_0^{A'}$, $\pl^{\bry_i}_{A'}\equiv q_{A'}^i$. We will often omit the sign $|_{y_i=0}$ as well as the arguments of the vertices, writing down only the corresponding symbols. 

\paragraph{$\boldsymbol{\mathcal{V}(\omega,\omega)}$ and higher spin algebra.} Given that the very first $L_\infty$-map is always trivial, the first nontrivial vertex $\mathcal{V}(\omega,\omega)$ defines a Lie algebra since the corresponding $L_\infty$-relation reduces to the Jacobi identity. It turns out that the relevant (Lie) higher spin algebra $\hs$ originates from an associative one, still denoted by $\hs$. The last fact leads to considerable simplifications. 

Indeed, there is a number of arguments that allow one to upgrade the $L_\infty$-structure to $A_\infty$: (a) Chiral Theory admits Yang--Mills gaugings of $U(N)$-type;\footnote{as well as $O(N)$ and $USp(N)$ \cite{Skvortsov:2020wtf}, which are simple reductions.} (b) within the AdS/CFT context Chiral Theory should be dual to a subsector of (Chern--Simons) vector models \cite{Ponomarev:2016lrm,Skvortsov:2018uru,Sharapov:2022awp}, where it is always possible to introduce further global symmetries of the same type. One way or another, the Lie algebra originates from an associative one, $\hs$, via the commutator and in order to account for (a) or (b) one needs to start with a bigger associative algebra $\hs \otimes \mathrm{Mat}_N$. In practice, for $N$ large enough one can recover the initial $\hs$ associative structure from the commutator, i.e., from $\mathcal{V}(\omega,\omega)$. Moreover, all the $L_\infty$-maps (vertices) result from the symmetrization of certain $A_\infty$-maps. It is the latter we will be looking for.

In what follows we assume that the higher spin algebra is of the form $\hs=A_\hhbar \otimes B$. Here, $A_\hhbar$ is the Weyl algebra, that is, the algebra of a polynomial functions $f(\hat y)$ in the  operators $\hat y_A$ subject to the canonical commutation relations $[\hat y_A,\hat y_B]=-2\hhbar \epsilon_{AB}$. Note that all $A_\hhbar$ are isomorphic to each other whenever $\hhbar\neq 0$ and the commutative limit $A_{\lambda=0}$ coincides with $\mathbb{C}[y^A]$. One can also understand $A_\hhbar$ as the result of deformation quantization of the polynomial algebra $A_0$, the quantum product being the Moyal--Weyl star-product. The symbol of the star-product is defined by
\begin{align}\label{hsalgebra}
    \mathcal{V}(f,g)&= \exp{[p_{01}+p_{02}+\hhbar\, p_{12}]}f({y}_1)\, g({y}_2)\Big|_{{y}_i=0} \equiv (f\star g)(y)\,.
\end{align}
The parameter $\hhbar$ has the meaning of the cosmological constant. All vertices of Chiral Theory depend smoothly on $\hhbar$. 

In principle, the factor $B$ may be any associative algebra. However, in order to have a proper $4d$ field theory interpretation, $B$ has to be $A_1\otimes \mathrm{Mat}_N$. Nevertheless, the construction below works for any associative noncommutative $B$ and other possible choices are discussed in Section \ref{sec:conclusions}. We also assume that the product (or trace, wherever needed) over $B$ is taken. For example, all vertices have the factorized form
\begin{align}
    \mathcal{V}(f_1,\ldots ,f_n)&= v(f_1'(y),\ldots , f_n'(y)) \otimes f_1''\ast\cdots \ast f_n''  \,,
\end{align}
where $f_i=f_i'(y) \otimes f_i''$, $f''_i\in B$, and $\ast$ denotes the product in $B$. In case $B=A_1\otimes \mathrm{Mat}_N$, all $\bry$-dependent factors are multiplied via the star-product:
\begin{align}
    f_1''(\bry)\star \cdots \star f_n''(\bry)&= \exp{\left[\sum_{0=i<j=n} q_i \cdot q_j\right]} f_1''(\bry_1)\cdots f_n''(\bry_n)\Big|_{\bry_i=0}\,.
\end{align}
Here $q$ for $\bry$ is the same as $p$ for $y$. Due to additional matrix factors, all $f''_i=\{f''_i(\bry){}\fud{\aAt}{\aBt}\}$ are also multiplied as matrices in the same order as with $\star$.  

As a vector space, the $A_\infty$-algebra of Chiral Theory is given by the sum $\mathbb{A}=\mathbb{A}_0\oplus \mathbb{A}_{-1}$. Coordinates on $\mathbb{A}_{-1}$ and $\mathbb{A}_0$ correspond to $\omega$ and $C$, respectively. Algebraically, the lowest $A_\infty$-relations imply (i) an associative algebra structure on $\mathbb{A}_{-1}$, which is the higher spin algebra $\hs$; (ii) an $\hs$-bimodule structure on $\mathbb{A}_0$.  

\paragraph{$\boldsymbol{\mathcal{U}(\omega,C)}$ and the dual module.} Thanks to the $A_\infty$-structure this bilinear vertex splits into the sum of two vertices\footnote{By a slight abuse of notation $\mathcal{V}(\omega,
\omega, C, \ldots ,C)$ and $\mathcal{U}(\omega, C, \ldots ,C)$ denote the whole collections of vertices/$A_\infty$-products at a given order that differ by the order of the arguments. When a detailed structure is discussed we enumerate various orderings by subscripts. }
\begin{align}
    \mathcal{U}(\omega,C)&=\mathcal{U}_1(\omega,C)+\mathcal{U}_2(C,\omega)
\end{align}
The $A_\infty$-relations imply that $\mathcal{U}_1(\omega,C)$ and $\mathcal{U}_2(C,\omega)$ define an $\hs$-bimodule structure on $C$. Action \eqref{niceaction} suggests that zero-forms $C$ take values in the space dual to the space of one-forms $\omega$, see \cite{Krasnov:2021nsq,Skvortsov:2022syz,Sharapov:2022faa,Sharapov:2022awp}. Therefore, we define the nondegenerate pairing
\begin{align}
    \langle \omega|C \rangle&=-\langle C|\omega \rangle= \exp[p_{12}]\,\omega(y_1)\,C(y_2) \big|_{y_i=0}
\end{align}
between the $\hs$-bimodule of fields $C$ and the higher spin algebra $\hs$ of fields $\omega$. With the help of this pairing we can define the bimodule structure by the following symbols:
\begin{equation}
    \begin{split}
        &\mathcal{U}_1(\omega,C)=+\exp{[\hhbar\, p_{01}+ p_{02}+p_{12}]}\, \omega({y}_1)\, C({y}_2)\Big|_{{y}_i=0}\,,\\
        &\mathcal{U}_2(C,\omega)=-\exp{[p_{01}-\hhbar\, p_{02}-p_{12}]}\, C({y}_1)\, \omega({y}_2)\Big|_{{y}_i=0}\,.
    \end{split}
\end{equation}
Consider, for example, the left action. At $\lambda=0$ the symbol corresponds to $\mathcal{U}_1(\omega,C)(y)=\omega(\pl_y) C(y)$, i.e., the commutative algebra $A_0=\mathbb{C}[y^A]$ acts on the dual space by differential operators.\footnote{For $\lambda=1$ one can recognize the twisted-adjoint action \cite{Vasiliev:1999ba}. The twisted-adjoint representation, however, does not admit the flat limit. It is also not very useful for Chiral Theory with cosmological constant: the zero-form should be treated differently, whereas the twisted-adjoint interpretation suggests to deal with $C$ as an element of $\hs$ and this immediately entails some problems with locality. } 

It is worth noting that the bilinear structure maps defined so far satisfy the boundary conditions imposed by the free limit \eqref{linearizeddata}. The next vertex will generate the trilinear term in \eqref{linearizeddata}, thereby, we do reproduce the $L_\infty$-algebra determined by the free action \eqref{niceaction}.

\paragraph{$\boldsymbol{\mathcal{V}(\omega,\omega,C)}$.} Since the $A_\infty$-algebra is concentrated in only two  degrees, $\mathbb{A}_0\oplus \mathbb{A}_{-1}$, there are three structure maps hidden in $\mathcal{V}(\omega,\omega,C)$:
\begin{align}
    \mathcal{V}(\omega,\omega,C)=\mathcal{V}_1(\omega,\omega,C)+\mathcal{V}_2(\omega,C,\omega)+\mathcal{V}_3(C,\omega,\omega)    \,.
\end{align}
For example, one of the $L_\infty$-relations reads
\begin{align}\label{Linfrel}
        &\mathcal{V}_1(\mathcal{V}(\omega,\omega),\omega,C)-\mathcal{V}(\omega,\mathcal{V}_1(\omega,\omega,C))+\mathcal{V}_1(\omega,\omega,\mathcal{U}_1(\omega,C))-\mathcal{V}_1(\omega,\mathcal{V}(\omega,\omega),C)=0\,.
\end{align}
It originates from the $A_\infty$-relation 
\begin{align}
        &\mathcal{V}_1(\mathcal{V}(a,b),c,u)-\mathcal{V}(a,\mathcal{V}_1(b,c,u))+\mathcal{V}_1(a,b,\mathcal{U}_1(c,u))-\mathcal{V}_1(a,\mathcal{V}(b,c),u)=0\,,
\end{align}
where $a,b,c\in \mathbb{A}_{-1}$ and $u\in \mathbb{A}_0$. It is, of course, much more constraining than the one of $L_\infty$. Indeed, in \eqref{Linfrel} the $\omega$'s, being one-forms, anti-symmetrize over the first three arguments. To solve the $A_\infty$-relation, it is useful to rewrite it in terms of symbols: 
\begin{align*}
0&=-\mathcal{V}_1(p_0+\hhbar\, p_1,p_2,p_3,p_4)e^{p_{01}}+\mathcal{V}_1(p_0,p_1+p_2,p_3,p_4)e^{\hhbar\, p_{12}}\\&\qquad -\mathcal{V}_1(p_0,p_1,p_2+p_3,p_4)e^{\hhbar\, p_{23}}+\mathcal{V}_1(p_0,p_1,p_2,\hhbar\, p_3+p_4)e^{p_{34}}
\end{align*}
and similarly for the rest of the $A_\infty$-relations, some of which mix $\mathcal{V}$ with different orderings of the arguments. The resulting equations are not difficult to solve directly \cite{Skvortsov:2022syz,Sharapov:2022awp}:
\begin{align*}
     \mathcal{V}_1(\omega,\omega,C)&=+p_{12}\, \int_{\Delta_2}\exp[\left(1-u\right) p_{01}+\left(1-v\right) p_{02}+u p_{13}+v p_{23} +\hhbar (1+u-v) p_{12} ]\,, \\
     \mathcal{V}_2(\omega,C,\omega)&=-p_{13}\, \int_{\Delta_2}\exp[\left(1-v\right) p_{01}+\left(1-u\right) p_{03}+v p_{12}-u p_{23}+\hhbar (1-u-v) p_{13}]\\
       &\phantom{=}\,-p_{13}\, \int_{\Delta_2}\exp[\left(1-u\right) p_{01}+\left(1-v\right) p_{03}+u p_{12}-v p_{23}+\hhbar (1-u-v) p_{13}]\,,
    \\
    \mathcal{V}_3(C,\omega,\omega)&=+p_{23}\, \int_{\Delta_2}\exp[\left(1-v\right) p_{02}+\left(1-u\right) p_{03}-v p_{12}-u p_{13}+\hhbar (1+u-v) p_{23} ]\,.
\end{align*}
Here $\Delta_2$ denotes the $2$-simplex $0\leq u\leq v \leq 1$. From the homological perturbation theory point of view, these vertices correspond to\footnote{We refer to Appendix \ref{app:homo} for  basic definitions and to \cite{Sharapov:2022faa,Sharapov:2022awp} for more details on how homological perturbation theory works. }
$$
   \mathcal{V}_1(\omega,\omega,C)=\omega(y) \star h[ \omega(y) \star \Lambda[C] ]|_{z=0}= \begin{tikzcd}[column sep=small,row sep=small]
   & {}& \\
    & \mu\arrow[u]  & \\
    \omega\arrow[ur]  & & \mu\arrow[ul, "h" ']   & \\
    & \omega \arrow[ur]& &\Lambda[C]\arrow[ul]
\end{tikzcd}
$$
its mirror image
$$
   \mathcal{V}_3(C,\omega,\omega)= h[\Lambda[C] \star  \omega(y)  ] \star \omega(y)|_{z=0} = \begin{tikzcd}[column sep=small,row sep=small]
   && {} \\
    && \mu\arrow[u]   \\
    &\mu\arrow[ur, "h"]  & & \omega \arrow[ul]    \\
    \Lambda[C] \arrow[ur]& &\omega\arrow[ul] &&
\end{tikzcd}
$$
and the middle vertex receives contributions from two graphs
\begin{align*}
    \mathcal{V}_2(\omega,C,\omega)&=\omega(y) \star h[  \Lambda[C] \star\omega(y) ]|_{z=0}+ h[  \omega(y)\star  \Lambda[C]] \star \omega(y)|_{z=0}=
\end{align*}
$$
=\begin{tikzcd}[column sep=small,row sep=small]
   & {}& \\
    & \mu\arrow[u]  & \\
    \omega\arrow[ur]  & & \mu\arrow[ul, "h" ']   & \\
    & \Lambda[C] \arrow[ur]& &\omega \arrow[ul]
\end{tikzcd} \oplus\begin{tikzcd}[column sep=small,row sep=small]
   && {} \\
    && \mu\arrow[u]   \\
    &\mu\arrow[ur, "h"]  & & \omega \arrow[ul]    \\
    \omega \arrow[ur]& &\Lambda[C]\arrow[ul] &&
\end{tikzcd}   
$$
Let us illustrate the process of evaluation of a tree on the example of $\mathcal{V}(a,b,c)$ with $a,b\in \mathbb{A}_{-1}$ and $c\in \mathbb{A}_0$, see also \cite{Sharapov:2022faa}. One begins with (here $\varkappa=\exp{(z^Ay_A)}$)
\begin{align}\label{LambdaGuy}
    \Lambda[c]&= dz^{A}z_{A} \int_0^1 t\,dt\, \varkappa(t z, y+p_3)\, c(y_3)\,.
\end{align}
Next, we evaluate the star-product:
\begin{align*}
    b(y)\star \Lambda[c]&= dz^{A}(z_{A}+p^2_{A})\, e^{y p_2} \int_0^1 t\,dt\, \varkappa(t z+t p_2, y+p_3+\hhbar\, p_2)\, b(y_2) c(y_3)\,.
\end{align*}
This is a one-form and we apply $h$ to it:
\begin{align*}
    h[b(y)\star \Lambda[c]]&= (z\cdot p_2)\, e^{y p_2}\int_0^1 dt'\, t\,dt\, \varkappa(tt' z+t p_2, y+p_3+\hhbar\, p_2)\, b(y_2) c(y_3)\,.
\end{align*}
In the last step we evaluate one more product and set $z=0$ to find
\begin{align*}
   a\star h[b\star \Lambda[c]]|_{z=0}&= p_{12}\, e^{y p_1+y p_2}\int_0^1 dt'\, t\,dt\, \varkappa(tt' p_1+t p_2, y+p_3+\hhbar\, p_1+\hhbar\, p_2)\, a(y_1) b(y_2) c(y_3).
\end{align*}
After renaming $y\rightarrow p_0$ and changing the integration domain to the $2d$ simplex $\Delta_2$, $u=tt'$, $v=t$, we arrive at
\begin{align*}
   \mathcal{V}_1(a,b,c)&= p_{12}\, e^{p_{01}+p_{02}} \int_{\Delta_2} \varkappa(u p_1+v p_2, p_0+p_3+\hhbar\, p_1+\hhbar\, p_2)\, a(y_1) b(y_2) c(y_3)\big|_{y_i=0}\,.
\end{align*}
This coincides with $\mathcal{V}_1$ on the previous page. We will derive the result for an arbitrary tree later on.

\paragraph{$\boldsymbol{\mathcal{U}(\omega,C,C)}$.} The next group of structure maps is
\begin{align}
    \mathcal{U}(\omega,C,C)=\mathcal{U}_1(\omega,C,C)+\mathcal{U}_2(C,\omega,C)+\mathcal{U}_3(C,C,\omega)   \,.
\end{align}
The $A_\infty$-relations can also be written down and solved directly \cite{Skvortsov:2022syz,Sharapov:2022awp}. It is remarkable that one does not have to do that. There is a canonical way to generate all $\mathcal{U}$-vertices from $\mathcal{V}$-vertices. We refer to this recipe as a duality map since it relies on the fact that $\mathbb{A}_0 =(\mathbb{A}_{-1})^*$, as $\hs$-bimodules. This is a particular manifestation of a (hidden) cyclicity of the underlying $A_\infty$-algebra $\hat{\mathbb{A}}$; we discuss it in Appendix  \ref{CY}.  Given a $\mathcal{V}$-vertex at some order, one can canonically pair it with $C$ to build a scalar. By cyclicity/duality,
\begin{equation}\label{DR}
    \langle \mathcal{V}(\omega,\omega,C,\ldots,C)|C\rangle=\langle \omega|\mathcal{U}(\omega, C,\ldots,C)\rangle\,.
\end{equation}
A consistent $\mathcal{U}$-vertex can be obtained by peeling off one $\omega$, which is again a canonical operation. 

The duality map gives automatically local $\mathcal{U}$-vertices, provided that the $\mathcal{V}$-vertices are local.\footnote{There is another canonical recipe \cite{Vasiliev:1988sa} in case $\mathbb{A}_0\simeq \mathbb{A}_{-1}$. However, this one gives nonlocal $\mathcal{U}$'s out of local $\mathcal{V}$'s. This recipe is built-in into \cite{Vasiliev:1990cm} and leads to one of the open problems pointed out in  \cite{Boulanger:2015ova}.} In the simplest case we find
\besubeqs
\begin{align}
    \mathcal{U}_1(p_0,p_1,p_2,p_3)&=+ \mathcal{V}_1(-p_3,p_0,p_1,p_2)\,,\\
    \mathcal{U}_2(p_0,p_1,p_2,p_3)&=- \mathcal{V}_2(-p_1,p_2,p_3,p_0)\,,\\
    \mathcal{U}_3(p_0,p_1,p_2,p_3)&= -\mathcal{V}_3(-p_1,p_2,p_3,p_0)\,.
\end{align}
\esubeqs
For example, the first one reads
\begin{align}\label{uone}
    \mathcal{U}_1&=p_{01} \int_{\Delta_2}\exp \left[\hhbar\left(1+u-v\right)   p_{01}+u p_{02}+\left(1-u\right) p_{03}+v p_{12}+\left(1-v\right) p_{13}\right]\,.
\end{align}
It is a local vertex because no $p_{23}$ enters the exponent. For completeness, the other two read
\begin{align*}
    \mathcal{U}_2&=-p_{02} \int_{\Delta_2}\exp \left[\left(1-v\right) p_{01}+v p_{03}-\left(1-u\right) p_{12}+u p_{23}-\hhbar \left(1-u-v\right)  p_{02}\right]\\
    &\phantom{=}-p_{02} \int_{\Delta_2}\exp \left[\left(1-u\right) p_{01}+u p_{03}-\left(1-v\right) p_{12}+v p_{23}-\hhbar\left(1-u-v\right)  p_{02}\right]\,,\\
    \mathcal{U}_3&= +p_{03} \int_{\Delta_2}\exp \left[\left(1-u\right) p_{01}+u p_{02}-\left(1-v\right) p_{13}-v p_{23}-\hhbar\left(1+u-v\right)   p_{03}\right]\,.
\end{align*}

\paragraph{$\boldsymbol{\mathcal{V}(\omega,\omega,C,C)}$.} The brute-force approach above is less efficient starting from this vertex. First of all, there are $6$ different orderings for $\omega^2C^2$. Secondly, the defining equations ($A_\infty$-algebra relations) are inhomogeneous w.r.t. the sought-for quartic vertices. A complete all-order solution follows immediately from homological perturbation theory \cite{Sharapov:2022faa,Sharapov:2022awp}.\footnote{The light-cone approach operates only with the physical degrees of freedom and, for this reason, may allow one to see certain structures that are not self-evident in a given covariant approach, see e.g. \cite{Metsaev:1991mt,Metsaev:1991nb,Ponomarev:2016lrm,Skvortsov:2018uru}. It was shown in \cite{Metsaev:1991mt,Metsaev:1991nb,Metsaev:2018xip} that the cubic vertices can be split into chiral and anti-chiral ones. The cubic vertices from the Lagrangian point of view have overlap with a great deal of the vertices, $\mathcal{V}(\omega,\omega)$, $\mathcal{U}(\omega,C)$, $\mathcal{V}(\omega,\omega,C)$, $\mathcal{U}(\omega,C,C)$ and even $\mathcal{V}(\omega,\omega,C,C)$. Indeed, we should be looking at all vertices that have any number of background insertions $\omega_0$ and are bilinear in the fluctuations. For example, $\mathcal{V}(\omega_0,\omega_0,C,C)$ is a kind of stress-tensor's contributions. For all these vertices, it should be possible to find a split into chiral and anti-chiral ones plus, possibly, other contributions that come from higher orders in the Lagrangian. Therefore, we expect that various truncations/subsectors like chiral/self-dual/holomorphic are closely related to each other, if not identical at these orders. In this regard it is worth mentioning some partial low order results in the literature \cite{Didenko:2018fgx,Didenko:2019xzz,Didenko:2020bxd,Gelfond:2021two}. } The vertices at this order read
{\allowdisplaybreaks
\begin{align*}
    \mathcal{V}_1(\omega,\omega,C,C)&=(p_{12})^2\int_{\mathcal{D}_1}\exp((1-u_1-u_2)p_{01}+(1-v_1-v_2)p_{02}+u_1 p_{13}+u_2 p_{14}+v_1 p_{23}+v_2 p_{24}+\\
    &+\lambda p_{12}(1+u_1+u_2-v_1-v_2+u_1 v_2-u_2 v_1))\,,\\
    \mathcal{V}_2(\omega,C,\omega,C)&=-(p_{13})^2\int_{\mathcal{D}_1}\exp(p_{01}(1-u_1-u_2)+(1-v_1-v_2)p_{03}+u_2 p_{12}+u_1 p_{14}-v_2 p_{23}+v_1 p_{34}+\\
    &+\lambda p_{13}(1+u_1-u_2-v_1-v_2-u_1 v_2+u_2 v_1))\\
    &-(p_{13})^2\int_{\mathcal{D}_1}\exp(p_{01}(1-u_1-u_2)+(1-v_1-v_2)p_{03}+u_1 p_{12}+u_2 p_{14}-v_1 p_{23}+v_2 p_{34}+\\
    &+\lambda p_{13}(1-u_1+u_2-v_1-v_2+u_1 v_2-u_2 v_1)) + \\
    &-(p_{13})^2\int_{\mathcal{D}_2}\exp((1-u^R-v^L)p_{01}+(1-u^L-v^R)p_{03}+v^L p_{12}+u^R p_{14}-u^L p_{23}+v^R p_{34}\\
    &+\lambda p_{13}(1-u^L+u^R-v^L-v^R-u^L u^R+v^L v^R)) \,,\\
    \mathcal{V}_3(\omega,C,C,\omega)&=(p_{14})^2\int_{\mathcal{D}_1}\exp((1-u_1-u_2)p_{01}+(1-v_1-v_2)p_{04}+u_2 p_{12}+u_1 p_{13}-v_2 p_{24}-v_1 p_{34}+\\
    &+\lambda p_{14}(1-u_1-u_2-v_1-v_2-u_1 v_2+u_2 v_1))-\\
    &+(p_{14})^2\int_{\mathcal{D}_1}\exp((1-v_1-v_2)p_{01}+(1-u_1-u_2)p_{04}+v_1 p_{12}+v_2 p_{13}-u_1 p_{24}-u_2 p_{34}+\\
    &+\lambda p_{14}(1-u_1-u_2-v_1-v_2-u_1 v_2+u_2 v_1))-\\
    &+(p_{14})^2\int_{\mathcal{D}_2}\exp((1-u^R-v^L)p_{01}+(1-u^L-v^R)p_{04}+v^L p_{12}+u^R p_{13}-u^L p_{24}-v^R p_{34}\\
    &+\lambda p_{14}(1-u^L-u^R-v^L-v^R-u^L u^R+v^L v^R)) \,,\\
    \mathcal{V}_4(C,\omega,\omega,C)&=(p_{23})^2\int_{\mathcal{D}_2}\exp((1-u^R-v^L)p_{02}+(1-u^L-v^R)p_{03}-v^L p_{12}-u^L p_{13}+u^R p_{24}+v^R p_{34}\\
    &+\lambda p_{23}(1+u^L+u^R-v^L-v^R-u^L u^R+v^L v^R))\,,\\
    \mathcal{V}_5(C,\omega,C,\omega)&=-(p_{24})^2\int_{\mathcal{D}_1}\exp((1-v_1-v_2)p_{02}+(1-u_1-u_2)p_{04}-v_2 p_{12}-u_2 p_{14}+v_1 p_{23}-u_1 p_{34}+\\
    &+\lambda p_{24}(1-u_1+u_2-v_1-v_2+u_1 v_2-u_2 v_1))+\\
    &-(p_{24})^2\int_{\mathcal{D}_1}\exp((1-v_1-v_2)p_{02}+(1-u_1-u_2)p_{04}-v_1 p_{12}-u_1 p_{14}+v_2 p_{23}-u_2 p_{34}+\\
    &+\lambda p_{24}(1+u_1-u_2-v_1-v_2-u_1 v_2+u_2 v_1))+\\
    &-(p_{24})^2\int_{\mathcal{D}_2}\exp((1-u^R-v^L)p_{02}+(1-u^L-v^R)p_{04}-v^L p_{12}-u^L p_{14}+u^R p_{23}-v^R p_{34}\\
    &+\lambda p_{24}(1+u^L-u^R-v^L-v^R-u^L u^R+v^L v^R))\,,\\
    \mathcal{V}_6(C,C,\omega,\omega)&=(p_{34})^2\int_{\mathcal{D}_1}\exp((1-v_1-v_2)p_{03}+(1-u_1-u_2)p_{04}-v_2 p_{13}-u_2 p_{14}-v_1 p_{23}-u_1 p_{24}+\\
    &+\lambda p_{34}(1+u_1+u_2-v_1-v_2+u_1 v_2-u_2 v_1))\,,\\
\end{align*}}\noindent
where we have introduced the integration variables
\begin{align*}
    u_1&\equiv\frac{t_1t_2(1-t_3)t_4}{1-t_1t_2t_3}\,, & v_1&\equiv\frac{t_1(1-t_2t_3)}{1-t_1t_2t_3} \,,\\
    u_2&\equiv\frac{(1-t_1t_2)t_3t_4}{1-t_1t_2t_3}\,, & v_2&\equiv\frac{(1-t_1)t_3}{1-t_1t_2t_3} \,,
\end{align*}
which correspond to the domain of integration $\mathcal{D}_1$ and
\begin{align*}
    u^L&\equiv\frac{t_1t_2(1-t_3)}{1-t_1t_2t_3t_4}\,, & v^L&\equiv\frac{t_1(1-t_2t_3t_4)}{1-t_1t_2t_3t_4}\,,\\
    u^R&\equiv\frac{(1-t_1)t_3t_4}{1-t_1t_2t_3t_4}\,, & v^R&\equiv\frac{t_3(1-t_1t_2t_4)}{1-t_1t_2t_3t_4} 
\end{align*}
for the domain $\mathcal{D}_2$. All times $t_i$ are integrated over $[0,1]$. In terms of $u$'s and $v$'s the domains of integration can be found by inverting the above relations. We start with $\mathcal{D}_1$:
\begin{align*}
    t_1&=\frac{u_2v_1(1-v_1-v_2)+u_1v_2(v_1+v_2)}{u_1v_2+u_2(1-v_1-v_2)} \,, & t_3&=\frac{v_2}{1-v_1}  \,, \\
    t_2&=\frac{u_1v_2}{u_2v_1(1-v_1-v_2)+u_1v_2(v_1+v_2)} \,, & t_4&=u_1+u_2\frac{1-v_1}{v_2} \,.
\end{align*}
The fact that the $t_i$'s take values in the interval $[0,1]$ translates into restrictions on the $u$ and $v$ variables. In Appendix \ref{app:domain}, we prove that these variables belong to a subinterval of $[0,1]$. Some of these restrictions merely confirm this. The other restrictions 
\begin{align*}
    0\leq v_2 \leq 1\,,\qquad 0\leq u_1\leq v_1\leq 1-v_2\,,\qquad \frac{u_1}{v_1}\leq \frac{u_2}{v_2}\leq \frac{1-u_1}{1-v_1}
\end{align*}
define the integration domain as
\begin{align*}
    \int_{\mathcal{D}_1}&\equiv\int_0^1 dv_2 \int_0^{1-v_2} dv_1 \int_0^{v_1} du_1 \int_{\frac{u_1 v_2}{v_1}}^{v_2\frac{1-u_1}{1-v_1}} du_2 \,.
\end{align*}
For $\mathcal{D}_2$ we obtain
\begin{align*}
    t_1&=\frac{u^Lu^R-v^Lv^R+v^L}{1-v^R} \,, & t_3&=\frac{u^Lu^R-v^Lv^R+v^R}{1-v^L} \,, \\
    t_2&=\frac{u^L}{u^Lu^R-v^Lv^R+v^L} \,, & t_4&=\frac{u^R}{u^Lu^R-v^Lv^R+v^R} \,.
\end{align*}
This gives  the restrictions
\begin{align*}
    &0 \leq u^L \leq 1 \,, & &0 \leq u^L \leq v^L \leq 1-u^R \,,\\
    &\frac{u^L}{v^L} \leq \frac{1-v^R}{1-u^R} \,, & &\frac{u^R}{v^R} \leq \frac{1-v^L}{1-u^L}\,,
\end{align*}
which determine the domain of integration to be
\begin{align*}
    \int _{\mathcal{D}_2}&\equiv \int_0^1 du^L \int_0^{1-u^L}du^R \int_{u^L}^{1-u^R}dv^L \int_{u^R\frac{1-u^L}{1-v^L}}^{1-\frac{u^L(1-u^R)}{v^L}} dv^R \,.
\end{align*}
Hence, both $\mathcal{D}_1$ and $\mathcal{D}_2$ are compact and the corresponding integrals converge. In the language of trees emerging from homological perturbation theory, there are only two nontrivial topologies given by
$$
   G_1=\omega \star h[ h[ \omega \star \Lambda[C] ] \star \Lambda[C]]|_{z=0}= \begin{tikzcd}[column sep=small,row sep=small]
   &{}&\\
    & \arrow[u] \mu  & \\
    \omega\arrow[ur]&&  \arrow[ul,"h"']\mu &  \\
    & \arrow[ur,"h"]\mu &&\arrow[ul]\Lambda[C] \\
    \arrow[ur]\omega & & \arrow[ul]\Lambda[C] &
\end{tikzcd}
$$
and
$$
   G_2= h[ \omega \star \Lambda[C] ]\star  h[ \omega \star \Lambda[C]]|_{z=0}= \begin{tikzcd}[column sep=small,row sep=small]
   &&&{}&&&\\
    &&& \arrow[u]\mu  &&& \\
    & \mu\arrow[urr,"h"]& && & \arrow[ull,"h"']\mu &  \\
    \omega\arrow[ur]&& \arrow[ul]\Lambda[C]  & &  \omega\arrow[ur]&&\arrow[ul]\Lambda[C]
\end{tikzcd}
$$
All other graphs can be derived from these by swapping incoming edges at any vertex. Evaluation of all diagrams leads to the quartic vertices above. For Chiral HiSGRA with vanishing cosmological constant all quartic vertices have been written down in \cite{Sharapov:2022faa}.

\paragraph{$\boldsymbol{\mathcal{U}(\omega,C,C,C)}$.} This group of structure maps can effortlessly be obtained via the duality map. For example, 
\begin{align*}
    \mathcal{U}_1(\omega,C,&C,C)(p_0,p_1,p_2,p_3,p_4)=G_1(-p_4,p_0,p_1,p_2,p_3)=\\
    &=(p_{01})^2\int_{\mathcal{D}_1}\exp\big[u_1 p_{02}+u_2 p_{03}+(1-u_1-u_2)p_{04}+v_1 p_{12}+v_2 p_{13}+(1-v_1-v_2)p_{14}\\
    &+\lambda (1+u_1+u_2-v_1-v_2+u_1 v_2-u_2 v_1)p_{01}\big]\,.
\end{align*}
All other $\mathcal{U}$-vertices can be derived in a similar manner. This completes the low order analysis, which can be useful for a number of reasons: to get an idea of how interaction vertices look like; to compute low order holographic correlation functions; to be compared with the all order analysis that follows. The rest of this section is occupied with the evaluation of all trees coming out of the homological perturbation theory. 

\subsection{All vertices with vanishing cosmological constant}
\label{sec:flat}
Thanks to the duality map, it suffices to work out vertices of type ${\mathcal{V}(\omega,C,\ldots,C,\omega,C,\ldots,C)}$, but we will provide a complete description of all non-zero vertices. Given the specific nuts and bolts of  homological perturbation theory it can be shown \cite{Sharapov:2022faa} that only a very limited class of trees makes nonvanishing contributions. They can be described as `trees with two branches'. Either branch has one leaf decorated by an element of $\mathbb{A}_{-1}$ and the other leaves by elements of $\mathbb{A}_0$. Such trees can be depicted as
$$
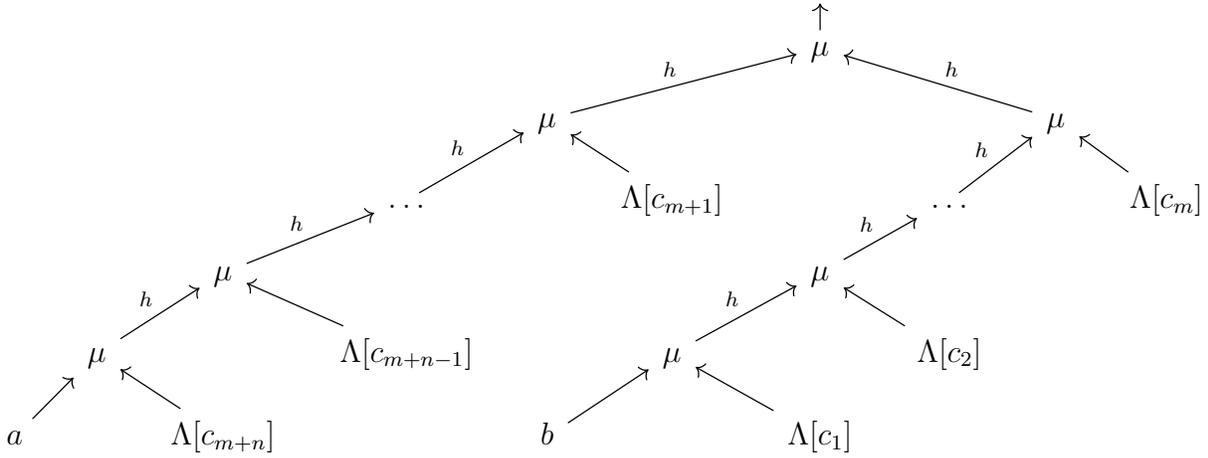
\begin{tikzcd}[column sep=small,row sep=small]
&            & &                   &                           &       & {} &       &               &\\
&            & &                   &                           &       & \arrow[u]\mu &       &               &\\
&            & &                   &\arrow[urr,"h"]\mu    &       &                           &       &\arrow[ull,"h"']\mu&\\
&            & &\dots\arrow[ur,"h"]    &                           &\arrow[ul]\Lambda[c_{m+1}]      &                           &\dots\arrow[ur,"h"]  &              &\arrow[ul]\Lambda[c_m]\\
&           &\mu\arrow[ur,"h"]      &   & & &\mu\arrow[ur,"h"] & &   &\\
&\mu\arrow[ur,"h"] & & \arrow[ul]\Lambda[c_{m+n-1}]                  &                           &       \mu\arrow[ur,"h"]&  &\arrow[ul]\Lambda[c_{2}]                 &               &\\
a\arrow[ur]&            &\arrow[ul]\Lambda[c_{m+n}]                & &                           b\arrow[ur]&       &       \arrow[ul]\Lambda[c_{1}]& & &                           
\end{tikzcd}
$$
with $c_i \in \mathbb{A}_0$ and $a,b \in \mathbb{A}_{-1}$. As a first step we need to understand what a single branch of arbitrary length looks like, after which we can join two such branches together to obtain a tree. In general, leaves with $c_i$ can be attached at the left or at the right, which results in a variety of trees for a certain choice of the length of the branches. Our approach is to construct trees with all these leaves attached on the right and then find a recipe to derive all permutations from this.  In this section, we are only concerned with integrands and do not care about the domains of integration in terms of the new variable.  We will return to the question of domain in Section \ref{sec:config}. Otherwise, the initial integration variables that emerge from homological perturbation theory, $t_i$, are integrated over $[0,1]$.

A single branch of length $n$ has the form
$$
B_n=h[\dots h[h[a\star \Lambda[c_1]]\star  \Lambda[c_2]]\star\dots\star \Lambda[c_n]]=\hspace{-2cm}\begin{tikzcd}[column sep=small,row sep=small]
& & & & {} & &\\
& & & & \arrow[u]\mu & &\\
& & & \arrow[ur,"h"]\mu && \arrow[ul]\Lambda[c_n] &\\
& & \arrow[ur,"h"] \dots && \arrow[ul]\Lambda[c_{n-1}] & &\\
& \arrow[ur,"h"]\mu && \arrow[ul]\Lambda[c_2] & & &\\
\arrow[ur]a && \arrow[ul]\Lambda[c_1] & & & &
\end{tikzcd}
$$
The low-order considerations  suggest that such a branch is evaluated as
\begin{align} \label{ansatz}
    B_n=&(zp_1)^n\int \exp\Big[(1-V_n)(yp_1)+U_n(zy)+\sum_{i=1}^{n}u_{n,i}(zp_{i+1})+\sum_{i=1}^{n}v_{n,i}p_{1,i+1}\Big]\,,
\end{align}
where 
\begin{align*}
    U_n&=\sum_{i=1}^{n}u_{n,i} \,, &    V_n&=\sum_{i=1}^{n}v_{n,i}\,,
\end{align*}
and $u_{n,i}$, $v_{n,i}$ are the integration variables with $i=1,\dots,n$. To verify this ansatz, we attach another leaf decorated by $c_{n+1}$ to the right of the branch, creating a branch of length $n+1$, which then reads
\begin{align} \label{B_{n+1}}
    \begin{aligned}
    B_{n+1}&=\frac{t_{2n+1} t_{2n+2}^n(1-t_{2n+1})^n(1-V_n)}{(1-t_{2n+1}U_n)^{n+3}}(zp_1)^{n+1}\times\\
    &\times\int\exp\Big[\frac{(1-t_{2n+1})(1-V_n)}{1-t_{2n+1}U_n}(yp_1)+\frac{(1-t_{2n+1})U_n+t_{2n+1}(1-U_n)}{1-t_{2n+1}U_n}t_{2n+2}(zy)+\\
    &+\frac{(1-t_{2n+2})t_{2n+2}}{1-t_{2n+1}U_n}\sum_{i=1}^nu_{n,i} (zp_{i+1})+\frac{1-U_n}{1-t_{2n+1}U_n}t_{2n+1}t_{2n+2}(zp_{n+2})+\\
    &+\sum_{i=1}^n (v_{n,i}-u_{n,i}\frac{t_{2n+1}(1-V_n)}{1-t_{2n+1}U_n})p_{1,i+1}+\frac{1-V_n}{1-t_{2n+1}U_n}t_{2n+1}p_{1,n+2}\Big]\,.
    \end{aligned}
\end{align}
One can bring this into much simpler form
\begin{align} \label{B_{n+1}newcoordinates}
    B_{n+1}&=(zp_1)^{n+1}\int\exp\Big[(1-V_{n+1})(yp_1)+U_{n+1}(zy)+\sum_{i=1}^{n+1}u_{n+1,i}(zp_{i+1})+\sum_{i=1}^{n+1}v_{n+1,i}p_{1,i+1}\Big]\,,
\end{align}
where the new integration variables are given by the recurrence relations
\begin{align} \label{newcoordinates}
    \begin{aligned}
    u_{n+1,i}&\equiv\frac{(1-t_{2n+1})t_{2n+2}}{1-t_{2n+1}U_n}u_{n,i} \,, & i=0,1,\dots,n \,, \\
    u_{n+1,n+1}&\equiv\frac{1-U_n}{1-t_{2n+1}U_n}t_{2n+1}t_{2n+2} \,, \\
    v_{n+1,i}&\equiv v_{n,i}-u_{n,i}\frac{t_{2n+1}(1-V_n)}{1-t_{2n+1}U_n} \,, & i=0,1,\dots,n \,, \\
    v_{n+1,n+1}&\equiv\frac{t_{2n+1}(1-V_n)}{1-t_{2n+1}U_n}  \,,
    \end{aligned}
\end{align}
where we have to set $U_0=V_0=0$ to match our initial values
\begin{align*}
    u_{1,1}&=t_1t_2 & v_{1,1}&=t_1 \,.
\end{align*}
All the $t_i$'s run from $0$ to $1$. In Appendix \ref{app:allorders}, we prove that the Jacobian associated with the change of variables from $\{u_{n,1},v_{n,1},\dots,u_{n,n},v_{n,n},t_{2n+1},t_{2n+2}\}$ to $\{u_{n+1,1},v_{n+1,1},\dots, u_{n+1,n+1},v_{n+1,n+1}\}$ is exactly the prefactor in \eqref{B_{n+1}}. Since \eqref{B_{n+1}newcoordinates} fits the ansatz \eqref{ansatz}, we conclude that the ansatz is correct for all branches. We also make the observation that the variables satisfy the remarkable chain of inequalities,
\begin{align*}
    \frac{u_{n,1}}{v_{n,1}} \leq \frac{u_{n,2}}{v_{n,2}} \leq \dots \leq \frac{u_{n,n}}{v_{n,n}} \leq \frac{1-U_{n}}{1-V_{n}} \,,
\end{align*}
which is proven in Appendix \ref{app:fulldomain}. This pattern allows one to easily retrieve the domain of integration associated to this choice of variables.

We can now compute a tree by evaluating the star-product of two branches with length $n$ and $m$. In order to obtain the most symmetric form, assume that the left branch contains only zero-forms attached to the left and we denote this branch by $\bar{B}_n$. This does not limit the generality: for $\lambda=0$ attaching zero-forms to the left or right gives the same result since the product is commutative. An important remark is that notation eventually becomes very cumbersome if we want the labels on $p_{ij}$ to consistently refer to the position of the elements $a,b \in \mathbb{A}_0$ and $c_1,\dots,c_n \in \mathbb{A}_{-1}$, read from left to right. Therefore, it is convenient to always assign $p_1,p_2$ and $a(y_1), b(y_2)$ to the first and second one-form, respectively, and assign $p_3,\dots,p_{n+2}$ and $c_1(y_3),\dots,c_{n}(y_{n+2})$ to elements of $\mathbb{A}_0$ based on the position on the branches that they originated from, starting from the bottom of the right branch, to the top and then from the top of the left branch to the bottom. We then leave the reshuffling of the labels according to the respective positions as seen in the tree as the last step in the recipe of finding vertices. Vertices should be $z$-independent, so we set $z=0$ at the end. This gives
\begin{align} \label{tree}
    \begin{aligned}
    \bar{B}_n\star B_m|_{z=0}&=\frac{(-1)^n(1-V^R_m)^n(1-V_n^L)^m}{(1-U_n^LU^R_m)^{n+m+2}}p_{12}^{n+m}\times\\
    &\times\int\exp\Big[\frac{(1-U^R_m)(1-V_n^L)}{1-U_n^LU^R_m}p_{01}+\frac{(1-V^R_m)(1-U_n^L)}{1-U_n^LU^R_m}p_{02}\\
    &+\frac{1-V_n^L}{1-U_n^LU^R_m}\sum_{i=1}^{m}u_{m,i}^R p_{1,2+i}+\sum_{i=1}^n\big( v_{n,i}^L-u_{n,i}^L\frac{U^R_m(1-V_n^L)}{1-U_n^LU^R_m}\big)p_{1,m+n+3-i}\\
    &+\sum_{i=1}^m\big( v_{m,i}^R-u_{m,i}^R\frac{U_n^L(1-V^R_m)}{1-U_n^LU^R_m}\big) p_{2,2+i} +\frac{1-V^R_m}{1-U_n^LU^R_m}\sum_{i=1}^n u_{n,i}^L p_{2,m+n+3-i}\Big]\,.
    \end{aligned}
\end{align}
Here we distinguish between variables coming from the left and the right branch by the superscripts $L$ and $R$, as both branches have their own set of recurrence relations \eqref{newcoordinates}, in which the $t_i$'s in the left branch run from $t_{m+1}$ to $t_{n+m}$, going from top to bottom. To simplify \eqref{tree} we introduce new variables
\begin{align} \label{newcoordinatestwobranches}
    \begin{aligned}
        r^L_{n,i}&\equiv \frac{1-V^R_m}{1-U_n^LU^R_m}u_{n,i}^L \,, & r^R_{m,i}&\equiv \frac{1-V_n^L}{1-U_n^LU^R_m}u_{m,i}^R \,, \\
        s^L_{n,i}&\equiv v_{n,i}^L-u_{n,i}^L\frac{U^R_m(1-V_n^L)}{1-U_n^LU^R_m} \,, & s^R_{m,i}&\equiv v_{m,i}^R-u_{m,i}^R\frac{U_n^L(1-V^R_m)}{1-U_n^LU^R_m}\,,
    \end{aligned}
\end{align}
which allows us to rewrite \eqref{tree} as
\begin{align} \label{twobranches}
    \begin{aligned}
    \bar{B}_n\star B_m|_{z=0}&=(-1)^n p_{12}^{n+m}\int \exp\Big[\big(1-\sum_{i=1}^n s^L_{n,i}-\sum_{i=1}^m r^R_{m,i}\big)p_{01}+\big(1-\sum_{i=1}^m s^R_{m,i}-\sum_{i=1}^n r^L_{n,i}\big)p_{02}+\\
    &+\sum_{i=1}^m r^R_{m,i}p_{1,2+i}+\sum_{i=1}^n s^L_{n,i}p_{1,m+n+3-i}+\sum_{i=1}^m s^R_{m,i}p_{2,2+i}+\sum_{i=1}^n r^L_{n,i}p_{2,m+n+3-i}\Big]\,.
    \end{aligned}
\end{align}
In Appendix \ref{app:allorders}, we show that the Jacobian resulting from the change of variables from the coordinates $\{u^L_{n,1},\dots,v^L_{n,n},u^R_{m,1},\dots,v^R_{m,m}\}$ to $\{r^L_{n,1},\dots,s^L_{n,n},r^R_{m,1},\dots,s^R_{m,m}\}$ is exactly the prefactor in \eqref{tree}.

In order to specify a domain of integration in (\ref{twobranches}), we rename the variables as
\begin{align}\label{rename}
\begin{aligned}
    \{u_1,\dots,u_m,u_{m+1},u_{m+2},\dots,u_{m+n},u_{m+n+1}\} &\\= \{r^R_{m,1},\dots,r^R_{m,m},&1-\sum_{i=1}^n s^L_{n,i}-\sum_{i=1}^{m} r^R_{m,i},s^L_{n,n},\dots,s^L_{n,1}\} \,, \\
    \{v_1,\dots,v_{m},v_{m+1},v_{m+2},\dots,v_{m+n},v_{m+n+1}\}&\\ = \{s^R_{m,1},\dots,s^R_{m,m}, &1-\sum_{i=1}^n r^L_{n,i}-\sum_{i=1}^m s^R_{m,i},r^L_{n,n},\dots,r^L_{n,1}\} \,,
    \end{aligned}
\end{align}
where $u_{m+n+1}=1-\sum_{i=1}^{m+n}u_i$ and $v_{m+n+1}=1-\sum_{i=1}^{m+n}v_i$. In Appendix \ref{app:fulldomain}, we prove that these variables satisfy the inequalities 
\begin{align}\label{inequalities}
    \frac{u_1}{v_1} \leq \frac{u_2}{v_2} \leq \dots \leq \frac{u_{m+n}}{v_{m+n}} \leq \frac{u_{m+n+1}}{v_{m+n+1}} \,.
\end{align}
Now \eqref{twobranches} takes the form
\begin{align} \label{twobranches2}
    \begin{aligned}
    \bar{B}_n\star B_m|_{z=0}&=(-1)^n (p_{12})^{n+m} \int \exp\Big[ u_{m+1} p_{01} + v_{m+1} p_{02} + (1-U_{m+n}) p_{1,m+n} + \\
    &+ (1-V_{m+n}) p_{2,m+n} +\sum_{i=1}^m u_i p_{1,2+i} + \sum_{i=1}^m v_i p_{2,2+i} + \sum_{i=1}^{n-1} u_{m+1+i} p_{1,m+2+i} +\\
    &+ \sum_{i=1}^{n-1} v_{m+1+i} p_{2,m+2+i} \Big]\,.
    \end{aligned}
\end{align}

\paragraph{Constructing vertices.}

There are still a few differences between the trees that we have constructed above and the vertices that solve for the $A_\infty$-relations. Above we associated $p_1$ and $p_2$ with the two leaves decorated by elements of $\mathbb{A}_{-1}$ and the other $p_i$ acted on the $c_i$ that were labeled from bottom right to bottom left on the branches. However, in the expressions for vertices the $p_i$'s are assigned from left to right. Moreover, we have only considered trees with elements of $\mathbb{A}_0$ attached to the left(right) on the left(right) branch. Obviously, general vertices are not restricted to this choice. As will become clear in the next section, the only change as compared to \eqref{twobranches2} is by relabeling of the $p_{ij}$'s when elements of $\mathbb{A}_0$ are attached differently in the absence of a cosmological term. 

\begin{figure}[h]
\centering
\begin{tikzpicture}[scale=1.1]
\draw[-Latex,thick] (6.5,1) to[out=100,in=-10]  (3.5,4);
\draw[thick]  (-0.5,0) -- (6.5,0);
\draw[-,thick] (3,3) -- (3,3.5);
\coordinate [label=above:$c_{m+1}$] (B) at (3,3.5);
\draw[-Latex,thick] (1,1) -- (1,0);
\draw[-]  (1,1) -- (3,3) -- (5,1);
\coordinate [label=below:$a$] (B) at (1,0);
\draw[-Latex]  (1.2,1.2) to[out=-20,in=90] (1.6,0);
\coordinate [label=below:$c_0$] (B) at (1.6,0);
\draw[-Latex]  (1.4,1.4) to[out=150,in=90] (0.35,0);
\coordinate [label=below:$c_{m+n}$] (B) at (0.35,0);
\draw[-Latex]  (1.6,1.6) to[out=-20,in=90] (2.2,0);
\coordinate [label=below:$c_{m+n-1}$] (B) at (2.4,0);
\draw[-Latex]  (1.8,1.8) to[out=150,in=90] (-0.4,0);
\coordinate [label=below:$c_*$] (B) at (-0.4,0);
\draw[-]  (2.0,2.0) to[out=-20,in=100] (2.8,1.2);
\draw[-]  (2.2,2.2) to[out=150,in=50] (0,1.6);
\draw[-Latex,thick] (5,1) -- (5,0);
\coordinate [label=below:$b$] (B) at (5,0);
\draw[-Latex]  (4.8,1.2) to[out=20,in=90] (5.6,0);
\coordinate [label=below:$c_1$] (B) at (5.6,0);
\draw[-Latex]  (4.6,1.4) to[out=200,in=90] (4.3,0);
\coordinate [label=below:$c_2$] (B) at (4.3,0);
\draw[-Latex]  (4.4,1.6) to[out=20,in=90] (6.2,0);
\coordinate [label=below:$c_3$] (B) at (6.2,0);
\draw[-Latex]  (4.2,1.8) to[out=200,in=90] (3.6,0);
\coordinate [label=below:$c_{4}$] (B) at (3.6,0);
\draw[-]  (3.8,2.2) to[out=200,in=90] (3.2,1.2);
\draw[-]  (3.4,2.6) to[out=20,in=140] (5.0,2.4);
\node[] at (3,0.7) {$T$};
\end{tikzpicture}\,
\begin{tikzpicture}[scale=1.1]
\node[] at (2,0.7) {$T_0$};
\draw[-Latex,thick] (6,1) to[out=100,in=-10]  (2.5,4);
\draw[thick]  (0.5,0) -- (6,0);
\draw[-,thick] (2,3) -- (2,3.5);
\coordinate [label=above:$c_0$] (B) at (2,3.5);
\draw[-]  (1,1) -- (2,3) -- (3,1);
\draw[-Latex,thick] (1,1) -- (1,0);
\coordinate [label=below:$a$] (B) at (1,0);
\draw[-Latex,thick] (3,1) -- (3,0);
\coordinate [label=below:$b$] (B) at (3,0);
\draw[-Latex]  (2.8,1.4) to[out=20,in=90] (3.6,0);
\coordinate [label=below:$c_1$] (B) at (3.6,0);
\draw[-Latex]  (2.6,1.8) to[out=20,in=90] (4.2,0);
\coordinate [label=below:$c_2$] (B) at (4.2,0);
\draw[-Latex]  (2.2,2.6) to[out=20,in=90] (5.4,0);
\coordinate [label=below:$c_{m+n}$] (B) at (5.4,0);
\draw[-]  (2.4,2.2) to[out=20,in=110] (4.6,1.2);
\end{tikzpicture}
\caption{\label{fig:trees} A generic tree $T$ in the left panel with elements of $\mathbb{A}_0$ attached left and right arbitrarily and the `base' tree $T_0$ in the right panel with only elements of $\mathbb{A}_0$ attached to the right on the right branch. $T$ can be obtained form $T_0$ through flipping $c_i$'s to the left of the right branch and/or shifting them to the left branch.}
\end{figure}
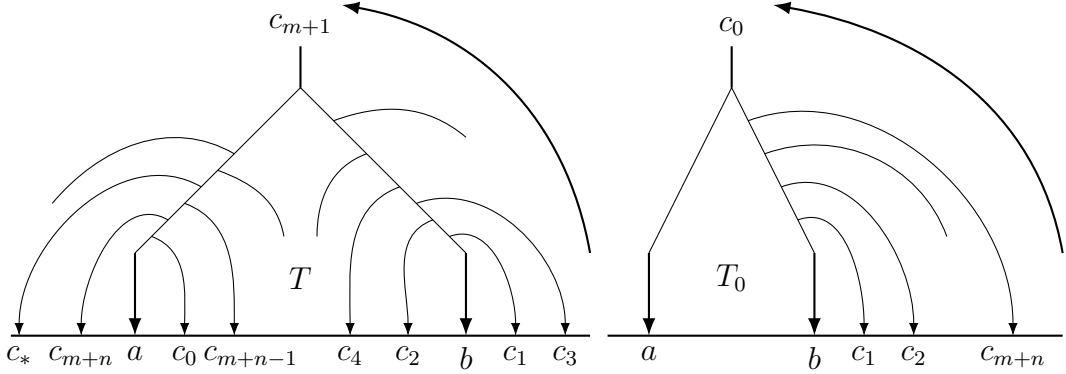

To simplify the procedure of obtaining expressions for trees, let us consider the trees in Fig.\ref{fig:trees}. We assign vectors $\vec a_i=(u_i,v_i)$, $\vec r_i  =  (p_{1,i},p_{2,i})$ to $c_i$ and $\vec r_0 = (p_{01},p_{02})$, $\vec a_0 = (1-\sum_{i=1}^{m+n}u_i, 1-\sum_{i=1}^{m+n} v_i)$ to $c_0$. We also introduce the matrices 
\begin{align}
    P&=(\vec 0,\vec 0,\vec r_1,\dots,\vec r_{m+n}, \vec r_0)\,, &
    Q&= (-\vec e_1,-\vec e_2,\vec a_1,\dots \vec a_{m+n},\vec a_0)\,,
\end{align}
where $\vec e_1 = \begin{pmatrix}
    1\\
    0
\end{pmatrix}$, $\vec e_2 = \begin{pmatrix}
    0\\
    1
\end{pmatrix}$. The expression for the symbol associated with  the basic tree $T_0$ can now be written as
\begin{align*}
    B_0 \star B_{m+n}|_{z=0} = (p_{12})^{m+n}\int_{\mathbb{V}_{m+n}}\exp(\text{tr}[PQ^t]);
\end{align*}
it is understood to yield the vertex $V(\omega,\omega,C,\dots,C)$ when acting on
\begin{align} \label{abc}
    a(y_1)b(y_2)c(y_3)\dotsc(y_{m+n+2})|_{y_i}=0 \,.
\end{align}
The configuration space $\mathbb{V}_{m+n}$ is given by the chain of inequalities in \eqref{inequalities}.
A generic tree $T$ can be obtained from $T_0$ through two types of operations: (i) flipping $c_i$ to the left of the right branch and (ii) a counterclockwise shift of all $c_i$'s along the cord connecting $a$ and $b$. Importantly, in the latter case $c_0$ also moves along the cord, while another $c_i$ takes its place. To express the symbol corresponding to $T$ we define $P_T = (\vec 0,\vec 0, \vec r_1,\dots,-\vec r_{m+1},\dots,\vec r_n, -\vec r_0)$. For vertices, the labels on $p_i$ and the corresponding arguments $y_i$ of $a$, $b$ and the $c_j$'s are read off from the tree from left to right. Since we have labeled them from bottom right to top left, we require a permutation $\sigma_T$ that relabels the $p_i$'s and $y_i$'s accordingly. Moreover, $\sigma_T$ also shuffles the elements in \eqref{abc} corresponding to their respective position in the tree $T$. In the absence of a cosmological constant, a generic tree $T$ contributes  to the vertex $\mathcal{V}(C,\dots,C,\omega,C,\dots,C,\omega,C,\dots,C)$ by
\begin{align*}
    s_T\sigma_T (p_{12})^{m+n}\int_{\mathbb{V}_{m+n}} \exp(\text{tr}[P_T Q^t]) a(y_1)b(y_2)c_1(y_3)\dots c_{y_{m+n}}(y_{m+n+2})|_{y_i=0} \,.
\end{align*}
Here, $s_T=(-1)^k$ and $k$ is the number of zero-forms $C$ in between the two $\omega$'s. The sign $s_T$ is the combination of the sign factor we get by evaluating the product of two branches with a sign coming from homological perturbation theory.

\subsection{All vertices with cosmological constant}
\label{sec:ads}
All the main properties discussed in the previous section remain true if we turn on the cosmological constant, which is a smooth deformation of Chiral Theory in flat space. Most importantly, the deformation maintains locality. In particular, we have to evaluate exactly the same graphs as before. It will turn out, as the low order examples illustrate, that switching on the cosmological constant adds one term to the exponent, e.g. $\hhbar p_{12} (\ldots)$ for $\mathcal{V}(\omega,\omega,C,\ldots,C)$ vertices. More specifically, a single branch takes the form
\begin{align}\label{onecosmologicalbranch}
    \begin{aligned}
    B_n&=(zp_1)^n\int\exp\Big[(1-V_n)yp_1+U_n(zy)+\lambda (zp_1)(U_n+\sum_{i,j=1}^n\text{sign}(j-i)u_{n,i}v_{n,j})+\\
    &+\sum_{i=1}^n u_{n,i}(zp_{i+1})+\sum_{i=1}^n v_{n,i}p_{1,i+1}\Big]\,.
    \end{aligned}
\end{align}
In the presence of the cosmological constant the construction of general vertices from the trees is more complicated than on the flat background. For example, attaching a leaf decorated by $c_{n+1}$ to the left of a branch of length $n$ yields
\begin{align*}
    \begin{aligned}
    h[\Lambda[c_{n+1}],B_n]&=(zp_1)^{n+1}\int\exp\Big[(1-V_{n+1})yp_1+U_{n+1} (zy)+\sum_{i=1}^{n+1} u_{n+1,i}(zp_{i+1})+\sum_{i=1}^{n+1} v_{n+1,i}p_{1,i+1}+\\
    &+\lambda (zp_1)(\sum_{i=1}^n u_{n+1,i}-u_{n+1,n+1}+\sum_{i,j=1}^n \text{sign}(j-i)u_{n+1,i}v_{n+1,j}-\sum_{i=1}^n u_{n+1,i}v_{n+1,n+1}+\\
    &+\sum_{j=1}^n u_{n+1,n+1}v_{n+1,j})\Big] \,,
    \end{aligned}
\end{align*}
i.e., the variables $u_{n+1,n+1}$ and $v_{n+1,n+1}$ enter the cosmological term with a minus sign as opposed to when the leaf is attached to the right. Otherwise, the expression remains the same. This coincides with the statement that the ordering of the leaves is irrelevant in the absence of the cosmological constant. Since the presence of the cosmological constant does not modify the piece of the expression we found in \eqref{twobranches2}, we will only consider the cosmological term for the following discussion. For a branch with leaves attached to the left and right arbitrarily, the cosmological term reads
\begin{align}\label{mostgeneralbranch}
\begin{aligned}
    \lambda (\tilde{U}_n+\sum_{i<j}^nu_{n,i}\tilde{v}_{n,j}-\sum_{j<i}^n\tilde{u}_{n,i}v_{n,j})(zp_1)\,,
\end{aligned}
\end{align}
where  $\tilde{x}\equiv \sigma_i x$,  $i$ corresponds to the label of the element of $\mathbb{A}_0$, and
\begin{align*}
    \sigma_{i}&\equiv
    \begin{cases}
    -1 &\text{if $\Lambda[C_{i}]$ is attached to the left (right) in the right (left) branch} \,,\\
    +1 &\text{if $\Lambda[C_i]$ is attached to the right (left) in the left (right) branch} \,.
    \end{cases}\
\end{align*}
The cosmological term `remembers' how the leaves were attached. In terms of the coordinates \eqref{newcoordinatestwobranches}, the cosmological term of a generic tree reads
\begin{align} \label{mostgeneraltree}
    \begin{aligned}
    &\lambda\Big(1+\sum_{i=1}^n \tilde{r}^L_{n,i}+\sum_{i=1}^m \tilde{r}^R_{m,i}-\sum_{i=1}^n s^L_{n,i}-\sum_{i=1}^m s^R_{m,i}-\sum_{i=1}^n r^L_{n,i}\sum_{j=1}^m r^R_{m,j}+\\
    &+\sum_{i=1}^n s^L_{n,i}\sum_{j=1}^m s^R_{m,j}+\sum_{i<j}^m r^R_{m,i}\tilde{s}^R_{m,j}-\sum_{j<i}^m \tilde{r}^R_{m,i}s^R_{m,j}+\sum_{i<j}^n r^L_{n,i}\tilde{s}^L_{n,j}-\sum_{j<i}^n \tilde{r}^L_{n,i}s^L_{n,j}\Big) p_{12}\,.
    \end{aligned}
\end{align}
In order to apply the change of coordinates \eqref{rename}, we need to differentiate between two cases: $n=0$ and $n>0$. In the former case we find the cosmological term to be
\begin{align*}
    \lambda  \Big(1+\tilde{U}_{m}-V_m+\sum_{i=1}^m \text{sign}(j-i)\sigma_{\max\{i,j\}}u_iv_j\Big) p_{12}\,,
\end{align*}
where we introduced
\begin{align*}
    \sigma_{\text{max}\{i,j\}} &= \left\{ \begin{array}{ll}
         &  \sigma_i \,, \quad \text{if } i>j\,, \\
         & \sigma_j \,, \quad \text{if } i<j \,.
    \end{array}\right.
\end{align*}
In case $n > 0$, we obtain
\begin{align*}
    \lambda \Big(&\sigma_{m+n}+\sum_{i=1}^m \sigma_i u_i+u_{m+1}+\sum_{i=1}^{n-1} \sigma_{m+i}u_{m+1+i} - \sigma_{m+n}\sum_{i=1}^{m+n}v_i + \sum_{i,j=1}^m \text{sign}(j-i)\sigma_{max\{i,j\}}u_iv_j+\\
    &+ \sum_{i=1}^m u_i v_{m+1}-\sum_{i=1}^m u_{m+1}v_i +\sum_{i=1}^{n-1} \sigma_{m+i} u_{m+1}v_{m+1+i}-\sum_{i=1}^{n-1}\sigma_{m+i}u_{m+1+i}v_{m+1}+\\
    &+\sum_{i=1}^m\sum_{j=1}^{n-1}\sigma_{m+j}u_iv_{m+1+j}-\sum_{i=1}^{n-1}\sum_{j=1}^m \sigma_{m+i}u_{m+1+i}v_j + \sum_{i,j=1}^{n-1}\text{sign}(j-i)\sigma_{max\{i,j\}}u_{m+1+i}v_{m+1+j}\Big)p_{12} \,.
\end{align*}
At the end of Section \ref{sec:flat}, we expressed the contribution to a vertex in terms of the matrices $P_T$ and $Q$. It turns out that, despite its complicated form, the cosmological term can be expressed in a similar fashion that is consistent with both aforementioned cases. We define a matrix $Q_T$ by filling up its columns, starting with $\vec e_1$, corresponding to $a$ in Fig.\ref{fig:trees} and from there on with $\vec a_i$  following through the tree counterclockwise. As an example, for the tree in the left panel of Fig.\ref{fig:trees} this looks like 
\begin{align}\label{qtmatrix}
    Q_T&=(-\vec e_1,\vec a_{m+n-1},\dots,\vec a_4,\vec a_2,-\vec e_2,\vec a_1,\vec a_3,\dots,\vec a_{m+1},\dots,\vec a_{m+n})\,.
\end{align}
The cosmological term for a generic tree is then given by $\lambda p_{12} |Q_T|$, where $|Q_T|$ is the sum of minors of $Q_T$. A generic tree with cosmological constant contributes to a vertex by
\begin{align}\label{compactform}
    s_T\sigma_T (p_{12})^{m+n}\int_{\mathbb{V}_{m+n}} \exp(\text{tr}[P_T Q^t]+\lambda p_{12}|Q_T|) a(y_1)b(y_2)c_1(y_3)\dots c_{y_{m+n}}(y_{m+n+2})|_{y_i=0} \,.
\end{align}
The simplest example is given by a single tree contributing to $\mathcal{V}(\omega,\omega,C,\ldots, C)$, see Eq. \eqref{besttree} below.

\subsection{Duality map and homological perturbation theory}
\label{sec:uvertices}
A very helpful idea put forward in \cite{Sharapov:2022faa,Sharapov:2022awp} is that of a duality map.  This map allows one to automatically generate all $\mathcal{U}$-vertices from $\mathcal{V}$-vertices. Moreover, the duality map manifestly preserves locality. Nevertheless, it is important to check that  homological perturbation theory leads to exactly the same $\mathcal{U}$-vertices as the duality map. Additionally, the duality map also allows one to relate various $\mathcal{V}$-vertices to each other.
\paragraph{$\mathcal{U}$-vertices.}
First of all, as it is shown in Appendix \ref{app:homo}, all the trees that contribute to $\mathcal{U}$-vertices are made up of a single branch (in contrast to the $\mathcal{V}$-vertices that consist of two-branch trees).  According to \eqref{circaction} the differentials $dz^A$ annihilate the module where the zero-forms $C$ take their values, so that  $dz^A\circ C\equiv0$ and the module action $\circ$ can only appear at the very last step. For example, for the vertex  $\mathcal{U}(\omega,C,C)$ with the symbol \eqref{uone} we should have
\begin{align}
    \mathcal{U}(\omega,C,C)= h[\omega \star \Lambda[C]] \circ C\,.
\end{align}
The reason is that any expression that acts on the bare $C$ has to be $dz$-independent to be different from zero and this can only occur at the end of the branch. The symbol of a branch $B_n$ of length $n$ is given by \eqref{onecosmologicalbranch}. There is a subtlety in computing the module action \eqref{moduleaction} for 
\begin{align}\label{BC}
     B_n\circ C&\equiv (B_n\star C^\tau)^\tau \equiv e^{zy}\Big[B_n(y,z) \star e^{zy} C(z)\Big]\Big|_{y\leftrightarrow z}\,,
\end{align}
$\tau$ being the involution defined by \eqref{tauinvol}. 
The point is that the expressions (\ref{BC}) involve star-products of nonpolynomial functions like  $e^{tz\cdot y}$, which, as is well-known, are ill-defined in general. For example, the product
$$
e^{tz\cdot y}\star e^{sz\cdot y}=\frac{e^{\frac{z\cdot y(t+s-2ts)}{1-ts}}}{(1-ts)^2}
$$
features a singularity as $t,s\rightarrow 1$. As a result, the integrals corresponding to the expressions (\ref{BC}) are not absolutely convergent. This, however, does not cause much trouble since $\circ$ appears only in the very last step and can easily be  resolved with any simple regularization. Specifically, we choose the following definition:
\begin{align}\label{regaction}
     B_n\circ C&\equiv (B_n\star C^{\tau_\varepsilon})^{\tau_\varepsilon}\equiv \lim_{\varepsilon\rightarrow +0}e^{(1-\varepsilon)zy}\Big(B_n(y,z) \star e^{(1-\varepsilon)zy} C(z)\Big)\Big|_{y\leftrightarrow z}\,,
\end{align} 
which just modifies the $\tau$-involution \eqref{tauinvol}. Plugging \eqref{onecosmologicalbranch} into \eqref{regaction}, we get after a straightforward calculation
\begin{align} \label{bigbranch}
    \begin{aligned}
    &\left(\frac{1-\varepsilon}{1-U_n(1-\varepsilon)}\right)^2 \left(\frac{\varepsilon}{1-U_n(1-\varepsilon)}\right)^n (p_{01})^n\times\\ &\int\exp\Big[\hhbar \frac{(1-V_n)(1-\varepsilon)+\varepsilon(U_n+\sum_{i,j=1}^n \text{sign}(j-i) u_{n,i}v_{n,j}
    )}{1-U_n(1-\varepsilon)}p_{01} +\\
    &\qquad+\varepsilon\sum_{i=1}^n \frac{u_{n,i}}{1-U_n(1-\varepsilon)}p_{0,i+1}+\frac{1-U_n}{1-U_n(1-\varepsilon)}p_{0,n+2} +\\
    &\qquad\qquad\sum_{i=1}^n(v_{n,i}-u_{n,i}\frac{(1-V_n)(1-\varepsilon)}{1-U_n(1-\varepsilon)})p_{1,i+1} +\frac{1-V_n}{1-U_n(1-\varepsilon)}p_{1,n+2}\Big] \,.
    \end{aligned}
\end{align}
Then, by analogy with the $\mathcal{V}$-vertices, we introduce the new variables
\begin{align}\label{changezeroform}
\begin{aligned}
    T_{n,i} &\equiv u_{n,i}\frac{\varepsilon }{1-U_n(1-\varepsilon)}\,, & S_{n,i}&\equiv v_{n,i}-u_{n,i}\frac{(1-V_n)(1-\varepsilon)}{1-U_n(1-\varepsilon)}\,, \\
    T_{n,n+1} &\equiv \frac{1-U_n}{1-U_n(1-\varepsilon)} \,, & S_{n,n+1} &\equiv \frac{1-V_n}{1-U_n(1-\varepsilon)}\,.
\end{aligned}
\end{align}
Again, the determinant of the Jacobian corresponding to this change of variables (note that we do not integrate $T_{n,n+1}$, $S_{n,n+1}$) cancels the exponential prefactor, see Appendix \ref{app:Jacobians}. In terms of the new coordinates, the symbol \eqref{bigbranch} takes the form
\begin{align} \label{bigbranchnewvar}
    \begin{aligned}
   (p_{01})^n  &\int \exp\Big[\hhbar\big(1+\sum_{i=1}^{n}T_{n,i}-\sum_{i=1}^{n}S_{n,i}+\sum_{i,j=1}^{n}\text{sign}(j-i)T_{n,i}S_{n,j}+\\
   &-\varepsilon S_{n,n+1}\big) p_{01}+\sum_{i=1}^{n+1} T_{n,i} p_{0,1+i}+\sum_{i=1}^{n+1} S_{n,i} p_{1,i+1} \Big] \,.
   \end{aligned}
\end{align}
At the same time, the duality map implies 
\begin{align} \label{V1U1duality}
    \langle \mathcal{V}_1(a,b,c_1,\dots,c_n)|c_{n+1}\rangle=\langle a|\mathcal{U}_1(b,c_1,\dots,c_n,c_{n+1})\rangle\,,
\end{align}
whence $\mathcal{U}_1(p_0,p_1,\dots,p_{n+1})=\mathcal{V}_1(-p_{n+1},p_0,p_1,\dots,p_n)$ with
\begin{align*}
    &\mathcal{V}_1(p_0,p_1,\dots,p_{n+2})=(p_{12})^n \exp\Big[(1-\sum_{i=1}^n u_{n,i})p_{01} +(1-\sum_{i=1}^n v_{n,i})p_{02}+\\
    &\sum_{i=1}^n u_{n,i}p_{1,i+2}+\sum_{i=1}^n v_{n,i}p_{2,i+2} +\hhbar \big (1+\sum_{i=1}^n u_{n,i}-\sum_{i=1}^n v_{n,i}+\sum_{i,j=1}^{n}\text{sign}(j-i)u_{n,i}v_{n,j}\big)p_{12}\Big]\,.
\end{align*}
We thus conclude that \eqref{bigbranchnewvar} approaches $\mathcal{U}_1(p_0,p_1,\dots,p_{n+1})$ as $\varepsilon\rightarrow +0$. Of course, the domain of integration for the $\mathcal{U}$-vertices is correct  and coincides with that for the $\mathcal{V}$-vertices. In the same way we can  evaluate $C \circ B_n$, which gives almost the same expression as before, up to a small change in the cosmological term. The final result reads
\begin{align*}
    \begin{aligned}
    (p_{01})^n  &\int \exp\Big[-\hhbar\big(1-\sum_{i=1}^{n}T_{n,i}-\sum_{i=1}^{n}S_{n,i}-\sum_{i,j=1}^{n}\text{sign}(j-i)T_{n,i}S_{n,j}+\\
   &-\varepsilon(1-\sum_{i=1}^n S_{n,i})\big) p_{01}+\sum_{i=1}^{n+1} T_{n,i} p_{0,1+i}+\sum_{i=1}^{n+1} S_{n,i} p_{1,i+1} \Big] \,.
    \end{aligned}
\end{align*}
In the same way as before, we get rid of the $\varepsilon$-dependent term by setting $\varepsilon=0$. This coincides with the result obtained from the duality map, i.e.
\begin{align*}
    \langle \mathcal{V}(a,c_{n+1},b,c_1,\dots,c_{n-1})|c_{n}\rangle = \langle a|\mathcal{U}(c_{n+1},b,c_1,\dots,c_n)\rangle \,.
\end{align*}
To obtain a generic branch, the other elements of $\mathbb{A}_0$ could also be attached on the left. Here, as before, we have adapted the convention of labeling the $p_{i}$'s from the bottom to the top of the branch and we need to perform a permutation $\sigma_T$ to rearrange them accordingly and shuffle the elements $a(y_1)c_1(y_2),\dots c_n(y_{n+1})|_{y_i=0}$. In the limit when $\varepsilon \rightarrow +0$, a generic contribution to $\mathcal{U}$-vertex with zero-forms attached left and right arbitrarily approaches
\begin{align*}
    \begin{aligned}
    \mathcal{U}(c_{i+1},\dots,c_{n+1},a,c_1,\dots,c_i) &= \sigma_T(p_{01})^n  \int \exp\Big[\hhbar \big(\sigma_{n+1}+\sum_{i=1}^{n}\tilde{T}_{n,i}-\sigma_{n+1}\sum_{i=1}^{n}S_{n,i}+\\
    &+\sum_{i<j}^{n}T_{n,i}\tilde{S}_{n,j}-\sum_{j<i}^n \tilde{T}_{n,i}S_{n,j}\big)p_{01}+\sum_{i=1}^{n+1} T_{n,i} p_{0,1+i}+\\
    &+\sum_{i=1}^{n+1} S_{n,i} p_{1,i+1} \Big]\times a(y_1)c_1(y_2),\dots c_n(y_{n+1})|_{{y_i}=0}\,.
    \end{aligned}
\end{align*}
Naturally, a generic $\mathcal{U}$-vertex is related to a class of $\mathcal{V}$-vertices by
\begin{align}\label{genericUVduality}
    \langle \mathcal{V}(a,c_1,\dots,c_i,b,c_{i+1},\dots,c_n)|c_{n+1}\rangle = \langle a|\mathcal{U}(c_1,\dots,c_i,b,c_{i+1},\dots,c_n)\rangle \,.
\end{align}
\paragraph{$\mathcal{V}$-$\mathcal{V}$-duality.}
The duality map also operates as a map between various $\mathcal{V}$-vertices via
\begin{align*}
    &\langle \mathcal{V}(c_{j+1},\dots,c_n,a,c_1,\dots,c_i,b,c_{i+1}\dots,c_{j-1}|c_j\rangle = \\
    &\qquad\qquad\langle \mathcal{V}(c_{j-k+1},\dots,c_n,a,c_1,\dots,c_i,b,c_{i+1},\dots,c_{j-k-1}|c_{j-k}\rangle  \,,
\end{align*}
where we rotated the arguments by $k$ units and $j \geq i+1$. Through this duality all $\mathcal{V}$-vertices with the same number of elements of $\mathbb{A}_0$ between $a$ and $b$ and the same total number of $\mathbb{A}_0$ elements are related to each other, which vastly reduces the number of vertices to be computed. In particular, it suffices to only determine $\mathcal{V}$-vertices of the type $\mathcal{V}(a,c_1,\dots,c_i,b,c_{i+1},\dots,c_n)$ and one can relate all $\mathcal{V}$-vertices in the class of vertices characterized by $(n,i)$. In hindsight, some hints of this duality were hidden in the expression for $\mathcal{V}$-vertices that we presented in \eqref{compactform}, namely (i) the overall sign is determined by the number of elements of $\mathbb{A}_0$ between $a$ and $b$, which is an invariant within a class, (ii) the matrix $Q_T$ in the cosmological term is constructed similarly for all vertices belonging to the same class and (iii) the configuration space of trees that share the same number of total elements of $\mathbb{A}_0$ is identical.
\paragraph{$\mathcal{U}$-$\mathcal{U}$-dualities.}
A natural generalization of the idea discussed above is to introduce dualities between $\mathcal{U}$-vertices themselves. However, this can only be done if a $\mathcal{U}$-vertex is contracted with an element of $\mathbb{A}_{-1}$ and subsequently the other element of $\mathbb{A}_{-1}$ is stripped off. This leaves only one duality relation for the $\mathcal{U}$-vertices, namely,
\begin{align}\label{uuduality}
    \langle a|\mathcal{U}_{n+2}(c_1,\dots,c_{n+1},b)\rangle &= \langle \mathcal{U}_1(a,c_1,\dots,c_{n+1})|b\rangle = - \langle b|\mathcal{U}_1(a,c_1,\dots,c_{n+1})\rangle \,.
\end{align}
Consistency of this duality can be checked either using the explicit expressions for the relevant vertices or through various dualities. The latter method is particularly easy to implement, as its consistency implies that the following diagram commutes: 
\begin{center}
\begin{tikzpicture}
  \matrix (m)
    [
      matrix of math nodes,
      row sep    = 3em,
      column sep = 4em
    ]
    {
      \mathcal{V}(a,b,c_1,\dots,c_n)              & \mathcal{U}(b,c_1,\dots,c_{n+1}) \\
      \mathcal{V}(c_1,\dots,c_n,a,b) & \mathcal{U}(c_1,\dots,c_{n+1},b)            \\
    };
  \path
    (m-1-1) edge [<->] node [left] {$\mathcal{V}$-$\mathcal{V}$} (m-2-1)
    (m-1-1.east |- m-1-2)
      edge [<->] node [above] {$\mathcal{V}$-$\mathcal{U}$} (m-1-2)
    (m-2-1.east) edge [<->] node [below] {$\mathcal{V}$-$\mathcal{U}$} (m-2-2)
    (m-1-2) edge [<->] node [right] {$\mathcal{U}$-$\mathcal{U}$} (m-2-2);
\end{tikzpicture}
\end{center}
(the type of duality is specified on the arrows).
\paragraph{$\mathbb{Z}_2$-transformation.}
When discussing the duality between $\mathcal{V}$- and $\mathcal{U}$-vertices we only considered taking out $a$, while for the duality among $\mathcal{V}$-vertices themselves we always took out a $c_i$ that appeared at the right of $b$. There is a natural pairing $\langle a(y)|c\rangle=-\langle c|a(-y)\rangle$ between $a \in \mathbb{A}_{-1}$ and $c \in \mathbb{A}_0$, which allows us to take out $b$ or $c_i$ to the left of $a$ in the aforementioned cases. As a consequence, some of the dualities can take place through two different routes, e.g.
\begin{align*}
    \langle \mathcal{V}_2(a,c_1,b)|c_2\rangle &= \langle a|\mathcal{U}_2(c_1,b,c_2) \rangle\,,\\
    \langle \mathcal{V}_2(a,c_1,b)|c_2\rangle &= -\langle \mathcal{U}_2(c_2,a,c_1)|b(-y)\rangle \,.
\end{align*}
Both cases evaluate to different expressions, but they are related to each other by a $\mathbb{Z}_2$-transformation that preserves the domain of integration, i.e.,
\begin{align*}
    \frac{u_1}{v_1} \leq \frac{u_2}{v_2} \leq \dots \leq \frac{u_n}{v_n} \leq \frac{u_{n+1}}{v_{n+1}} \,.
\end{align*}
This maps $\{v_{n+1},\dots v_1\} \rightarrow \{u_1,\dots,u_{n+1}\}$ and $\{u_{n+1},\dots u_1\} \rightarrow \{v_1,\dots,v_{n+1}\}$. 

To summarize, we have directly checked that our $A_\infty$-algebra $\mathbb{A}$ has the remarkable property we called the duality map in \cite{Sharapov:2022faa,Sharapov:2022awp}. This implies that the $A_\infty$-algebra $\mathbb{A}$ underlying Chiral Theory is a pre-Calabi--Yau algebra \cite{IYUDU202163, kontsevich2021pre}, see Appendix \ref{CY} for more detail. In practical terms, this implies that there are few independent multi-linear products with a given number of arguments.

\section{Configuration space}
\label{sec:config}
By construction, each contracting homotopy $h$ entering an interaction vertex brings one integration variable $t_i\in[0,1]$, so that the whole integration domain appears to be the hypercube $[0,1]^{2n}$. However, in terms of `times' $t_i$ the `propagators' in front of $p_{ij}$ as well as the pre-exponential factors look ugly (see \cite{Sharapov:2022faa,Sharapov:2022awp} and the change of variables in the previous section). In addition, it is not immediately obvious that the integrals converge. In terms of the new variables $u$'s and $v$'s all integrands are obviously smooth functions and the  
question of convergence reduces to the compactness of the new integration domain. In Appendix \ref{app:domain}, we prove that the domain is compact indeed.

With the help of the new integration variables the vertices simplify a lot. In particular, the propagators are linear except for the only $\hhbar$-term in the exponent where it is no more than bilinear and the pre-exponential factor  is completely eliminated by the Jacobians of the coordinate transformations. These drastic simplifications should convince one that the variables we have chosen above are the preferred ones. It is time to describe the integration domain in more detail. Let us concentrate on vertices of type $\mathcal{V}(\omega,\omega,C,\ldots, C)$, of which the symbol is given by \cite{Sharapov:2022awp}
\begin{align}\label{besttree}
\begin{aligned}
       G=&(p_{12})^n \exp\Big[ (1-\sum_i u_i) p_{01} +(1-\sum_i v_i) p_{02} +\sum_i u_i p_{1,i+2}+\sum_i v_i p_{2,i+2}+ \\
     &\qquad \qquad \qquad \qquad +\hhbar\, \Big(1+\sum_i (u_i-v_i) +\sum_{i,j} u_iv_j \sign(j-i) \Big ) p_{12} \Big]\,. 
\end{aligned}
\end{align}
We will first consider this family of vertices at lower orders in Section \eqref{sec:configstart}, then provide a straightforward generalization to all orders with details left to Appendix \ref{app:fulldomain}. A more formal description of the configuration space together with its relation to Grassmannians is presented in Section \ref{sec:swallowtail}.

\subsection{Order by order analysis}
\label{sec:configstart}
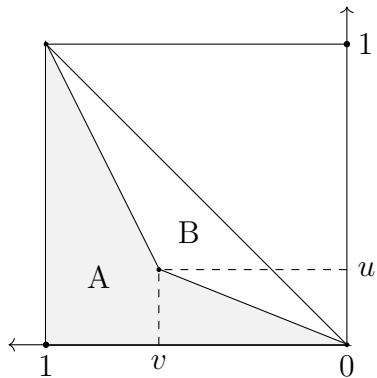
\begin{wrapfigure}{r}{0.32\textwidth}
\begin{tikzpicture}
\draw[thick]  (0,0) -- (0,4)--(1.5,1)--(4,0)--cycle;
\draw[-]  (0,4) -- (4,4);
\draw[-]  (0,4) -- (4,0);
\fill[black!5!white] (0,0) -- (0,4)--(1.5,1)--(4,0)--cycle;
\filldraw (0,4) circle (0.7 pt);
\filldraw (1.5,1) circle (0.7 pt);
\filldraw (0,0) circle (1 pt);
\filldraw (4,4) circle (1 pt);
\filldraw (4,0) circle (0.7 pt);
\coordinate [label=below:$1$] (B) at (0,0);
\coordinate [label=right:$1$] (B) at (4,4);
\coordinate [label=below:$0$] (B) at (4,0);
\coordinate [label=below:$v$]  (B) at (1.5,0);
\coordinate [label=right:$u$] (B) at (4,1);
\draw[dashed] (4,1) -- (1.5,1); 
\draw[dashed] (1.5,0) -- (1.5,1); 
\draw[->](4,0)--(4,4.5);
\draw[->](4,0)--(-0.5,0);
\node[] at (0.7,0.9) {A};
\node[] at (1.9,1.5) {B};
\end{tikzpicture}
\caption{Cubic order.}\label{F1}
\end{wrapfigure}
Let us start from the cubic vertex $\mathcal{V}(\omega,\omega,C)$, for which the integration domain has been identified as the simplex $0<u<v<1$ in the Cartesian plane, see Fig. \ref{F1}. The configuration space is constituted by points lying below the diagonal of the unit square. A simple plane geometry exercise identifies the  multiplier $1+u-v$ of the cosmological constant as twice the area of the shaded region $A$. The volume of the configuration space is $1/2$ since any point below the diagonal is admissible. 

At the quartic order,  $\mathcal{V}(\omega,\omega,C,C)$, the integration domain is defined  by more complicated  inequalities:
$$
0\leq v_2\leq 1\,,\qquad 0\leq u_1\leq v_1\leq 1-v_2\,,\qquad \frac{u_1}{v_1}\leq \frac{u_2}{v_2}\leq \frac{1-u_1}{1-v_1}\,.
$$
In order to clarify their geometric  meaning it is convenient to introduce the pair of new variables $v_3$ and $u_3$ subject to the relations 
\begin{equation}\label{uv}
u_1+u_2+u_3=1\,,\qquad v_1+v_2+v_3=1\,.
\end{equation}
With these variables  we can rewrite the inequalities above in a more symmetric form:
$$
0 \leq v_2\leq 1\,,\qquad 0\leq u_1\leq v_1\leq 1-v_2\,,\qquad \frac{u_1}{v_1}\leq \frac{u_2}{v_2}\leq \frac{u_3}{v_3}\,.
$$
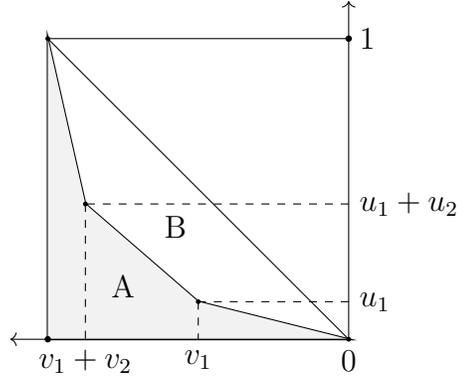
\begin{wrapfigure}{r}{0.4\textwidth}
\begin{tikzpicture}
\draw[thick]  (0,0) -- (0,4) -- (0.5,1.8) --(2,0.5)--(4,0)--cycle;
\draw[-]  (0,4) -- (4,4);
\draw[-]  (0,4) -- (4,0);
\fill[black!5!white] (0,0) -- (0,4) -- (0.5,1.8) --(2,0.5)--(4,0)--cycle;
\filldraw (0,4) circle (0.7 pt);
\filldraw (0.5,1.8) circle (0.7 pt);
\filldraw (2,0.5) circle (0.7 pt);
\filldraw (0,0) circle (1 pt);
\filldraw (4,4) circle (1 pt);
\filldraw (4,0) circle (0.7 pt);
\coordinate [label=right:$1$] (B) at (4,4);
\coordinate [label=below:$v_1+v_2$]  (B) at (0.5,0);
\coordinate [label=below:$v_1$]  (B) at (2.0,0);
\coordinate [label=right:$u_1$] (B) at (4,0.5);
\coordinate [label=right:$u_1+u_2$] (B) at (4,1.8);
\coordinate [label=below:$0$] (B) at (4,0);
\draw[dashed] (4,0.5) -- (2,0.5);
\draw[dashed] (4,1.8) -- (0.5,1.8);
\draw[dashed] (2,0.5) -- (2,0); 
\draw[dashed] (0.5,1.8) -- (0.5,0); 
\draw[->](4,0)--(4,4.5);
\draw[->](4,0)--(-0.5,0);
\node[] at (1.0,0.7) {A};
\node[] at (1.7,1.5) {B};
\end{tikzpicture}
\caption{Quartic order.}\label{F2}
\end{wrapfigure}

The last group of inequalities implies that the corresponding segments, see Fig. \ref{F2}, form a concave shape (the upper boundary of region $A$). In other words, the quadrilateral $B$ is convex and one of its edges coincides with the diagonal of the unit  square.
Again, the multiplier of the cosmological constant, $1+u_1+u_2-v_1-v_2+u_1 v_2-u_2 v_1$, can be recognized as twice the area of the shaded region $A$.  For an obvious reason we will call such concave polygons $A$ {\it swallowtails}. It is easy to see that the volume of this four-dimensional configuration space is equal to  $1/24$. 

Now the generalization to all orders is straightforward, see Appendix \ref{app:fulldomain} for the proof: vertex $\mathcal{V}(\omega,\omega,C,\ldots,C)$ with $n$ zero-forms $C$ is given by $2n$-tuple integral over the configuration space of swallowtails with $n+3$ vertices, three of which are fixed to be the corners of the unit square. 
Since the integration domain is obviously compact, the interaction vertices are well-defined at least as $A_\infty$ structure maps. The positions of the $n$ points inside the wedge, which are the actual degrees of freedom of a swallowtail, correspond to coefficients in front of $p_{1,i+2}$ and $p_{2,i+2}$ that connect the two one-forms $\omega$ to $n$ zero-forms $C$. In case $\hhbar\neq 0$, the coefficient of the cosmological term $\hhbar p_{12}$ is given by twice the area of the swallowtail. As discussed at length in \cite{Sharapov:2022faa,Sharapov:2022awp}, the fact that no other differential operators $p_{ij}$ appear that would connect pairs of zero-forms implies space-time locality. 

Regarding trees with other topologies, first of all the configuration space is exactly the same as above, see Appendix \ref{app:fulldomain}. This, among other things, implies that the homological perturbation theory, even though yielding a solution, does not reveal all hidden symmetries of the vertices. 

\begin{figure}
\centering\label{fig:flip}
\begin{tikzpicture}
\begin{scope}[shift={(-4.5,0)}]
\draw[thick]  (0,0) -- (0,4) -- (0.5,1.8) --(2,0.5)--(4,0)--cycle;
\draw[-]  (0,4) -- (4,4);
\draw[-]  (0,4) -- (4,0);
\fill[black!5!white] (0,0) -- (0,4) -- (0.5,1.8) --(2,0.5)--(4,0)--cycle;
\filldraw (0,4) circle (0.7 pt);
\filldraw (0.5,1.8) circle (0.7 pt);
\filldraw (2,0.5) circle (0.7 pt);
\filldraw (0,0) circle (1 pt);
\filldraw (4,4) circle (1 pt);
\filldraw (4,0) circle (0.7 pt);
\coordinate [label=right:{$1=u_1+u_2+u_3$}] (B) at (4,4);
\coordinate [label=below:$v_1+v_2$]  (B) at (0.5,0);
\coordinate [label=below:$v_1$]  (B) at (2.0,0);
\coordinate [label=right:$u_1$] (B) at (4,0.5);
\coordinate [label=right:$u_1+u_2$] (B) at (4,1.8);
\coordinate [label=below:$0$] (B) at (4,0);
\coordinate [label=above:{$1=v_1+v_2+v_3$}] (B) at (0,4);
\draw[dashed] (4,0.5) -- (2,0.5);
\draw[dashed] (4,1.8) -- (0.5,1.8);
\draw[dashed] (2,0.5) -- (2,0); 
\draw[dashed] (0.5,1.8) -- (0.5,0); 
\draw[->](4,0)--(4,4.5);
\draw[->](4,0)--(-0.5,0);
\node[] at (1.0,0.7) {A};
\node[] at (1.7,1.5) {B};
\coordinate [label=above:$_{a_3}$] (B) at (0.55,2.5);
\coordinate [label=above:$_{a_2}$] (B) at (1.6,0.9);
\coordinate [label=above:$_{a_1}$] (B) at (3.3,0.12);
\end{scope}

\begin{scope}[shift={(0,-6)}]
\draw[thick]  (3.5,0) -- (4,0) -- (3.5,2.2) -- cycle;
\draw[thick]  (0,0) -- (3.5,0) -- (3.5,-1.8) -- (2,-0.5) -- cycle;
\draw[-] (4,4) -- (0,4);
\draw[-] (0,0) -- (0,4);
\draw[-] (0,0) -- (0,-4);
\draw[-] (0,-4) -- (4,-4);
\draw[-] (4,-4) -- (4,0);
\fill[black!5!white] (0,0) -- (3.5,0) -- (3.5,-1.8) -- (2,-0.5) -- cycle;
\fill[black!5!white] (3.5,0) -- (4,0) -- (3.5,2.2) -- cycle;
\filldraw (0,4) circle (0.7 pt);
\filldraw (0,-4) circle (0.7 pt);
\filldraw (4,-4) circle (0.7 pt);
\filldraw (0,0) circle (1 pt);
\filldraw (4,4) circle (1 pt);
\filldraw (4,0) circle (0.7 pt);
\filldraw (3.5,0) circle (0.7 pt);
\filldraw (2,-0.5) circle (0.7 pt);
\filldraw (3.5,-1.8) circle (0.7 pt);
\filldraw (3.5,2.2) circle (0.7 pt);
\coordinate [label=right:$1$] (B) at (4,4);
\coordinate [label=below left:$1$]  (B) at (0,0);
\coordinate [label=above:$v_2+v_3$]  (B) at (2.0,0);
\coordinate [label=above left:$v_3$] (B) at (3.5,0);
\coordinate [label=right:$u_3$] (B) at (4,2.2);
\coordinate [label=right:$0$] (B) at (4,0);
\coordinate [label=right:$-u_1$] (B) at (4,-0.5);
\coordinate [label=right:$-u_1-u_2$] (B) at (4,-1.8);
\coordinate[label=right:$-1$] (B) at (4,-4);
\draw[dashed] (4,-0.5) -- (2,-0.5);
\draw[dashed] (4,-1.8) -- (3.5,-1.8);
\draw[dashed] (2,-0.5) -- (2,0); 
\draw[dashed] (3.5,2.2) -- (4,2.2); 
\draw[->](4,0)--(4,4.5);
\draw[->](4,0)--(-0.5,0);
\draw[-](4,-4)--(4,-4.5);
\node[] at (3.7,0.4) {$+$};
\node[] at (3,-0.8) {$-$};
\coordinate [label=above:$_{a_3}$] (B) at (3.85,1.5);
\coordinate [label=above:$_{a_2}$] (B) at (2.6,-1.6);
\coordinate [label=above:$_{a_1}$] (B) at (0.9,-0.7);
\end{scope}

\begin{scope}[shift={(-9,-6)}]
\draw[thick]  (2,0) -- (4,0) -- (2.5,1.3) -- (2,3.5) -- cycle;
\draw[thick]  (0,0) -- (2,0) -- (2,-0.5) -- cycle;
\draw[-] (4,4) -- (0,4);
\draw[-] (0,0) -- (0,4);
\draw[-] (0,0) -- (0,-4);
\draw[-] (0,-4) -- (4,-4);
\draw[-] (4,-4) -- (4,0);
\fill[black!5!white] (2,0) -- (4,0) -- (2.5,1.3) -- (2,3.5) -- cycle;
\fill[black!5!white] (0,0) -- (2,0) -- (2,-0.5) -- cycle;
\filldraw (0,4) circle (0.7 pt);
\filldraw (0,-4) circle (0.7 pt);
\filldraw (4,-4) circle (0.7 pt);
\filldraw (0,0) circle (1 pt);
\filldraw (4,4) circle (1 pt);
\filldraw (4,0) circle (0.7 pt);
\filldraw (2,0) circle (0.7 pt);
\filldraw (2,-0.5) circle (0.7 pt);
\filldraw (2,3.5) circle (0.7 pt);
\filldraw (2.5,1.3) circle (0.7 pt);
\coordinate [label=right:$1$] (B) at (4,4);
\coordinate [label=below left:$1$]  (B) at (0,0);
\coordinate [label=above:$v_2+v_3$]  (B) at (1.8,0);
\coordinate [label=below:$v_2$] (B) at (2.5,0);
\coordinate [label=right:$u_2+u_3$] (B) at (4,3.5);
\coordinate [label=right:$0$] (B) at (4,0);
\coordinate [label=right:$-u_1$] (B) at (4,-0.5);
\coordinate [label=right:$u_2$] (B) at (4,1.3);
\coordinate[label=right:$-1$] (B) at (4,-4);
\draw[dashed] (4,-0.5) -- (2,-0.5);
\draw[dashed] (4,1.3) -- (2.5,1.3);
\draw[dashed] (2,3.5) -- (4,3.5); 
\draw[dashed] (2.5,1.3) -- (2.5,0); 
\draw[->](4,0)--(4,4.5);
\draw[->](4,0)--(-0.5,0);
\draw[-](4,-4)--(4,-4.5);
\node[] at (2.7,0.45) {$+$};
\node[] at (1.75,-0.25) {$-$};
\coordinate [label=above:$_{a_3}$] (B) at (2.55,2.5);
\coordinate [label=above:$_{a_2}$] (B) at (3.5,0.5);
\coordinate [label=above:$_{a_1}$] (B) at (0.9,-0.7);
\end{scope}
\end{tikzpicture}

\caption{Quartic order with various orderings of the zero-forms. On the top panel we have the swallowtail that determines $\mathcal{V}(\omega,\omega,C,C)$. The coefficient of the cosmological term is twice the area of region $A$, which is made of two segments of unit length followed by $a_1$, $a_2$, $a_3$. On the bottom right panel we flipped the position of two zero forms, which makes a contribution to $\mathcal{V}(\omega,C,C,\omega)$. Accordingly, the order of the segments is changed: $a_1$ and $a_2$ are inserted in between the first one and the second one that are of unit length. The coefficient of the cosmological term is twice the oriented area: the area below the mid-line contributes with minus sign. Similarly, on the bottom left panel one zero-form is flipped giving  a contribution to $\mathcal{V}(\omega,C,\omega,C)$. The edge $a_1$ is placed in between the segments of unit length and the coefficient of the cosmological constant again follows from twice the oriented area. }\label{quarticflipped}
\end{figure}
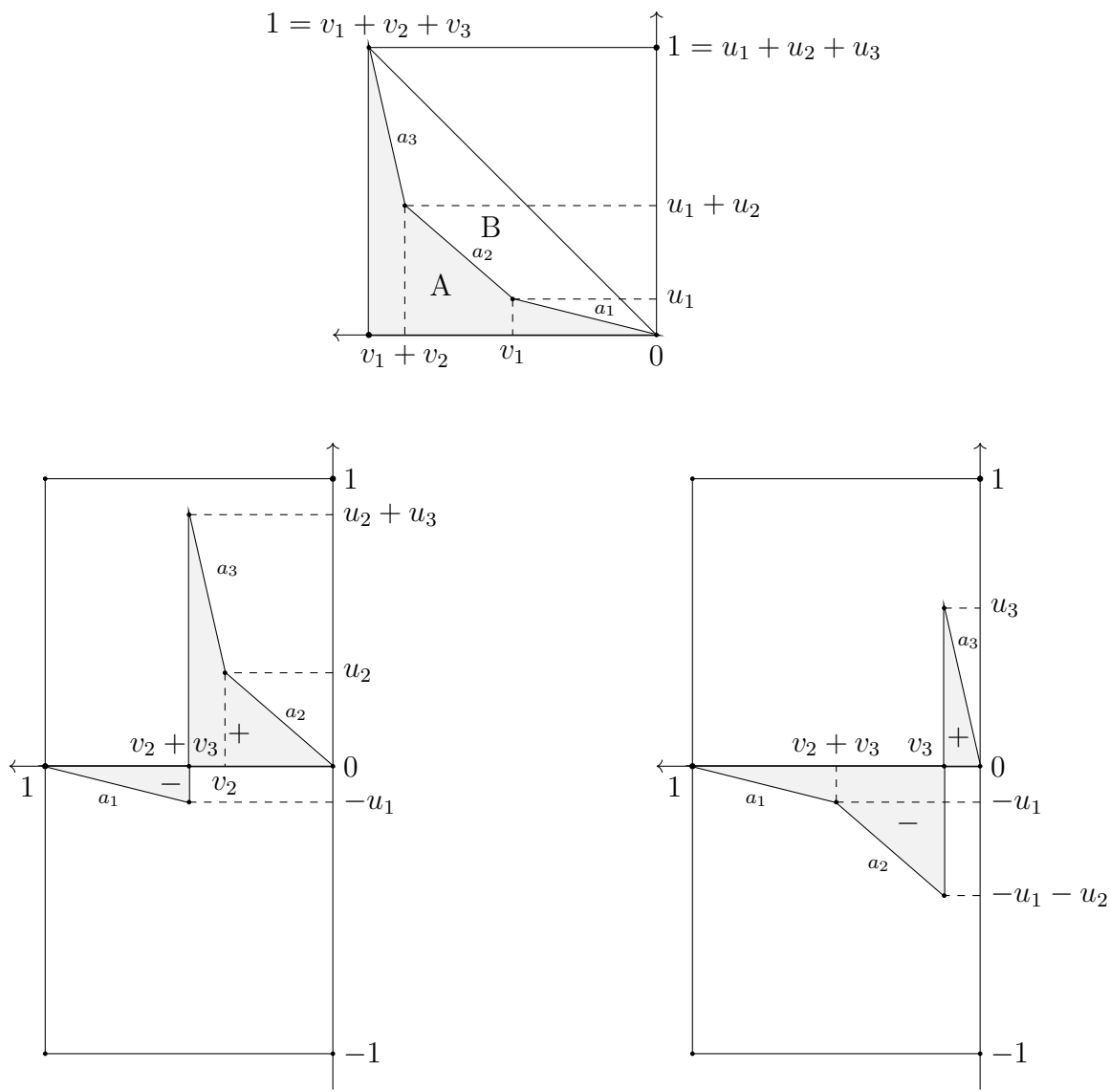

Trees with different ordering of zero-forms on either branch within the same topology have the same configuration space. {The term in the exponential proportional to the cosmological constant changes however by flipping one or more signs. For instance, changing the order of both zero-forms in $\mathcal{V}(\omega,\omega,C,C)$ gives a tree for  the vertex $\mathcal{V}(\omega,C,C,\omega)$. The multiplier of the cosmological constant reads $1-u_1-u_2-v_1-v_2-u_1v_2+u_2v_1$ and is equal to twice the area of region `$+$' minus region `$-$' in Fig. \ref{quarticflipped}}.
{The edges whose coordinates correspond to the flipped zero-forms create a new structure, which turns out to be a swallowtail itself. Meanwhile, these vectors are removed from the original swallowtail, which preserves the defining features of a swallowtail. Thus, for a tree with mixed ordering of its zero-forms, the term proportional to the cosmological constant is related to the difference between the area of two swallowtails, which is an oriented area. Also notice that for a mixed ordering this term can become negative.} There is a simple algebraic interpretation of these manipulations as the sum $|Q_T|$ of minors of matrix $Q_T$, \eqref{qtmatrix}, see also below.
Now we proceed to a more formal discussion of the configuration space and its relation to Grassmannians. 

\subsection{Measuring swallowtails }
\label{sec:swallowtail}
Consider a Euclidean plane $\mathbb{E}^2$ with its natural metric topology. It will be convenient on occasion to forget about Euclidean structure and treat $\mathbb{E}^2$ as an affine space with the automorphism group $\mathrm{Aff}(2,\mathbb{R})=GL(2,\mathbb{R})\ltimes \mathbb{R}^2$. By the Jordan curve theorem each simple polygon chain separates $\mathbb{E}^2$ into two disconnected regions, called exterior and interior. 
Consequently,  to each vertex of a simple polygon  one can assign exterior and interior angles.
We say that a vertex is convex (concave) if its interior angle is $\leq \pi$ ($>\pi$). 
A polygon is called convex if all its vertices are convex. By definition, a concave polygon has at least one concave vertex. 

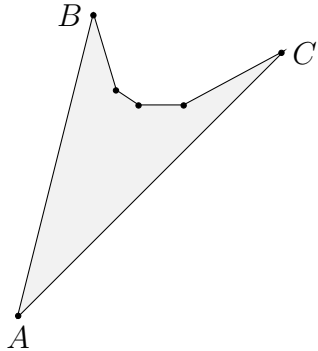
\begin{figure}[h]
\center{
\begin{tikzpicture}
\draw[thick]  (0,0) -- (1,4)--(1.3,3)--(1.6,2.8)--(2.2,2.8)--(3.5,3.5)--cycle;
\fill[black!5!white] (0,0) -- (1,4)--(1.3,3)--(1.6,2.8)--(2.2,2.8)--(3.5,3.5)--cycle;
\filldraw (0,0) circle (1.0 pt);
\filldraw (1,4) circle (1.0 pt);
\filldraw (1.3,3) circle (1.0 pt);
\filldraw (1.6,2.8) circle (1.0 pt);
\filldraw (2.2,2.8) circle (1.0 pt);
\filldraw (3.5,3.5) circle (1.0 pt);
\coordinate [label=below:$A$] (B) at (0,0);
\coordinate [label=left:$B$] (B) at (1,4);
\coordinate [label=right:$C$] (B) at (3.5,3.5);
\end{tikzpicture}
}
\caption{A simple concave $6$-gon. The vertices $A$, $B$, and $C$ are convex, the other three are concave.}\label{st}
\end{figure}

It is clear that for a simple concave polygon the minimal number of convex vertices is equal to $3$ (hence, every triangle is convex).
We are interested in simple concave polygons with exactly three convex vertices that go one after another. As in the previous section, these will be referred to as {\it swallowtails}, see Fig.\ref{st}. 
 It is known that convexity is an affine property, meaning that the affine transformations of $\mathrm{Aff}(2,\mathbb{R})$ map swallowtails to swallowtails. We say that two swallowtails are equivalent to each other if they are related by an affine transformation. 
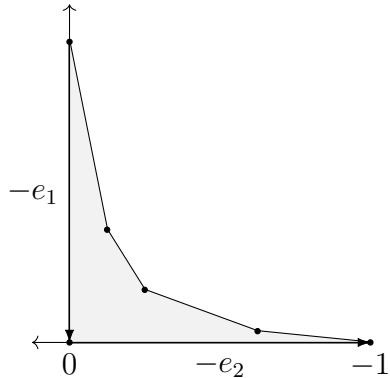
\begin{figure}[h]
\center{
\begin{tikzpicture}
\draw[thick]  (0,0) -- (0,4)--(0.5,1.5)--(1,0.7)--(2.5,0.15)--(4,0)--cycle;
 \fill[black!5!white] (0,0) -- (0,4)--(0.5,1.5)--(1,0.7)--(2.5,0.15)--(4,0)--cycle;
\filldraw (0,4) circle (1.0 pt);
\filldraw (0.5,1.5) circle (1.0 pt);
\filldraw (1,0.7) circle (1.0 pt);
\filldraw (2.5,0.15) circle (1.0 pt);
\filldraw (0,0) circle (1 pt);
\filldraw (4,0) circle (1.0 pt);
\coordinate [label=below:$-1$] (B) at (4,0);
\coordinate [label=below:$0$] (B) at (0,0);
\coordinate [label=left:$-e_1$] (B) at (0,2);
\coordinate [label=below:$-e_2$] (B) at (2,0);
\draw[-Latex](0,4)--(0,0);
\draw[-Latex](0,0)--(4,0);
\draw[->](0,0)--(0,4.5);
\draw[->](4,0)--(-0.5,0);
\end{tikzpicture}
}
\caption{A canonical swallowtail with six vertices.}\label{cst}
\end{figure}

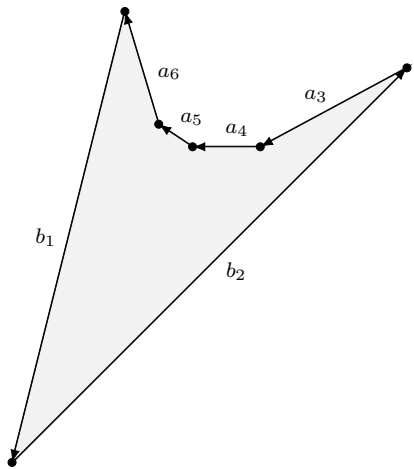
\begin{figure}[h]
\center{
\begin{tikzpicture}[scale=1.5]
\draw[thick]  (0,0) -- (1,4)--(1.3,3)--(1.6,2.8)--(2.2,2.8)--(3.5,3.5)--cycle;
\fill[black!5!white] (0,0) -- (1,4)--(1.3,3)--(1.6,2.8)--(2.2,2.8)--(3.5,3.5)--cycle;

\draw[-Latex] (1,4) -- (0,0);
\draw[-Latex] (1.3,3) -- (1,4);
\draw[-Latex] (1.6,2.8) -- (1.3,3);
\draw[-Latex] (2.2,2.8) -- (1.6,2.8);
\draw[-Latex] (3.5,3.5) -- (2.2,2.8);
\draw[-Latex] (0,0) -- (3.5,3.5);

\filldraw (0,0) circle (1.0 pt);
\filldraw (1,4) circle (1.0 pt);
\filldraw (1.3,3) circle (1.0 pt);
\filldraw (1.6,2.8) circle (1.0 pt);
\filldraw (2.2,2.8) circle (1.0 pt);
\filldraw (3.5,3.5) circle (1.0 pt);

\coordinate [label=left:$_{b_1}$] (B) at (0.5,2);
\coordinate [label=right:$_{b_2}$] (B) at (1.8,1.7);
\coordinate [label=above:$_{a_6}$] (B) at (1.4,3.3);
\coordinate [label=above:$_{a_5}$] (B) at (1.6,2.9);
\coordinate [label=above:$_{a_4}$] (B) at (2,2.8);
\coordinate [label=above:$_{a_3}$] (B) at (2.7,3.1);
\end{tikzpicture}
}
\caption{An oriented swallowtail.}\label{ost}
\end{figure}

In order to describe the  equivalence classes of swallowtails modulo affine transformations we fix an origin $0$ and an orthonormal basis $(e_1,e_2)$ in $\mathbb{E}^2$. Then, we translate the middle of the three convex vertices to the origin $0\in \mathbb{E}^2$. Finally, applying a linear transformation of $GL(2,\mathbb{R})$, we can match the edges forming the convex vertex with the (reversed for convenience) unit basis vectors $-e_1$ and $-e_2$. In such a way each swallowtail appears to be equivalent to one of the forms depicted in Fig. \ref{cst}. Although the last step does not specify the linear transformation uniquely, the only ambiguity  concerns the permutation of the basis vectors $e_1$ and $e_2$.  To fix this ambiguity one needs to choose an orientation in $\mathbb{E}^2$. 
We will indicate each of two  possible orientations by putting arrows on the edges of polygons as in Fig. \ref{ost}. The affine transformations that preserve either orientation form a subgroup $\mathrm{Aff}^+(2,\mathbb{R})=GL^+(2,\mathbb{R})\ltimes \mathbb{R}^2$ of the full affine group $\mathrm{Aff}(2,\mathbb{R})$. We will denote the space of all nonequivalent oriented swallowtails with $n$ vertices by $\mathbb{V}_n$.

Considering now  the oriented edges of an $n$-gon as affine vectors in $\mathbb{E}^2$, we can arrange their  coordinates into a $2\times n$ array $P\in \mathrm{Mat}_{\mathbb{R}}(2,n)$; in so doing, the order of vectors corresponds to the order of edges.\footnote{By altering the order of vectors/edges one can easily get a polygon with crossed edges as in Fig. \ref{quarticflipped}} For instance, the array corresponding to the swallowtail in Fig.\ref{ost} looks as 
$$
P=(b_1, b_2, a_3,a_4,a_5,a_6)=\left(\begin{array}{cccccc}b_{1}^1 &b_2^1&a_3^1&a_4^1&a_5^1&a_6^1\\
b_{1}^2 &b_2^2&a_3^2&a_4^2&a_5^2&a_6^2
\end{array}
\right)\,.
$$
Clearly, each array $P$ determines the corresponding polygon up to translations in $\mathbb{E}^2$ and cyclic permutations of its columns does not affect the polygon.  For oriented swallowtails we can fix the order completely by writing the coordinates of the right edge of the middle convex vertex in the  first column. Applying the transformation  $$
G=-\left(\begin{array}{cc}
    b_1^1 &b_2^1  \\
    b_2^1 & b_2^2 
\end{array}\right)^{-1}\in GL^+(2,\mathbb{R})
$$
brings the matrix  $P$ into the canonical form 
$$
GP=(-e_1,-e_2,a_3,a_4,a_5, a_6)=\left(\begin{array}{cccccc}-1 &0&a_3^1&a_4^1&a_5^1&a_6^1\\
0&-1&a_3^2&a_4^2&a_5^2&a_6^2
\end{array}
\right)
$$
that corresponds to the swallowtail in Fig. \ref{cst}. We will refer to such swallowtails as canonical representatives. Notice that the remaining entries $a$'s are not arbitrary. First of all, the closeness of the polygon chain implies that the sum of column vectors is equal to zero, i.e.,
\begin{equation}\label{ba}
b_1+b_2+a_3+\cdots+a_n=0\,.
\end{equation}
This allows us to express one of the vectors $a_i$ as the sum of the others. The concavity condition imposes further restrictions on $a$'s. Let $[a,b]$ denote the determinant of a $2\times 2$-matrix $(a,b)$. Then an array 
$$P=(b_1,b_2,a_3,\ldots,a_n)\in \mathrm{Mat}_{\mathbb{R}}(2,n)$$ 
defines a oriented swallowtail iff its entries satisfy Eq.(\ref{ba}) together with the following inequalities: 
\begin{equation}\label{aabb}
\begin{array}{c}
 \pluk_{12}=[b_1,b_2]>0\,,\qquad \pluk_{1i}= [b_1,a_i]<0\,,\qquad \pluk_{2i}=[b_2,a_i]>0\,,\\[5mm]\pluk_{ij}=[a_i,a_j]<0\,,\qquad 3\leq i<j\leq n \,.
 \end{array}
\end{equation}
The introduced variables $\pluk_{ij}$ are convenient to express the area of a swallowtail:
$$
\mathrm{Area}(P)=\frac12\sum_{i<j}\pluk_{ij}\,.
$$

Eqs. (\ref{ba}) and (\ref{aabb}) define $\mathbb{V}_n$ -- the space of all nonequivalent oriented swallowtails with $n>3$ vertices -- as a bounded domain  in $\mathbb{R}^{2(n-3)}$. The space $\mathbb{V}_n$ enjoys a natural measure given by the volume form
\begin{equation}\label{m1}
\omega_n=\prod_{k=3}^{n-1} da^1_k\wedge da^2_k\,,
\end{equation}
where the coordinates $(a^1_k, a^2_k)$ correspond to a canonical representative $P$ with $b_1=-e_1$ and $b_2=e_2$ as in Fig. \ref{cst}. With this measure one can easily find that $\mathrm{Vol}(\mathbb{V}_4)=1/2$ and $\mathrm{Vol}(\mathbb{V}_5)=1/24$.

Geometrically, there are two natural ways to look at a $2\times n$-array: either as a set of $n$ vector in $\mathbb{R}^2$ or as a pair of vectors in $\mathbb{R}^n$. So far we have followed the former interpretation; now let us try the latter. By definition, taking the quotient of full rank matrices of  $\mathrm{Mat}_{\mathbb{R}}(2,n)$ 
by the left action of $GL^+(2;\mathbb{R})$ gives the oriented Grassmannian $\widetilde G_{\mathbb{R}}(2,n)$. It  can also be visualized as the space of all oriented $2$-planes in $\mathbb{R}^n$.\footnote{More generally, one defines $\widetilde G_{\mathbb{R}}(k,n)$ to be the space of all oriented $k$-planes in $\mathbb{R}^n$. Topologically, $\widetilde G_{\mathbb{R}}(k,n)$ is just the universal double cover of the Grassmann manifold $G_{\mathbb{R}}(k,n)$.} This allows us to think of $\mathbb{V}_n$ as a subset of the oriented Grassmannian $\widetilde G_{\mathbb{R}}(2,n)$. The subset is defined by the linear equation (\ref{ba}) and inequalities (\ref{aabb}). From this perspective the variables $\{\pluk_{ij}\}$, where $i,j=1,\ldots, n$ and $i<j$, are nothing but the Pl\"ucker coordinates defining the embedding of $\widetilde G_{\mathbb{R}}(2,n)$ into the oriented projective space $\widetilde{\mathbb{P}}^{N}=\widetilde G_{\mathbb{R}}(1,N)$ of dimension $N=\frac12n(n-1)-1$. (As a smooth manifold $\widetilde{\mathbb{P}}^N$ is diffeomorphic to the standard $N$-sphere, which is the universal covering space  of $\mathbb{P}^N$.) It is known that the image of the Pl\"ucker embedding $i: \widetilde G_{\mathbb{R}}(2,n)\rightarrow \widetilde{\mathbb{P}}^N$ is given by the intersection of the projective quadrics 
\begin{equation}\label{PR}
Q_{ijkl}: \quad \pluk_{ij}\pluk_{kl}-\pluk_{ik}\pluk_{jl}+\pluk_{jk}\pluk_{il}=0\,,\qquad \forall i<j<k<l\,.
\end{equation}
These are known as the Pl\"ucker relations. Among other things the relations say that the 
inequalities (\ref{aabb}), which single out an open domain in the intersection $\bigcap Q_{ijk}$,  are highly redundant. For instance, the relation $$\pluk_{13}\pluk_{24}=\pluk_{12}\pluk_{34}+\pluk_{23}\pluk_{14}$$ implies that $\pluk_{13}<0$ whenever 
$$
\pluk_{24}>0\,,\qquad \pluk_{12}<0\,,\qquad \pluk_{23}<0\,,\qquad \pluk_{34}>0\,,\qquad  \pluk_{14}>0\,.
$$
The above geometric interpretation in terms of swallowtails suggests that it would be enough to specify the signs of 
only consecutive minors $\pluk_{i, i+1}$ and $\pluk_{1n}$ provided Eq. (\ref{ba}) holds. 
As to the remaining relation  (\ref{ba}), it is clearly equivalent to the pair of linear equations 
\begin{equation}\label{ppp}
\pluk_{12}=-\sum_{i=3}^n\pluk_{1i}=\sum_{i=3}^n\pluk_{2i}\,,
\end{equation}
which define a plane $\Pi$ of codimension two in $\widetilde{\mathbb{P}}^N$. Summarizing all of the above, we can identify the space of swallowtails $\mathbb{V}_n$ with an open region in the intersection of the projective codimension-two plane (\ref{ppp}) with the projective quadrics (\ref{PR}); the region is specified by prescribing signs (\ref{aabb}) to the Pl\"ucker coordinates.  In terms of the projective coordinates $\pluk_{ij}$ the volume form (\ref{m1}) on $\mathbb{V}_n\subset \widetilde{\mathbb{P}}^N$ is obtained as the  restriction of the form 
$$
\Omega_n=\prod_{i=3}^{n-1}\frac{d\pluk_{1i}\wedge d\pluk_{2i}}{(\pluk_{12})^2}
$$
of degree $2(n-3)$ on $\widetilde{\mathbb{P}}^N$. The closure $\overline{\mathbb{V}}_n\subset \widetilde{\mathbb{P}}^N$ defines the integration domain for the
interaction vertices of order $n$. Topologically, $\overline{\mathbb{V}}_n$ is a smooth manifold with corners. Hence, it admits a smooth stratification. For example, the stratum of codimension one corresponds to degenerate canonical swallowtails where exactly one concave (or convex) internal angle becomes $\pi$ (or $0$).

In the last decade, much attention has been paid to the so-called {\it positive Grassmannians} because of their remarkable  applications in statistical physics, integrable models, and  scattering amplitudes. For a recent account of the subject we refer the reader to \cite{PGr, williams2021positive}. By definition, a positive Grassmannians is just an open  region of  a real Grassmann manifold where all Pl\"ucker coordinates are  strictly positive. Our considerations show that other distributions of signs among the Pl\"ucker coordinates may also be of interest, at least  for some field-theoretical problems.

\section{Discussion and Conclusions}
\label{sec:conclusions}
In this  paper, we have obtained all vertices of Chiral Theory with and without cosmological constant. As it was already pointed out in \cite{Sharapov:2022faa,Sharapov:2022awp} the final form of the vertices is remarkably simple: the exponents become linear in the new variables (or  quadratic for nonzero cosmological constant) and complicated exponential prefactors are eliminated by the corresponding Jacobians. 
Another result is an explicit description of the configuration space. It is given by what we call swallowtails --  concave polygons that have two edges coinciding with two adjacent edges of the unit square. Another way to describe the same geometric shape is to consider the space of convex polygons that can be inscribed into a unit square with one edge coinciding with the diagonal. The area of the swallowtail also has a meaning and determines the coefficient of the cosmological term.

There is an intriguing relation \cite{Sharapov:2017yde,Sharapov:2022eiy} to the formality theorems, in particular to Shoikhet--Tsygan--Kontsevich formality \cite{Kontsevich:1997vb, Shoikhet:2000gw}. This indicates that with the help of the simple configuration space we have now the $A_\infty/L_\infty$-relations can be proved via Stokes theorem, which we will address elsewhere. A more intriguing question is whether the configuration space we identified can be generalized and extended into the `bulk'. Indeed, the Poisson structure $\pi$ we begin with is just $\epsilon^{AB}$, i.e., symplectic and constant. For this reason, all Kontsevich--Shoikhet's graphs where $\pi$ is hit by derivatives  disappear. What remains of undifferentiated $\pi$ is the Moyal--Weyl star-product and the Feigin--Felder--Shoikhet cocycle \cite{FFS} that justifies the existence of cubic vertices as well as higher order vertices. These structures are also closely related to the deformation quantization of Poisson Orbifolds \cite{Sharapov:2022eiy,Sharapov:2022phg}. There should also exist an extension of our construction to Feigin's $gl_\lambda$ \cite{Feigin}. Another direction is due to a surprising appearance of pre-Calabi--Yau algebras \cite{IYUDU202163, kontsevich2021pre}, see Appendix \ref{CY}. Eventually, all of this should admit a description in terms of a certain two-dimensional topological field theory. 

Another interesting direction is to uncover what is special about the multi-linear products we found as compared to other representatives of the same $A_\infty/L_\infty$-algebra. From the viewpoint of a sigma-model $d\Phi=Q(\Phi)$, different choices of coordinates for the underlying homological vector field $Q$ translate into redefinitions of fields $\Phi$, most of which are too nonlocal to give meaningful interactions. In other words, most of coordinates for $Q$ violate the equivalence theorem. It is tempting to say that there should always exist a coordinate system that leads to maximally local interactions. For every field theory, one can think of $Q$ as a deformation of a `free' homological vector field $Q_0$ that defines a certain graded Lie algebra (via the bilinear maps of the associated $L_\infty$-algebra) and the first order deformation corresponds to a certain Chevalley--Eilenberg cocycle. Therefore, the maximal locality requirement selects one specific representative of the Chevalley--Eilenberg cohomology. It is easy to see that the vertices we found are maximally local (any field redefinition can only increase the number of derivatives). It would be interesting to find out exactly which property of the Chevalley--Eilenberg cohomology is equivalent to maximal locality in the field theory language.

The immediate  applications of the obtained results are obvious: (a) it would be interesting to look for exact solutions building upon the general tools \cite{Sezgin:2005pv,Didenko:2009td,Aros:2017ror}  worked out in the context of formal HiSGRA;\footnote{By a {\it formal HiSGRA} we mean the sigma-models above, $d\Phi=Q(\Phi)$, without taking locality into account.   Interestingly, the equations may have nicely looking solutions even for physically nonsensical vertices hidden in $Q$ (see \cite{Didenko:2021vdb,Didenko:2021vui} for the careful treatment of a black brane solution).  } (b) it is important to compute holographic correlation functions as to compare with (Chern--Simons) vector models (Chiral Theory should be dual to a closed subsector of Chern--Simons vector models \cite{Sharapov:2022awp}); (c) presymplectic AKSZ actions along the lines of \cite{Sharapov:2021drr} can be constructed as well as possible counterterms and anomalies can be classified \cite{Sharapov:2020quq}; (d) the study of integrability of Chiral Theory \cite{Ponomarev:2017nrr} and its relation to twistors \cite{Tran:2021ukl} should also be a fruitful direction.  

In the regard to item (d) let us point out that the $A_\infty$-algebra of Chiral Theory $\hat{\mathbb{A}}$ naturally defines a two-dimensional theory, which should be closely related to an important observation made in \cite{Ponomarev:2017nrr} that the equations of Chiral Theory in flat space can be cast into the form of the principal chiral model. Indeed, the higher spin algebra $\hs$ is given by the tensor product $A_\hhbar \otimes A_1 \otimes \mathrm{Mat}_N$, where $A_\lambda$ is the Weyl algebra, with $\hhbar$ being the parameter of noncommutativity (effective cosmological constant). Clearly, all the higher products  of $\hat{\mathbb{A}}$ owe their existence to the first factor $A_\hhbar$ and its bimodule $A^\ast_\lambda$, while the rest part, $B=A_1 \otimes \mathrm{Mat}_N$, enters via the usual associative product. What makes the system four-dimensional is the functional dimension of $\hs$. If we simply drop $A_1$ and take $B=\mathrm{Mat}_N$ (or any other associative algebra with zero functional dimension), we can write the same sigma-model $d\Phi=Q(\Phi)$,  but on a two-dimensional space.\footnote{The functional dimension of $A_\hhbar$ implies that the theory is off-shell in $2d$ or, perhaps similarly to \cite{Alkalaev:2019xuv}, can be understood as an on-shell one for infinitely-many fields. In $3d$ the theory would be on-shell to begin with. } The factor $A_\hhbar$ implies that $AdS_2$ is a natural vacuum for such a system. According to \cite{Ponomarev:2017nrr} this system (as well as the whole Chiral Theory) should be integrable. Its exact solutions can perhaps be obtained by adapting the techniques from \cite{Sharapov:2019vyd} and it would be interesting to compare it with the standard techniques from integrable models. With $B=\mathrm{Mat}_2$ one can get a $3d$ interpretation. The functional dimension of {$A_\lambda$}, which is $2$, corresponds to off-shell equations in $2d$ and to on-shell in $3d$, which seem to be the most natural dimensions for the theory underlying the Chiral one. It would be interesting to uncover the properties of this parent theory.

Lastly, let us present the Chiral HiSGRA equations of motion in a concise form.\footnote{In this regard one can mention the very recent Didenko equations \cite{Didenko:2022qga} that are claimed to give a local theory in $AdS_4$. Provided the vertices are explicitly extracted from \cite{Didenko:2022qga} it would be interesting to compare them with Chiral Theory in $AdS_4$. A closely related interesting open question is whether there are more than one local higher spin gravity in $AdS_4$. Without taking locality into account there are infinitely many formal deformations at higher orders \cite{Vasiliev:1999ba,Sharapov:2020quq}. Also, similar ambiguities are present for low spin theories. Therefore, the question of (non-)uniqueness of local theories remains open, which is also relevant for the study of quantum consistency of Chiral Theory. } As it has been discussed, the $\mathcal{V}$-vertices come from trees with two branches and $\mathcal{U}$-vertices originate from trees with just one branch. The expression for the most general branch $B_n[C,\dots, \omega,\ldots C]$ is given in \eqref{compactform}. Let us introduce the sum 
$$
B[\omega, C]=\sum_{n=0}^\infty B_n[C,\dots, \omega,\ldots C]
$$
over all possible branches and orderings of zero-forms $C$ therein. 
With this we can write the equations of motion as
\begin{align*}
    d\omega=B[\omega, C]\star B[\omega,C]\Big|_{z=0}\,,\qquad
    dC=B[\omega, C]\circ C-C\circ B[\omega, C]\,.
\end{align*}
As is seen, upon switching  on interaction, the one-form field $\omega$ on the right is just replaced  with $B=\omega + O(C)$. One  can regard the full branch $B$ as an effective  field $\omega$ `dressed' by $C$. The equations can also be understood as a Poisson sigma-model, see Appendix \ref{CY}.

\section*{Acknowledgments}
\label{sec:Aknowledgements}
The work of E. S. and R. van D. was partially supported by the European Research Council (ERC) under the European Union’s Horizon 2020 research and innovation programme (grant agreement No 101002551) and by the Fonds de la Recherche Scientifique --- FNRS under Grant No. F.4544.21. A. Sh. gratefully acknowledges the financial support of the Foundation for the Advancement of Theoretical Physics and Mathematics “BASIS”. The work of A. Su. was supported by the Russian Science Foundation grant 18-72-10123 in association with the Lebedev Physical Institute. E.S. expresses his gratitude to the Erwin Schr{\"o}dinger Institute in
Vienna for the hospitality during the program ``Higher Structures and Field Theory'' while this work was in progress.

\appendix

\section{Homological perturbation theory: a recipe}
\label{app:homo}
In \cite{Sharapov:2022awp,Sharapov:2022faa}, a detailed description was given of how to construct vertices in Chiral HiSGRA from homological perturbation theory. Here we present only the practical steps required for explicit calculations and focus upon the case of  nonzero cosmological constant as the flat limit naturally arises from this.

As a first step, one needs to find a suitable multiplicative resolution  of the higher spin algebra.
 To this end,  we introduce the algebra $\mathbb{C}[y^A,z^A]$ of complex polynomial functions in $y^1$, $y^2$, $z^1$, $z^2$.\footnote{As a historical comment, it should be pointed out that a highly nontrivial idea of getting vertices via solving simple equations with respect to an appropriately introduced $z$-extension  was put forward in \cite{Vasiliev:1990cm, Vasiliev:1999ba}. However, an important physical condition to have well-defined interactions instead of just formally consistent ones was not imposed in \cite{Vasiliev:1990cm,Vasiliev:1999ba}, see e.g. \cite{Giombi:2009wh,Giombi:2010vg,Boulanger:2015ova,Skvortsov:2015lja} for explicit checks that revealed this fact, which is a difference between an ansatz for interactions and an actual theory. It was also understood \cite{Bekaert:2015tva,Maldacena:2015iua,Sleight:2017pcz,Ponomarev:2017qab} that the whole class of theories aimed for in \cite{Vasiliev:1990cm,Vasiliev:1999ba} is subtle due to featuring stronger nonlocalities than the standard field theory techniques allow for. This class is supposed to be dual to Chern--Simons vector models \cite{Klebanov:2002ja,Sezgin:2003pt,Leigh:2003gk,Giombi:2011kc}, which makes it a very interesting target. The main point of \cite{Sharapov:2022awp,Sharapov:2022faa} and of the present paper is that Chiral Theory is a well-defined  (i.e., local and smoothly depending on the cosmological constant) theory, which for $\hhbar\neq 0$ should be a closed subsector of Chern--Simons vector models' dual theory that is yet to be found. A detailed discussion of the key differences between the proposals of \cite{Vasiliev:1990cm,Vasiliev:1999ba,Didenko:2018fgx,Didenko:2019xzz,Didenko:2020bxd,Gelfond:2021two} and the present one can be found in \cite{Sharapov:2022awp}. For instance, according to \cite{Vasiliev:1999ba} there is no flat limit for higher-spin interactions, which is certainly not the case for Chiral Theory. } The algebra is equipped with the Weyl--Moyal star-product 
\begin{align}\label{superstar}
    (f\star g) (Y)=\exp(Y^a\partial^1_a+Y^a\partial^2_a+\Omega^{ab}\partial^1_a\partial^2_b)f(Y_1)g(Y_2)|_{Y_{1,2}=0} \,,
\end{align}
where $Y^a\equiv (y^A,z^A)$ and the matrix $\Omega^{ab}$ is given by
\begin{align*}
    \Omega^{ab}=-\begin{pmatrix}
        \lambda \epsilon & \epsilon \\
        -\epsilon & 0
    \end{pmatrix} \,.
\end{align*}
The symbol of the star-product operator is given by
\begin{align*}
    \exp(p_{01}+p_{02}+r_{01}+r_{02}+p_1\cdot r_2-r_1\cdot p_2+\lambda p_{12}) \,,
\end{align*}
where $r$ to $z$ is the same as $p$ to $y$. One can also write the star-product in the integral form 
\begin{align}\label{IS}
    (f\star g)(y,z)&= \int du\,dv\,dp\,dq\,f(y+u,z+v) g(y+q,z+p) \exp(v\cdot q-u\cdot p +\lambda\, p\cdot v)\,.
\end{align}
The generators of the algebra, $y^A$ and $z^A$, act as
\begin{align*}
   & \begin{aligned}
    y_{A}\star f &=(y_{A}  -\hhbar\, \pl^y_{A}-\pl^z_{A})f\,, \\
    f \star y_{A} &=(y_{A}  +\hhbar\, \pl^y_{A}-\pl^z_{A})f\,, 
    \end{aligned}   &  z_{A}\star f&=f\star z_A= (z_{A}+\pl^y_{A}) f\,.
\end{align*}
Recall that all indices $A$ are raised and lowered with the help of the $\epsilon$-symbol. 
With these relations one can find that the function $\varkappa= \exp(z^A y_A)$ satisfies the relations\footnote{Here we tacitly extend the star-product from polynomials to some analytic functions of $z$'s and $y$'s for which the integral (\ref{IS}) makes sense.}
\begin{align}\label{klein}
    z_A\star\varkappa=\varkappa\star z_A=0 \,.
\end{align}

Then we extend the star-product algebra $\mathbb{C}[y^A,z^A]$ to the exterior algebra of polynomial differential forms $\mathfrak{R}=\mathbb{C}[y^A,z^A,dz^A]$ endowed with the exterior differential $d_z$ in $z$'s. The product in $\mathfrak{R}$ is the combination of the star-product and the usual exterior product of the basis differentials $dz^A$. The latter will be denoted by the  dot product, not by the wedge $\wedge$.  The Poincar\'e Lemma gives then solutions to the equations $d_z f^{(1)}=f^{(2)}$ and $d_z f^{(0)}=f^{(1)}$ for any closed one-form $f^{(1)}=dz^A f^{(1)}_A(z)$ and a two-form $f^{(2)}=\frac{1}{2}\epsilon_{AB}f^{(2)}(z)dz^Adz^B$. They read
\begin{align*}
    f^{(1)}&=h[f^{(2)}]=dz^A z_{A}\int_0^1 t dt f^{(2)}(tz) \,, & f^{(0)}&=h[f^{(1)}]=z^A\int_{0}^{1} dt f_{A}^{(1)}(tz) \,.
\end{align*}
We also set $h[f^{(0)}]=0$ for any zero-form $f^{(0)}$. These relations define $h$ as the standard contracting homotopy for the de Rham complex of polynomial differential forms:
\begin{equation}\label{dh}
d_zh+hd_z=1-\pi\,,
\end{equation}
$\pi$ being the natural projection onto the subspace of $z$ independent zero-forms. The form degree and the exterior differential $d_z$ give $\mathfrak{R}$ the structure of a differential graded algebra (or dg-algebra for short). 
Rel. (\ref{dh}) implies that the cohomology of the dg-algebra $(\mathfrak{R}, d_z)$ is concentrated in degree zero and is described by $z$- and $dz$-independent polynomials. Hence, $H(\mathfrak{R}, d_z)\simeq A_\lambda$ and $(\mathfrak{R}, d_z)$ define a multiplicative resolution (aka  {\it model}) of the algebra $A_\lambda$. Starting with the differential graded algebra  $\mathfrak{R}$ one can systematically construct resolutions for many other algebras. For example, taking the tensor product of $\mathfrak{R}$ with an associative algebra  $B$ yields the dg-algebra $\mathfrak{R}\otimes B$, where $d_z$ extends to $B$ by zero. 
The algebra $\mathfrak{R}\otimes B$ defines then a model of the tensor product algebra $A_\lambda\otimes B$. Another possibility is to consider the trivial extension of $\mathfrak{R}$ by a differential $\mathfrak{R}$-bimodule ${M}$ concentrated in a single degree.  The result is given by a dg-algebra $\mathfrak{R}\oplus{M}$ with the product 
\begin{equation}\label{prod}
(b,a)(\tilde b,\tilde a)=(b\tilde b, b\tilde a+a\tilde b)\qquad \forall b,\tilde b\in \mathfrak{R}, \quad \forall a,\tilde a\in M\,.
\end{equation}
Since the differential necessarily annihilates ${M}$, the algebra $\mathfrak{R}\oplus{M}$ defines a model for the trivial extension $A_\lambda\oplus {M}$. In application to Chiral Theory we combine both the operations $\otimes$ and $\oplus$. Specifically, we take $B=A_1\otimes \mathrm{Mat}_N$ and define the bimodule structure on the space of formal power series $M=\mathbb{C}[[y^A]]$ by setting\footnote{The quickest way to check the bimodule axioms is with the $\tau$-involution introduced in \cite[App. A]{Sharapov:2022awp}.}
\begin{equation}\label{circaction}
\begin{aligned}
     y^A\circ a=(-\partial_y^A+\hhbar y^A)a\,, && z^A\circ a=a\circ z^A=0\,,\\
     a\circ y^A=(-\partial_y^A-\hhbar y^A)a\,,  &&\qquad dz^A\circ a=a\circ dz^A=0
\end{aligned}
\end{equation}
for all $a\in M$. As is seen the left and right actions of $\mathfrak{R}$ on ${M}$ are different unless $\hhbar\neq 0$. The quickest way to check the bimodule axioms is with the $\tau$-involution introduced in \cite[App. A]{Sharapov:2022awp}. For any function $a(y,z)$ we set
\begin{equation}\label{tauinvol}
a^\tau(y,z)=a(z,y)e^{z^Ay_A}\,.
\end{equation}
Clearly, $\tau^2=1$. Then one can equivalently define  the above $\circ$-product by the relation
\begin{equation}\label{moduleaction}
b\circ a\circ \tilde b= (b\star a^\tau\star \tilde b)^\tau\,,\qquad \forall b,\tilde b\in \mathfrak{R}, \quad a\in M\,,
\end{equation}
and the condition that $dz^A \circ a=0=a\circ dz^A$.  In this form, the bimodule axioms for the $\circ$-product hold due to the associativity of the star-product.  

The elements of the bimodule $M$ are assigned the degree one. Then the differential graded algebra 
$$
\mathcal{R}=(\mathfrak{R}\oplus M)\otimes A_1\otimes \mathrm{Mat}_N=\mathfrak{R}\otimes A_1\otimes\mathrm{Mat}_N\bigoplus M\otimes A_1\otimes\mathrm{Mat}_N=\mathbf{R}\bigoplus \mathfrak{M}
$$
defines a multiplicative resolution of the algebra 
\begin{equation}\label{HSA}
\mathcal{A}=H(\mathfrak{R},d_z)=A_\lambda\otimes A_1\otimes\mathrm{Mat}_N\bigoplus M\otimes A_1\otimes\mathrm{Mat}_N=\mathfrak{A}\bigoplus\mathfrak{M}\,.
\end{equation}
The left summand $\mathfrak{A}$ is given by the matrix extension of the higher spin algebra $\hs=A_\lambda\otimes A_1$, the algebra where the one-form field $\omega$ assumes its values. The right summand $\mathfrak{M}$ defines then a bimodule over the algebra $\hs\otimes \mathrm{Mat}_N$, the target space of the zero-form field $C$. 
Recall that the differential in the algebra $\mathcal{R}$ is given by the trivial extension of the exterior differential $d_z$. As observed in \cite{Sharapov:2022awp}, the differential $d_z$ admits a nontrivial perturbation by another differential $\delta$ of degree one. The latter is defined as 
\begin{equation}\label{del}
\delta(b,a)=(\delta a, 0)\,,\qquad \delta a = a^\tau dz^1 dz^2\qquad\forall b\in \mathbf{R},\quad \forall a\in\mathfrak{M}\,.
\end{equation}
It is clear that $\delta^2=0$ and  $d_z\delta=-\delta d_z=0$. Eq. (\ref{klein}) ensures the graded Leibniz identity for the differential (\ref{del}) and the product (\ref{prod}). Therefore, the sum $D=d_z+\delta$ endows the algebra $\mathcal{R}$ with a new differential of degree one. It is not hard to see that the cohomology of the perturbed differential is given by the same algebra (\ref{HSA}), that is, $H(\mathcal{R},D)\simeq H(\mathcal{R}, d_z)=\mathcal{A}$. 
Having the same cohomology, the dg-algebras $(\mathcal{R}, d_z)$ and $(\mathcal{R}, D)$ are not quasi-isomorphic to each other: the former algebra  is formal, whereas the latter is not. The last fact implies that in addition to the binary product $m_2$ (induced by that in $\mathcal{R}$) the cohomology space  $H(\mathcal{R}, D)$ enjoys higher  multi-linear products $m_k$ making it into an $A_\infty$-algebra. (For the definition of an $A_\infty$-algebra see e.g. \cite{stasheff2018linfty}, \cite{IYUDU202163}.) This $A_\infty$-algebra, let us denote it by $\hat{\mathbb{A}}$, is called the {\it minimal model} of the dg-algebra $(\mathfrak{R}, D)$. By definition, the binary product $m_2$ coincides with the associative product in $\mathcal{A}$ and the triple product $m_3$ is given by a nontrivial  Hochschild cocycle representing a class of $HH^3(\mathcal{A},\mathcal{A})$.


Homological perturbation theory (which details can be found in Refs. \cite{HK,GLS, merkulov1999strongly}) provides explicit formulas for the multi-linear products $m_2, m_3, m_4, \ldots$ of the $A_\infty$-algebra $\hat{\mathbb{A}}$. All the products are constructed  as compositions of two basic operations: the contracting homotopy $h$ and the associative product in the multiplicative resolution $\mathcal{R}$. The latter gives rise to the coderivation $\mu$ defined by
$$
\mu (b,\tilde b)=(-1)^{\deg b-1}b\star \tilde b\,,\qquad \mu(b,a)=(-1)^{\deg b-1}b\circ a\,,\qquad \mu(a,b)=-a\circ b\,,
$$
for all $b,\tilde b\in \mathbf{R}$ and $a\in \mathfrak{M}$.
Suitable compositions are conveniently depicted by rooted planar trees. Each such a tree graph consists of  vertices, internal edges,  and external edges.  Both ends of an internal edge are on two vertices. All edges are oriented and orientation is indicated by an arrow. Each vertex has two incoming and one outgoing edge. An external edge has one end on a vertex and another end is free. The graphs are supposed to be connected. All the vertices correspond to the product $\mu$, whereas the internal edges depict the action of the contracting homotopy $h$:
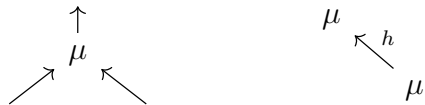
$$ \begin{tikzcd}[column sep=small,row sep=small]
   & {}& \\
    & \mu\arrow[u]  & \\
    \arrow[ur]  & & \arrow[ul]   & 
\end{tikzcd} \;\;\;\;\;\;\;\;\;\;\;\;
\begin{tikzcd}[column sep=small,row sep=small]
    \mu & \\
    &\mu\arrow[ul, "h" ']   
\end{tikzcd}
$$
By definition, the algebra $\mathfrak{A}$ and the $\mathfrak{A}$-bimodule $\mathfrak{M}$ have degrees $-1$ and $0$, respectively, whereas all the products $m_k$ of the $A_\infty$-algebra $\hat{\mathbb{A}}$ are of degree one\footnote{In \cite{Sharapov:2022wpz} we used a different convention according to which all $m_k$'s are of degree $-1$. In that case, the elements of $\mathfrak{M}$ have still  degree $0$, whereas $\mathfrak{A}$ is placed in degree $1$.}. By degree considerations, each nonzero product $m_k$ may have either one or two  arguments in $\mathfrak{A}$ and the other in $\mathfrak{M}$. In the first case the image of $m_k$ belongs to the algebra $\mathfrak{A}$, whereas in the second to the bimodule $\mathfrak{M}$.  In field-theoretical terms, these two components of the product $m_k$ correspond to the interaction vertices of $\mathcal{V}$ and $\mathcal{U}$ types. Let us consider them separately. 

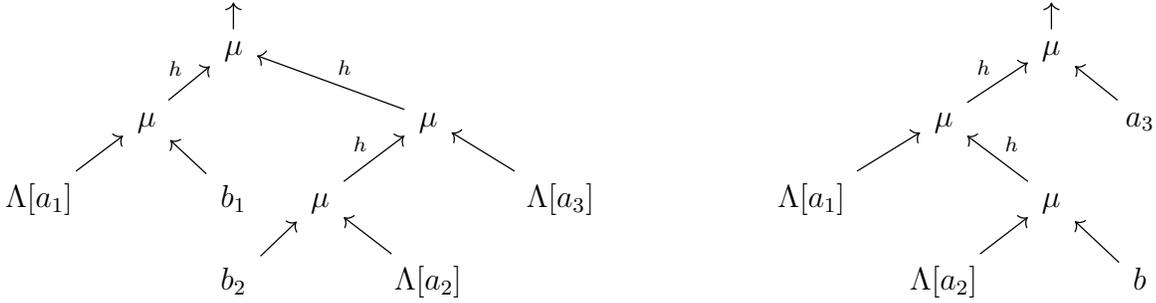
\begin{figure}
    \begin{tikzcd}[column sep=small,row sep=small]
   &&{}&&&&\\
    && \arrow[u]\mu  &&& \\
    & \mu\arrow[ur,"h"]& &&  \arrow[ull,"h"']\mu & & \\
    \Lambda[a_1]\arrow[ur]&& \arrow[ul]b_1 &   \mu\arrow[ur, "h" ]&&\arrow[ul]\Lambda[a_3]&\\
    &&b_2\arrow[ur]&&\Lambda[a_2]\arrow[ul]&&
\end{tikzcd}
   \begin{tikzcd}[column sep=small,row sep=small]
   &&&&&{}&&&&\\
   && &&&\mu \arrow[u]  &&& \\
    &&&&\mu\arrow[ru, "h" ]&&a_3\arrow[lu]&\\
    &&&\Lambda[a_1]\arrow[ur]&&\mu\arrow[lu, "h"']&&&\\
    &&&&\Lambda[a_2]\arrow[ur]&&b\arrow[ul]&&&
\end{tikzcd}
\caption{A planar rooted tree on the left panel corresponds to the analytical expression $h(h\delta a_1\star b_1)\star h(h(b_2\star h\delta a_2) \star h\delta a_3)|_{z=0}$ contributing to  $m_5(a_1,b_1,b_2,a_2,a_3)$; here $b_1,b_2\in \mathfrak{A}$ and $a_1,a_2,a_3\in \mathfrak{M}$. The right panel shows a planar  tree for the expression $h(h\delta a_1\star h(h\delta a_2\star b))\circ a_3|_{z=0}$, which contributes to $m_4(a_1,a_2,b,a_3)$; here $b\in \mathfrak{A}$ and $a_1,a_2,a_3\in \mathfrak{M}$. Notice the `bare' argument $a_3$.
}\label{PT}
\end{figure}

\paragraph{Two arguments in $\mathfrak{A}$.} The corresponding component of $m_k$ is described by the sum of trees with two branches, see left panel in Fig.  \ref{PT}. 
The incoming external edges (or leaves) correspond to the arguments of  $m_k$. More precisely, the arguments $b_1, b_2\in \mathfrak{A}$ may decorate only the four end leaves on {\it different} branches. The other leaves are decorated by the expressions 
$\Lambda[a_i]=h\delta a_i$ for $a_i\in \mathfrak{M}$. The only outgoing external edge (or root) corresponds to the value of the product $m_{k}(a_1,\ldots,b_1, \ldots, b_2, \ldots, a_{k-2})$ that arises after setting $z=0$.  The order of arguments is determined  by the natural order of incoming edges at each vertex of a planar tree. The contributions of different trees are added up (with unit weight) to obtain the desired $m_k$.

\paragraph{One argument in $\mathfrak{A}$.} The product $m_k(a_1,\ldots, b, \ldots,a_{k-1})$ is obtained by summing  up the one-branch trees; an example of such a tree is shown in the right panel of Fig. \ref{PT}. The only argument $b$ of  $\mathfrak{A}$ decorates one of the two end leaves, whereas the leaf incoming the root vertex is decorated by a `bare' element $a\in \mathfrak{M}$.  As  above, the order of arguments is determined  by the natural order of incoming edges at each vertex and the root edge symbolizes setting $z=0$ in the final expression for $m_k$. Unlike the previous case, the integrals defining the corresponding analytical expressions require a minor regularization as explained in the main text.

Finally, performing graded symmetrization  of the arguments of the products $m_k$  makes our $A_\infty$-algebra $\hat{\mathbb{A}}$ into a minimal $L_\infty$-algebra $\mathbb{L}$. At the level of interaction vertices such symmetrization is automatically achieved by substituting the form fields $\omega$ and $C$ instead of the arguments $b$'s and $a$'s. It is the $L_\infty$-algebra $\mathbb{L}$ that governs the interaction in  Chiral Theory. 

\section{Pre-Calabi--Yau algebras and duality map}\label{CY}

The above construction of the $A_\infty$-algebra $\hat{\mathbb{A}}$ by means of homological perturbation theory is  absolutely insensitive to the  choice of the tensor factor $B=A_1\otimes \mathrm{Mat}_N$. For any associative algebra $B$ we get $\hat{\mathbb{A}}=\mathbb{A}\otimes B$, where the minimal $A_\infty$-algebra $\mathbb{A}$ extends the binary product in $A_\lambda\oplus M$. Furthermore, the $A_\lambda$-bimodule $M$ is actually dual to the algebra $A_\lambda$ viewed as the natural bimodule over itself, i.e., $M\simeq A_\lambda^\ast$.  The corresponding nondegenerate pairing is given by 
\begin{align}\label{cpar}
     \langle a| u\rangle =e^{p_{12}}a(y_1)u(y_2)|_{y_i=0}\,,\qquad \forall a\in A_\lambda\,,\quad \forall u\in M\,.
\end{align}
One can easily verify that $ \langle b\star a\star c| u\rangle =  \langle a| c\circ u\circ b\rangle$. Recall that the elements of the algebra $A_\lambda$ are prescribed, by definition, the degree $-1$, whereas the elements of the bimodule $M$ live in degree $0$. With this convention all the products $m_k$ in $\mathbb{A}$ have degree one. By the above isomorphism, we can write\footnote{Dualization inverts the $\mathbb{Z}$-degree, while the symbol $[1]$ shifts the degree of the dual module by one. } $A_\lambda\oplus M\simeq A_\lambda\oplus A^\ast_\lambda[1]$. The pairing (\ref{cpar}) gives rise to a canonical symplectic form $\omega$ on the graded vector space $A_\lambda\oplus A^\ast_\lambda[1]$. This is defined as
\begin{align}\label{sf}
   \omega ( a+u ,\tilde a +\tilde u)= \langle a| \tilde u\rangle - \langle \tilde{a}|u\rangle \,.
\end{align}
Clearly, $\deg \omega =1$. Define the sequence of multi-linear forms 
\begin{align}\label{Sk}
    S_k(\alpha_0, \alpha_1,\ldots,\alpha_k)=\omega\big(\alpha_0, m_{k}(\alpha_1,\ldots,\alpha_k)\big)\,,\qquad k=2,3,\ldots\,,
\end{align}
where $\alpha=a+u\in A_\lambda\oplus A^\ast_\lambda[1]$. By definition, the $A_\infty$-algebra $\mathbb{A}$  is called {\it cyclic} (w.r.t. $\omega$) if
\begin{align}\label{Scycle}
      S_k(\alpha_0,\alpha_1,\ldots,\alpha_k)=(-1)^{\bar{\alpha}_0(\bar\alpha_1+\cdots+\bar\alpha_{k})}   S_k(\alpha_1,\ldots,\alpha_{k},\alpha_0)\,,
\end{align}
where $\bar\alpha=\deg \alpha-1$. A direct verification shows that the above identities are indeed satisfied. Hence, $\mathbb{A}$ is a cyclic $A_\infty$-algebra.  The other two properties of $\mathbb{A}$ -- shifted duality $M=A^\ast_\lambda[1]$ and the fact that $A_\lambda$ is a subalgebra of $\mathbb{A}$ -- allows us to classify   $\mathbb{A}$ as a $2$-pre-Calabi--Yau algebra \cite{IYUDU202163},
\cite{kontsevich2021pre}. The general definition is as follows. 
\begin{definition}
A $d$-pre-Calabi--Yau structure on an $A_\infty$-algebra $A$ is a cyclic $A_\infty$-structure on $A\oplus A^\ast[1-d]$, associated with the natural pairing between $A$ and $A^\ast[d-1]$,  such that $A$ is an $A_\infty$-subalgebra in $A\oplus A^\ast[1-d]$.
\end{definition}

In our case, $d=2$ and the role of an $A_\infty$-algebra $A$ is played by the associative algebra $A_\lambda$.  The latter is clearly a subalgebra in $\mathbb{A}$.  The cyclicity property (\ref{Scycle}) relates various structure maps $m_k$ among themselves. In particular, it connects the components of  the $m_k$'s with one and two arguments in $A_\lambda$:
$$
\langle a_1|m_{k+1}(u_1,\ldots, a_2,\ldots, u_k)\rangle=-\langle m_{k+1}( u_2,\ldots, a_2,\ldots, u_{k},a_1)|u_1\rangle\,.
$$
In the main text, we use these relations to express the $\mathcal{U}$-vertices via $\mathcal{V}$-vertices. 

In the case that the associative algebra $B$ enjoys a trace, one can easily extend the 2-pre-Calabi--Yau structure from $\mathbb{A}$ to the tensor product $\hat{\mathbb{A}}=\mathbb{A}\otimes B$. The symplectic structure  extends as
\begin{equation}\label{ww}
\Omega(\alpha\otimes b,\tilde \alpha\otimes\tilde b)=\omega(\alpha,\tilde\alpha)\mathrm{Tr}(b\tilde b)\qquad \forall \alpha,\tilde \alpha\in \mathbb{A}\,,\quad \forall b,\tilde b\in B\,,
\end{equation}
and the multi-linear functions (\ref{Sk}) take the form
\begin{equation}\label{SSk}
{\bf{S}}_k(\alpha_0\otimes b_0,\ldots, \alpha_k\otimes b_k)=S_k(\alpha_0,\ldots,\alpha_k)\mathrm{Tr}(b_0\cdots b_k)\,.
\end{equation}
The cyclic invariance (\ref{Scycle}) of the ${\bf S}_k$'s is obvious. 

Following the ideas of noncommutative geometry \cite{Kontsevuch:2006jb}, one can regard the cyclic forms (\ref{SSk}) as functions on a noncommutative manifold associated with $\hat{\mathbb{A}}$. The constant symplectic structure (\ref{ww}) gives then rise to a kind of Gerstenhaber bracket on the space of such functions, called {\it necklace bracket} \cite{kontsevich2021pre}.  This can be viewed as a noncommutative counterpart of the Schouten--Nijenhuis bracket on polyvector fields. It is convenient to combine the functions (\ref{SSk})  into a single non-homogeneous function $\mathbf {S}=\sum_{k=2}^\infty \mathbf{S}_k$. With the help of the necklace bracket all  $A_\infty$-structure relations for $\hat{\mathbb{A}}$ can be compactly encoded by the equation
$
    [\mathbf{S},\mathbf{S}]_{\mathrm{nec}}=0
$\,.
On passing from the $A_\infty$-algebra $\hat{\mathbb{A}}$ to the associated $L_\infty$-algebra $\mathbb{L}$, the last equation turns into the Batalin--Vilkovisky equation for the `classical master action' $\mathbf{S}(\omega, C)$ of ghost number 2 on the target space of form fields $\omega$ and $C$.
Geometrically, one can regard $\mathbf{S}(\omega, C)$ as a Poisson bivector on the space of fields $C$. Upon this interpretation the field equations  (\ref{eq:chiraltheory}) define a  Poisson sigma-model in four dimensions. Schematically, 
\begin{align}
    d C^i &= \pi^{ij}(C)\, \omega_j\,, &d\omega_k&= \tfrac12\pl_k \pi^{ij}(C)\, \omega_i\, \omega_j\,,
\end{align}
where the Poisson bivector $\pi^{ij}(C)$ is read off from $\mathbf{S}=\pi^{ij}(C)\omega_i\omega_j$.

\section{All order vertices}
\label{app:allorders}

\subsection{Jacobians}
\label{app:Jacobians}
Here we compute the Jacobians introduced in sections \ref{sec:flat} and \ref{sec:uvertices}.
\paragraph{Single branch.} Eqs. \eqref{B_{n+1}} and \eqref{B_{n+1}newcoordinates} are related by a change of variables \newline $\{u_{n,1},v_{n,1},\dots,u_{n,n},v_{n,n},t_{2n+1},t_{2n+2}\}$ to $\{u_{n+1,1},v_{n+1,1},\dots u_{n+1,n+1},v_{n+1,n+1}\}$ with the Jacobian
\begin{align} \label{det}
    |J_n|=
    \begin{vmatrix}
        \tfrac{(1-t_{2n+1})t_{2n+2}}{1-t_{2n+1}U_n}\delta_{ij} & 0 & 0 & u_{n,i} \delta_{jj} \\
        -\tfrac{t_{2n+1}(1-V_n)}{1-t_{2n+1}U_n}\delta_{ij} & \delta_{ij} & 0 & 0 \\
        \tfrac{t_{2n+1}t_{2n+2}(t_{2n+1}-1)}{(1-t_{2n+1}U_n)^2} \delta_{jj} & 0 & \tfrac{t_{2n+2}(1-U_n)}{(1-t_{2n+1}U_n)^2} & \frac{t_{2n+1}(1-U_n)}{1-t_{2n+1}U_n} \\
        \tfrac{t_{2n+1}^2 (1-V_n)}{(1-t_{2n+1}U_n)^2}\delta_{jj} & -\tfrac{t_{2n+1}}{1-t_{2n+1}U_n}\delta_{jj} & \tfrac{1-V_n}{1-t_{2n+1}U_n} & 0
    \end{vmatrix} \,,
\end{align}
where $i,j=1,\dots,n$. Keeping in mind that some entries are vectors or matrices, Gaussian elemination allows one to find a diagonal form. To give an example of the steps taken during this process, one can multiply the matrix in the second row of \eqref{det} by $\frac{t_{2n+1}}{1-t_{2n+1}U_n}$ and add each of its rows to the last row in \eqref{det}. After a few manipulations, one arrives at
\begin{align*}
    |J_n|&=\left| \text{diag}\Big(\tfrac{(1-t_{2n+1})t_{2n+2}}{1-t_{2n+1}U_n}\delta_{ij} + \tfrac{(1-t_{2n+1})t_{2n+1}t_{2n+2}}{(1-t_{2n+1}U_n)^2} u_{n,i}\delta_{jj},\;\delta_{ij}+\tfrac{t_{2n+1}}{1-t_{2n+1}U_n}u_{n,i}\delta_{jj},\;\tfrac{1-V_n}{1-t_{2n+1}U_n}, \;t_{2n+1}\Big)\right| \,.
\end{align*}
Notice that the matrix is not completely diagonal as not all of its blocks are proportional to $\delta_{ij}$. We obtain
\begin{align*}
    |J_n|&=\frac{t_{2n+1}(1-V_n)}{1-U_n}\text{det}\Big(\tfrac{(1-t_{2n+1})t_{2n+2}}{1-t_{2n+1}U_n}\delta_{ij} + \tfrac{(1-t_{2n+1})t_{2n+1}t_{2n+2}}{(1-t_{2n+1}U_n)^2} u_{n,i}\delta_{jj}\Big) \text{det}\Big(\delta_{ij}+\tfrac{t_{2n+1}}{1-t_{2n+1}U_n}u_{n,i}\delta_{jj}\Big) \,.
\end{align*}
Applying Sylvester's determinant theorem, $\text{det}(I+xy^T)=1+x^Ty$, gives
\begin{align*}
    |J_n|=\frac{t_{2n+1}}{(1-t_{2n+1}U_n)^2}\left(\frac{(1-t_{2n+1})t_{2n+2}}{1-t_{2n+1}U_n}\right)^n\frac{1-V_n}{1-t_{2n+1}U_n}\,,
\end{align*}
which is exactly the prefactor in \eqref{B_{n+1}}.
\paragraph{Two branches.}
Another change of coordinates was applied to go from  \eqref{tree} to \eqref{twobranches}. Here the coordinates  $\{u^L_{n,1},\dots,v^L_{n,n},u^R_{m,1},\dots,v^R_{m,m}\}$ were replaced with $\{r^L_{n,1},\dots,s^L_{n,n},r^R_{m,1},\dots,s^R_{m,m}\}$. The corresponding Jacobian reads
\begin{align*}
    |J_n|=
    \left\lvert
    \begin{matrix}
        \tfrac{1-V_m}{1-U_mU_n}\delta_{ij}+\tfrac{(1-V_m)U_m}{(1-U_mU_n)^2}u_{n,i}\delta_{jj} & 0 & \tfrac{(1-V_m)U_n}{1-U_mU_n}u_{n,i}\delta_{jj} &                              -\tfrac{u_{n,i}\delta_{jj}}{1-U_mV_n} \\
        -\tfrac{U_m(1-V_n)}{1-U_mU_n} \delta_{ij} & \delta_{ij} & 0 & 0 \\
        \tfrac{U_m(1-V_n)}{(1-U_mU_n)^2}u_{m,i}\delta_{jj} & -\tfrac{u_{m,i}\delta_{jj}}{1-U_mU_n} & \tfrac{1-V_n}{1-U_mU_n}\delta_{ij}+\tfrac{U_n(1-V_n)}{(1-U_mU_n)^2}u_{m,i}\delta_{jj} & 0 \\
        0 & 0 & -\tfrac{U_n(1-V_m)}{1-U_mU_n}\delta_{ij} & \delta_{ij}
    \end{matrix}
    \right\rvert\,.
\end{align*}
Gaussian elimination allows one to rewrite this as
\begin{align*}
    |J_n|=
    \begin{vmatrix}
    A & 0 & 0 & 0 \\
    B & C & 0 & 0 \\
    0 & 0 & D & 0 \\
    0 & 0 & E & F
    \end{vmatrix}
    =|A||C||D||F| \,,
\end{align*}
where
\begin{align*}
    A&=\frac{1-V_m}{1-U_mU_n}\delta_{ij}+\frac{(1-V_m)U_m}{(1-U_mU_n)^2}u_{n,i}\delta_{jj}\,, &  D&=\frac{1-V_n}{1-U_mU_n}\delta_{ij}+\frac{U_n(1-V_n)}{(1-U_mU_n)^2}u_{m,i}\delta_{jj}\,,\\
    B&=-\frac{U_m(1-V_n)}{1-U_mU_n} \delta_{ij} \,, &E&=-\frac{U_n(1-V_m)}{1-U_mU_n}\delta_{ij}\,,  \\
    C&=\delta_{ij}\,, &   F&=\delta_{ij}\,.
\end{align*}
Sylvester's determinant theorem now states that
\begin{align*}
    |J_n|=|A||D|=\frac{1}{(1-U_mU_n)^2}\left(\frac{1-V_m}{1-U_mU_n}\right)^n\left(\frac{1-V_n}{1-U_mU_n}\right)^m\,,
\end{align*}
which is the prefactor in \eqref{tree} up to the alternating minus sign.

\paragraph{$\mathcal{U}$-vertices.} The determinant of the Jacobian corresponding to the change of variables \eqref{changezeroform} reads
\begin{align*}
    |J| &=
    \begin{vmatrix}
        +\frac{\epsilon}{1-U_n(1-\epsilon)}\delta_{ij}+\frac{\epsilon(1-\epsilon)}{(1-U_n(1-\epsilon))^2}u_{n,i}\delta_{jj} & 0 \\
        -\frac{(1-V_n)(1-\epsilon)}{1-U_n(1-\epsilon)}\delta_{ij} - \frac{(1-V_n)(1-\epsilon)^2}{(1-U_n(1-\epsilon))^2}u_{n,i}\delta_{jj} & \delta_{ij} + \frac{1-\epsilon}{1-U_n(1-\epsilon)}u_{n,i}\delta_{jj}
    \end{vmatrix} \,.
\end{align*}
Using Sylvester's determinant theorem yields
$|J| = (\frac{1}{1-U_n(1-\epsilon)})^2(\frac{\epsilon}{1-U_n(1-\epsilon)})^n$. This is identified with the prefactor in \eqref{bigbranch} in the limit $\varepsilon \rightarrow 0$.

\subsection{Compactness of integration domain}
\label{app:domain}
\paragraph{Single branch.} The change of variable \eqref{newcoordinates} determines the domain of integration in \eqref{B_{n+1}newcoordinates}. We are interested in knowing if this domain is compact or not. It is useful to start with deriving some properties of $U_n$ and $V_n$. The $t_i$'s run from $0$ to $1$, hence
\begin{align*} 
    1-U_{n+1}& \geq \frac{(1-t_{2n+1})(1-U_n)}{1-t_{2n+1}U_n} \geq 0
\end{align*}
whenever $U_n \leq 1$. Since $U_1=t_1t_2 \leq 1$, it follows that $U_n \leq 1$ for all $n \geq 1$. Using this result, we find
\begin{align*}
    U_{n+1}&=\frac{(1-t_{2n+1})U_n+(1-U_n)t_{2n+1}}{1-t_{2n+1}U_n} t_{2n+2} \geq 0 \,.
\end{align*}
Similarly,
\begin{align*}
    1-V_{n+1}&=\frac{(1-V_n)(1-t_{2n+1})}{1-t_{2n+1}U_n} \geq 0
\end{align*}
if $V_n \leq 1$. Since $V_1=t_1$, we conclude that $V_n \leq 1$ for all $n \geq 1$.
\begin{align*}
    V_{n+1}&=\frac{V_n(1-t_{2n+1})+t_{2n+1}(1-U_n)}{1-t_{2n+1}U_n} \geq 0
\end{align*}
for $V_n \geq 0$. As $V_1=t_1$ we conclude that $V_n \geq 0$ for all $n \geq 1$. Using the above result we see that
\begin{align*}
    V_{n+1}-U_{n+1}& = \frac{(V_n-U_n t_{2n+1})(1-t_{2n+1})+t_{2n+1}(1-U_n)(1-t_{2n+2})}{1-t_{2n+1}U_n} \geq 0
\end{align*}
provided that $V_{n} \geq U_n$. Since $U_1=t_1t_2$, $V_1=t_1$, and  $V_1 \geq U_1$, we conclude by induction that $V_n \geq U_n$ for all $n \geq 1$.

Now the restrictions on the individual variables should be more obvious. It is useful to think of the variable $u_{n+m,n}$, with $m \geq 1$, as originating from $u_{n,n}$ when the first relation in \eqref{newcoordinates} is applied  $m$ times. The same is true for $v_{n+m,n}$. It is therefore convenient to first study the properties of $u_{n,n}$ and $v_{n,n}$. It is easy to see that
\begin{align*}
    0 &\leq u_{n+1,n+1}=\frac{1-U_n}{1-t_{2n+1}U_n}t_{2n+1}t_{2n+2} \leq 1
\end{align*}
and
\begin{align*}
   0 &\leq  v_{n+1,n+1}=\frac{1-V_n}{1-t_{2n+1}U_n}t_{2n+1} \leq 1 \,.
\end{align*}
Then
\begin{align*}
    0 &\leq u_{n+m,n}=\frac{1-U_{n+m-1}}{1-t_{2(n+m)-1}U_{n+i-1}} u_{n+m-1,n} \leq 1
\end{align*}
whenever $0 \leq u_{n+m-1,n} \leq 1$. By induction we find that $0 \leq u_{n+m,n} \leq 1$, as 0 $\leq u_{n,n} \leq 1$. As a result all $u$ variables belong to the interval $[0,1]$. For the $v$ variables we find
\begin{align*}
    v_{n+m,n}&=v_{n+m-1,n}-u_{n+m-1,n}\frac{t_{2(n+m)-1}(1-V_{n+m-1})}{1-t_{2(n+m)-1}U_{n+m-1}} \leq v_{n+m-1,n} \,.
\end{align*}
Again,  proceeding by induction and using the fact that $v_{n,n} \leq 1$ we conclude that $v_{n+m,n} \leq 1$. To prove that these variables are also nonnegative requires a bit more work. We will use the relation
\begin{align} \label{centersofmass}
    \frac{1-U_n}{1-U_{n-1}}& \geq \frac{1-t_{2n-1}}{1-t_{2n-1}U_{n-1}} = \frac{1-V_n}{1-V_{n-1}} \,.
\end{align}
We have
\begin{align*}
    v_{n+m,n}&\geq v_{n+m-1,n}-u_{n+m-1}\frac{1-V_{n+m-1}}{1-U_{n+m-1}}=\\
    &=v_{n+m-2,n}-u_{n+m-2}(\frac{t_{2(n+m)-3}(1-V_{n+m-2})}{1-t_{2(n+m)-3}U_{n+m-2}}+\frac{(1-t_{2(n+m)-3})t_{2(n+m)-2}}{1-t_{2(n+m)-3}U_{n+m-2}}\frac{1-V_{n+m-1}}{1-U_{n+m-1}}) \geq \\
    & \geq v_{n+m-2,n}-u_{n+m-2,n}\frac{1-V_{n+m-2}}{1-U_{n+m-2}} \geq \dots \geq v_{n,n}-u_{n,n}\frac{1-V_n}{1-U_n}\,.
\end{align*}
The equalities arise from setting $t_i=1$ for even $i$ and going from the second to the third line we used \eqref{centersofmass}. It only remains to show that
\begin{align*}
    v_{n,n}-u_{n,n}\frac{1-V_n}{1-U_n}& \geq \frac{t_{2n-1}(1-V_{n-1})}{1-t_{2n-1}U_{n-1}}(1-t_{2n+2}) \geq 0 \,,
\end{align*}
which proves that $v_{n+m,n} \geq 0$. Ultimately, we have shown that all $u$ and $v$ variables belong to the interval $[0,1]$, although they obey even stricter restrictions, which will be discussed in the next section. Moreover, the $U_n$ and $V_n$ are restricted to the interval $[0,1]$ as well and the domain of integration for a single branch is thus a subspace of the hypercube $[0,1]^{2n}$.

\paragraph{Trees.} Another change of variables is proposed in \eqref{newcoordinatestwobranches}. As we know that all $u$ and $v$ variables and their sums $U_n$ and $V_n$ belong to the interval $[0,1]$ and that $V_n \geq U_n$, it is not hard to see that
\begin{align*}
    0 &\leq r^L_{n,i}=\frac{1-V^L_n}{1-U^L_{n}U^R_m}u^L_{n,i} \leq 1
\end{align*}
and
\begin{align*}
    s^L_{n,i} \leq v^L_{n,i} \leq 1 \,.
\end{align*}
We also find that
\begin{align*}
    s^L_{n,i}&\geq v^L_{n,i}-u^L_{n,i}\frac{1-V^L_n}{1-U^L_n} \geq 0 \,,
\end{align*}
where the latter relation coincides with $v_{n+1,i} \geq 0$ for a single branch. Obviously, the same properties hold for $r^R_{m,i}$ and $s^R_{m,i}$ and consequently the domain of integration for a tree consisting of two branches with length $n$ and $m$ belongs to the interval $[0,1]^{2(n+m)}$.

\subsection{Full domain of integration}
\label{app:fulldomain}
In Section \ref{sec:vertices}, we explicitly constructed the configuration space for the quartic vertices. We will establish an analogous configuration space for branches of arbitrary length and eventually for all trees.

\paragraph{A single branch.}
Let us start by considering Rel. \eqref{newcoordinates}. Keeping in mind that all $u$ and $v$ variables and their sums $U_n$ and $V_{n}$ belong to the interval $[0,1]$, some relations may be derived. It is, however, hard to prove any relations between the variables at the same level $n$. It is therefore useful to think of $u_{n,i}$ as originating from $u_{i,i}$ and having moved up $n-i$ levels using the first relation in \eqref{newcoordinates}. The same is true for $v_{n,i}$. Thus, we first start by evaluating
\begin{align*}
    \frac{u_{n+1,n+1}}{v_{n+1,n+1}}&=\frac{t_{2n+2}(1-U_n)}{1-V_n} \leq \frac{1-U_n}{1-V_n} \leq \frac{1-U_{n+1}}{1-V_{n+1}} \,,
\end{align*}
where we have used \eqref{centersofmass} and equality is obtained for $t_{2n+2}=1$. Next, we consider
\begin{align} \label{eq0}
    \frac{v_{n+1,i}}{u_{n+1,i}} = \frac{1-t_{2n+1}U_n}{(1-t_{2n+1})t_{2n+2}}\frac{v_{n,i}}{u_{n,i}}-\frac{t_{2n+1}(1-V_n)}{1-t_{2n+1}U_n} \geq \frac{1}{t_{2n+2}}\frac{v_{n,i}}{u_{n,i}}
\end{align}
and inverting this gives
\begin{align*}
    \frac{u_{n+1,i}}{v_{n+1,i}} \leq t_{2n+2}\frac{u_{n,i}}{v_{n,i}} \leq \frac{u_{n,i}}{v_{n,i}} \,.
\end{align*}
By induction we find
\begin{align*}
    \frac{u_{n+1,i}}{v_{n+1,i}}& \leq t_{2n+2}\frac{u_{n,i}}{v_{n,i}} \leq \dots \leq t_{2n+2} \frac{u_{i,i}}{v_{i,i}} \leq \\
    &\leq t_{2n+2} \frac{1-U_{i-1}}{1-V_{i-1}} \leq t_{2n+2} \frac{1-U_n}{1-V_{n}} = \frac{u_{n+1,n+1}}{v_{n+1,n+1}}\,,
\end{align*}
where again we made use of \eqref{centersofmass}. Equality is obtained if $t_{2k+1}=0$ and $t_{2k}=1$ for all $k\in[i,n]$. Now we consider the relation between $\frac{u_{n+1,i}}{v_{n+1,i}}$ and $\frac{u_{n+1,j}}{v_{n+1,j}}$ for $i<j<n+1$. Following \eqref{eq0} we can bring the latter down to the level where it emanated from, which can be written schematically as
\begin{align} \label{eq1}
    \frac{v_{n+1,j}}{u_{n+1,j}} &= A \frac{v_{j,j}}{u_{j,j}} - B = \frac{A}{t_{2j}} \frac{1-V_{j-1}}{1-U_{j-1}} - B\,,
\end{align}
with $A \geq 1$ and $B \geq 0$. We then bring the former to the same level, which reads
\begin{align} \label{eq2}
    \frac{v_{n+1,i}}{u_{n+1,i}} = A \frac{v_{j,i}}{u_{j,i}} - B \,.
\end{align}
Since $i<j$, we have not reached the lowest level yet, so continuing this process yields
\begin{align*}
    \frac{v_{j,i}}{u_{j,i}} \geq \frac{1}{t_{2j}} \frac{v_{i,i}}{u_{i,i}} \geq \frac{1}{t_{2j}} \frac{1-V_{i-1}}{1-U_{i-1}} \geq \frac{1}{t_{2j}}\frac{1-V_{j-1}}{1-U_{j-1}}\,,
\end{align*}
and altogether
\begin{align*}
    \frac{v_{n+1,i}}{u_{n+1,i}} & \geq \frac{A}{t_{2j}}\frac{1-V_{j-1}}{1-U_{j-1}}  - B = \frac{v_{n+1,j}}{u_{n+1,j}} \,.
\end{align*}
Thus, we find
\begin{align*}
    \frac{u_{n+1,i}}{v_{n+1,i}} \leq \frac{u_{n+1,j}}{v_{n+1,j}} \,, \quad i<j \,.
\end{align*}
In particular, the equality sign occurs when $t_{2k}=1$ and $t_{2k+1}=0$ for all $k \in [i,j-1]$. Summarizing the above results, we can write
\begin{align*}
    \frac{u_{n+1,1}}{v_{n+1,1}} \leq \frac{u_{n+1,2}}{v_{n+1,2}} \leq \dots \leq \frac{u_{n+1,n}}{v_{n+1,n}} \leq \frac{u_{n+1,n+1}}{v_{n+1,n+1}} \leq \frac{1-U_{n+1}}{1-V_{n+1}} \,.
\end{align*}
Lastly, we derive a relation between the first $u$ and $v$ variable at each level. Consider
\begin{align} \label{fractions}
    v_{n+1,i}-u_{n+1,i}&=v_{n,i}-u_{n,i}(\frac{t_{2n+1}(1-V_n)}{1-t_{2n+1}U_n}+\frac{(1-t_{2n+1}t_{2n+2})}{1-t_{2n+1}U_n}) \geq\\
    & \geq v_{n,i}-u_{n,i}\frac{1-t_{2n+1}V_n}{1-t_{2n+1}U_n} \geq v_{n,i}-u_{n,i} \,.
\end{align}
Hence, if $v_{n,i} \leq u_{n,i}$, then $v_{n+1,i} \geq u_{n+1,i}$. From the initial values we know that $v_{1,1} \geq u_{1,1}$, which then extends through first terms to all orders, i.e., $u_{n+1,1} \leq v_{n+1,1}$. Together with \eqref{fractions} this determines the domain of integration for a branch of arbitrary length, analogous to the domain of integration $\mathcal{D}_1$ in Section \ref{sec:vertices}.
\paragraph{Trees.}
For the construction of trees we performed the coordinate transformation \eqref{newcoordinatestwobranches}. In the following discussion the statements for $r^L_{n,i}, s^L_{n,i}$ and $r^R_{m,i}, s^R_{m,i}$ are mostly the same. When both sets of variables obey a similar relation, we will mention only the former. In Appendix \ref{app:domain}, we have already shown that $r^L_{n,i} \leq 1$ and $s^L_{n,i} \leq 1$, so we can introduce new variables $r^L_n,s^L_n$ that satisfy
\begin{align*}
    \sum_{i=1}^{n} r^L_{n,i} + r^L_n &= 1 \,, &  \sum_{i=1}^{n} s^L_{n,i} + s^L_n &= 1 \,.
\end{align*}
From the analysis of a single branch we know that $\frac{v_{n,i}}{u_{n,i}} \geq \frac{v_{n,j}}{u_{n,j}}$ if $i < j$. Hence
\begin{align*}
    \frac{s^L_{n,i}}{r^L_{n,i}} &= \frac{1-U^L_n U^R_m}{1-V^R_m} \frac{v^L_{n,i}}{u^L_{n,i}} - U^R_m \frac{1-V^L_n}{1-V^R_m} \geq \frac{s^L_{n,j}}{r^L_{n,j}} \,, \quad \text{if } i<j \,, 
\end{align*}
with equality occurred for $\frac{v_{n,i}}{u_{n,i}} = \frac{v_{n,j}}{u_{n,j}}$.  Setting $v_0 \equiv r^L_n s^R_m$, $u_0 \equiv r^R_m s^L_n$, we find
\begin{align*}
    \frac{s^L_{n,n}}{r^L_{n,n}} &= \frac{1-U^L_n U^R_m}{1-V^R_m} \frac{1}{t_{2n}} \frac{1-V^L_{n-1}}{1-U^L_{n-1}} - U^R_m \frac{1-V^L_n}{1-V^R_m} \geq \frac{1-U^L_n U^R_m}{1-V^R_m} \frac{1-V^L_{n}}{1-U^L_{n}} - U^R_m \frac{1-V^L_n}{1-V^R_m} = \\
    &=\frac{(1-U^R_m)(1-V^L_n)}{(1-U^L_n)(1-V^R_m)} = \frac{1-\sum_{i=1}^m r^R_{m,i}-\sum_{i=1}^n s^L_{n,i}}{1-\sum_{i=1}^m s^R_{m,i}-\sum_{i=1}^n r^L_{n,i}} \,,
\end{align*}
with equality for $t_{2n}=1$, and
\begin{align*}
    \frac{s^R_{m,m}}{r^R_{m,m}} \geq \frac{1-\sum_{i=1}^m r^R_{m,i}-\sum_{i=1}^n s^L_{n,i}}{1-\sum_{i=1}^m s^R_{m,i}-\sum_{i=1}^n r^L_{n,i}} \,.
\end{align*}
Combining the above results yields
\begin{align}
    \frac{u_1}{v_1} \leq \frac{u_2}{v_2} \leq \dots \leq \frac{u_{m+n}}{v_{m+n}} \leq \frac{u_{m+n+1}}{v_{m+n+1}} \,.
\end{align}
where we used the variables defined in \eqref{rename}. Moreover, $u_1=r_{m,1}^R$, $v_1=s_{m,1}^R$, so we have
\begin{align*}
    \frac{s^R_{m,1}}{r^R_{m,1}} = \frac{1-U^L_n U^R_m}{1-V^L_n} \frac{v^R_{m,1}}{u^R_{m,1}}-\frac{U^L_{n}(1-V^R_m)}{1-V^L_n} \geq \frac{1-U^L_n(1+U^R_m-V^R_m)}{1-V^L_n} \geq \frac{1-U^L_n}{1-V^L_n} \geq 1 \,,
\end{align*}
where in the first inequality we used that $u^R_{m,1} \leq v^R_{m,1}$ and in the second we used $V^R_m \geq U^R_m$, which were both previously derived. This leads to the inequalities $0 \leq u_1 \leq v_1 \leq 1$. This collection of inequalities defines the configuration space for a tree. Notice that the configuration space of a general tree looks very similar to the configuration space of a `single-branch' tree. In fact, up to relabeling, the configuration space of a `two-branch' tree with the lengths of branches $n_1$ and $ n_2$ coincides with the configuration space of a single branch of length $n_1+n_2$. It follows that the domain of integration of trees can be related between different topologies by relabeling of variables.

\newpage
\footnotesize
\providecommand{\href}[2]{#2}\begingroup\raggedright\endgroup


\begin{thebibliography}{10}

\bibitem{Bekaert:2022poo}
X.~Bekaert, N.~Boulanger, A.~Campoleoni, M.~Chiodaroli, D.~Francia,
  M.~Grigoriev, E.~Sezgin, and E.~Skvortsov, ``{Snowmass White Paper: Higher
  Spin Gravity and Higher Spin symmetry},''
  \href{http://arxiv.org/abs/2205.01567}{{\ttfamily arXiv:2205.01567
  [hep-th]}}.

\bibitem{Boulanger:2015ova}
N.~Boulanger, P.~Kessel, E.~D. Skvortsov, and M.~Taronna, ``{Higher spin
  interactions in four-dimensions: Vasiliev versus Fronsdal},'' {\em J. Phys.}
  {\bfseries A49} no.~9, (2016) 095402,
\href{http://arxiv.org/abs/1508.04139}{{\ttfamily arXiv:1508.04139 [hep-th]}}.

\bibitem{Bekaert:2015tva}
X.~Bekaert, J.~Erdmenger, D.~Ponomarev, and C.~Sleight, ``{Quartic AdS
  Interactions in Higher-Spin Gravity from Conformal Field Theory},'' {\em
  JHEP} {\bfseries 11} (2015) 149,
\href{http://arxiv.org/abs/1508.04292}{{\ttfamily arXiv:1508.04292 [hep-th]}}.

\bibitem{Maldacena:2015iua}
J.~Maldacena, D.~Simmons-Duffin, and A.~Zhiboedov, ``{Looking for a bulk
  point},'' {\em JHEP} {\bfseries 01} (2017) 013,
\href{http://arxiv.org/abs/1509.03612}{{\ttfamily arXiv:1509.03612 [hep-th]}}.

\bibitem{Sleight:2017pcz}
C.~Sleight and M.~Taronna, ``{Higher-Spin Gauge Theories and Bulk Locality},''
  {\em Phys. Rev. Lett.} {\bfseries 121} no.~17, (2018) 171604,
\href{http://arxiv.org/abs/1704.07859}{{\ttfamily arXiv:1704.07859 [hep-th]}}.

\bibitem{Ponomarev:2017qab}
D.~Ponomarev, ``{A Note on (Non)-Locality in Holographic Higher Spin
  Theories},'' {\em Universe} {\bfseries 4} no.~1, (2018) 2,
\href{http://arxiv.org/abs/1710.00403}{{\ttfamily arXiv:1710.00403 [hep-th]}}.

\bibitem{Metsaev:1991mt}
R.~R. Metsaev, ``{Poincare invariant dynamics of massless higher spins: Fourth
  order analysis on mass shell},''
{\em Mod. Phys. Lett.} {\bfseries A6} (1991) 359--367.

\bibitem{Metsaev:1991nb}
R.~R. Metsaev, ``{$S$ matrix approach to massless higher spins theory. 2: The
  Case of internal symmetry},''
{\em Mod. Phys. Lett.} {\bfseries A6} (1991) 2411--2421.

\bibitem{Ponomarev:2016lrm}
D.~Ponomarev and E.~D. Skvortsov, ``{Light-Front Higher-Spin Theories in Flat
  Space},'' {\em J. Phys.} {\bfseries A50} no.~9, (2017) 095401,
\href{http://arxiv.org/abs/1609.04655}{{\ttfamily arXiv:1609.04655 [hep-th]}}.

\bibitem{Skvortsov:2018jea}
E.~D. Skvortsov, T.~Tran, and M.~Tsulaia, ``{Quantum Chiral Higher Spin
  Gravity},'' {\em Phys. Rev. Lett.} {\bfseries 121} no.~3, (2018) 031601,
\href{http://arxiv.org/abs/1805.00048}{{\ttfamily arXiv:1805.00048 [hep-th]}}.

\bibitem{Skvortsov:2020wtf}
E.~Skvortsov, T.~Tran, and M.~Tsulaia, ``{More on Quantum Chiral Higher Spin
  Gravity},'' {\em Phys. Rev.} {\bfseries D101} no.~10, (2020) 106001,
\href{http://arxiv.org/abs/2002.08487}{{\ttfamily arXiv:2002.08487 [hep-th]}}.

\bibitem{Blencowe:1988gj}
M.~Blencowe, ``{A Consistent Interacting Massless Higher Spin Field Theory in
  $D$ = (2+1)},''
{\em Class.Quant.Grav.} {\bfseries 6} (1989) 443.

\bibitem{Bergshoeff:1989ns}
E.~Bergshoeff, M.~P. Blencowe, and K.~S. Stelle, ``{Area Preserving
  Diffeomorphisms and Higher Spin Algebra},''
{\em Commun. Math. Phys.} {\bfseries 128} (1990) 213.

\bibitem{Campoleoni:2010zq}
A.~Campoleoni, S.~Fredenhagen, S.~Pfenninger, and S.~Theisen, ``{Asymptotic
  symmetries of three-dimensional gravity coupled to higher-spin fields},''
  {\em JHEP} {\bfseries 1011} (2010) 007,
\href{http://arxiv.org/abs/1008.4744}{{\ttfamily arXiv:1008.4744 [hep-th]}}.

\bibitem{Henneaux:2010xg}
M.~Henneaux and S.-J. Rey, ``{Nonlinear $W_{\infty}$ as Asymptotic Symmetry of
  Three-Dimensional Higher Spin Anti-de Sitter Gravity},'' {\em JHEP}
  {\bfseries 1012} (2010) 007,
\href{http://arxiv.org/abs/1008.4579}{{\ttfamily arXiv:1008.4579 [hep-th]}}.

\bibitem{Pope:1989vj}
C.~N. Pope and P.~K. Townsend, ``{Conformal Higher Spin in (2+1)-dimensions},''
{\em Phys. Lett.} {\bfseries B225} (1989) 245--250.

\bibitem{Fradkin:1989xt}
E.~S. Fradkin and V.~{\relax Ya}. Linetsky, ``{A Superconformal Theory of
  Massless Higher Spin Fields in $D$ = (2+1)},'' {\em Mod. Phys. Lett.}
  {\bfseries A4} (1989) 731.
[Annals Phys.198,293(1990)].

\bibitem{Grigoriev:2019xmp}
M.~Grigoriev, I.~Lovrekovic, and E.~Skvortsov, ``{New Conformal Higher Spin
  Gravities in $3d$},'' {\em JHEP} {\bfseries 01} (2020) 059,
\href{http://arxiv.org/abs/1909.13305}{{\ttfamily arXiv:1909.13305 [hep-th]}}.

\bibitem{Grigoriev:2020lzu}
M.~Grigoriev, K.~Mkrtchyan, and E.~Skvortsov, ``{Matter-free higher spin
  gravities in 3D: Partially-massless fields and general structure},''
  \href{http://dx.doi.org/10.1103/PhysRevD.102.066003}{{\em Phys. Rev. D}
  {\bfseries 102} no.~6, (2020) 066003},
  \href{http://arxiv.org/abs/2005.05931}{{\ttfamily arXiv:2005.05931
  [hep-th]}}.

\bibitem{Segal:2002gd}
A.~Y. Segal, ``{Conformal higher spin theory},'' {\em Nucl. Phys.} {\bfseries
  B664} (2003) 59--130,
\href{http://arxiv.org/abs/hep-th/0207212}{{\ttfamily arXiv:hep-th/0207212
  [hep-th]}}.

\bibitem{Tseytlin:2002gz}
A.~A. Tseytlin, ``{On limits of superstring in $AdS_5\times S^5$},'' {\em
  Theor. Math. Phys.} {\bfseries 133} (2002) 1376--1389,
  \href{http://arxiv.org/abs/hep-th/0201112}{{\ttfamily arXiv:hep-th/0201112
  [hep-th]}}.
[Teor. Mat. Fiz.133,69(2002)].

\bibitem{Bekaert:2010ky}
X.~Bekaert, E.~Joung, and J.~Mourad, ``{Effective action in a higher-spin
  background},'' {\em JHEP} {\bfseries 02} (2011) 048,
\href{http://arxiv.org/abs/1012.2103}{{\ttfamily arXiv:1012.2103 [hep-th]}}.

\bibitem{deMelloKoch:2018ivk}
R.~de~Mello~Koch, A.~Jevicki, K.~Suzuki, and J.~Yoon, ``{AdS Maps and Diagrams
  of Bi-local Holography},'' {\em JHEP} {\bfseries 03} (2019) 133,
\href{http://arxiv.org/abs/1810.02332}{{\ttfamily arXiv:1810.02332 [hep-th]}}.

\bibitem{Aharony:2020omh}
O.~Aharony, S.~M. Chester, and E.~Y. Urbach, ``{A Derivation of AdS/CFT for
  Vector Models},''
\href{http://arxiv.org/abs/2011.06328}{{\ttfamily arXiv:2011.06328 [hep-th]}}.

\bibitem{Sperling:2017dts}
M.~Sperling and H.~C. Steinacker, ``{Covariant 4-dimensional fuzzy spheres,
  matrix models and higher spin},'' {\em J. Phys.} {\bfseries A50} no.~37,
  (2017) 375202,
\href{http://arxiv.org/abs/1704.02863}{{\ttfamily arXiv:1704.02863 [hep-th]}}.

\bibitem{Tran:2021ukl}
T.~Tran, ``{Twistor constructions for higher-spin extensions of (self-dual)
  Yang-Mills},'' \href{http://dx.doi.org/10.1007/JHEP11(2021)117}{{\em JHEP}
  {\bfseries 11} (2021) 117}, \href{http://arxiv.org/abs/2107.04500}{{\ttfamily
  arXiv:2107.04500 [hep-th]}}.

\bibitem{Steinacker:2022jjv}
H.~Steinacker and T.~Tran, ``{A Twistorial Description of the IKKT-Matrix
  Model},'' \href{http://arxiv.org/abs/2203.05436}{{\ttfamily arXiv:2203.05436
  [hep-th]}}.

\bibitem{Weinberg:1964ew}
S.~Weinberg, ``{Photons and Gravitons in S Matrix Theory: Derivation of Charge
  Conservation and Equality of Gravitational and Inertial Mass},''
{\em Phys. Rev.} {\bfseries 135} (1964) B1049--B1056.

\bibitem{Coleman:1967ad}
S.~R. Coleman and J.~Mandula, ``{All Possible Symmetries of the S Matrix},''
{\em Phys. Rev.} {\bfseries 159} (1967) 1251--1256.

\bibitem{Skvortsov:2018uru}
E.~Skvortsov, ``{Light-Front Bootstrap for Chern-Simons Matter Theories},''
  {\em JHEP} {\bfseries 06} (2019) 058,
\href{http://arxiv.org/abs/1811.12333}{{\ttfamily arXiv:1811.12333 [hep-th]}}.

\bibitem{Sezgin:2002rt}
E.~Sezgin and P.~Sundell, ``{Massless higher spins and holography},'' {\em
  Nucl.Phys.} {\bfseries B644} (2002) 303--370,
\href{http://arxiv.org/abs/hep-th/0205131}{{\ttfamily arXiv:hep-th/0205131
  [hep-th]}}.

\bibitem{Klebanov:2002ja}
I.~R. Klebanov and A.~M. Polyakov, ``{AdS dual of the critical $O(N)$ vector
  model},'' {\em Phys. Lett.} {\bfseries B550} (2002) 213--219,
\href{http://arxiv.org/abs/hep-th/0210114}{{\ttfamily arXiv:hep-th/0210114}}.

\bibitem{Sezgin:2003pt}
E.~Sezgin and P.~Sundell, ``{Holography in 4D (super) higher spin theories and
  a test via cubic scalar couplings},'' {\em JHEP} {\bfseries 0507} (2005) 044,
\href{http://arxiv.org/abs/hep-th/0305040}{{\ttfamily arXiv:hep-th/0305040
  [hep-th]}}.

\bibitem{Leigh:2003gk}
R.~G. Leigh and A.~C. Petkou, ``{Holography of the N=1 higher spin theory on
  AdS(4)},'' {\em JHEP} {\bfseries 0306} (2003) 011,
\href{http://arxiv.org/abs/hep-th/0304217}{{\ttfamily arXiv:hep-th/0304217
  [hep-th]}}.

\bibitem{Maldacena:2011jn}
J.~Maldacena and A.~Zhiboedov, ``{Constraining Conformal Field Theories with A
  Higher Spin Symmetry},''
\href{http://arxiv.org/abs/1112.1016}{{\ttfamily arXiv:1112.1016 [hep-th]}}.

\bibitem{Boulanger:2013zza}
N.~Boulanger, D.~Ponomarev, E.~D. Skvortsov, and M.~Taronna, ``{On the
  uniqueness of higher-spin symmetries in AdS and CFT},''
  \href{http://dx.doi.org/10.1142/S0217751X13501625}{{\em Int. J. Mod. Phys.}
  {\bfseries A28} (2013) 1350162},
\href{http://arxiv.org/abs/1305.5180}{{\ttfamily arXiv:1305.5180 [hep-th]}}.

\bibitem{Alba:2013yda}
V.~Alba and K.~Diab, ``{Constraining conformal field theories with a higher
  spin symmetry in d=4},''
\href{http://arxiv.org/abs/1307.8092}{{\ttfamily arXiv:1307.8092 [hep-th]}}.

\bibitem{Alba:2015upa}
V.~Alba and K.~Diab, ``{Constraining conformal field theories with a higher
  spin symmetry in $d> 3$ dimensions},''
\href{http://arxiv.org/abs/1510.02535}{{\ttfamily arXiv:1510.02535 [hep-th]}}.

\bibitem{Skvortsov:2020gpn}
E.~Skvortsov and T.~Tran, ``{One-loop Finiteness of Chiral Higher Spin
  Gravity},''
\href{http://arxiv.org/abs/2004.10797}{{\ttfamily arXiv:2004.10797 [hep-th]}}.

\bibitem{Sharapov:2022awp}
A.~Sharapov and E.~Skvortsov, ``{Chiral Higher Spin Gravity in (A)dS${}_4$ and
  secrets of Chern--Simons Matter Theories},''
  \href{http://arxiv.org/abs/2205.15293}{{\ttfamily arXiv:2205.15293
  [hep-th]}}.

\bibitem{Metsaev:2018xip}
R.~R. Metsaev, ``{Light-cone gauge cubic interaction vertices for massless
  fields in AdS(4)},''
  \href{http://dx.doi.org/10.1016/j.nuclphysb.2018.09.021}{{\em Nucl. Phys.}
  {\bfseries B936} (2018) 320--351},
\href{http://arxiv.org/abs/1807.07542}{{\ttfamily arXiv:1807.07542 [hep-th]}}.

\bibitem{Tran:2022tft}
T.~Tran, ``{Toward a twistor action for chiral higher-spin gravity},''
  \href{http://arxiv.org/abs/2209.00925}{{\ttfamily arXiv:2209.00925
  [hep-th]}}.

\bibitem{Skvortsov:2022syz}
E.~Skvortsov and R.~Van~Dongen, ``{Minimal models of field theories: Chiral
  Higher Spin Gravity},'' \href{http://arxiv.org/abs/2204.10285}{{\ttfamily
  arXiv:2204.10285 [hep-th]}}.

\bibitem{Sharapov:2022faa}
A.~Sharapov, E.~Skvortsov, A.~Sukhanov, and R.~Van~Dongen, ``{Minimal model of
  Chiral Higher Spin Gravity},''
  \href{http://arxiv.org/abs/2205.07794}{{\ttfamily arXiv:2205.07794
  [hep-th]}}.

\bibitem{Sharapov:2022wpz}
A.~Sharapov, E.~Skvortsov, and R.~Van~Dongen, ``{Chiral Higher Spin Gravity and
  Convex Geometry},'' \href{http://arxiv.org/abs/2209.01796}{{\ttfamily
  arXiv:2209.01796 [hep-th]}}.

\bibitem{kontsevich2021pre}
M.~Kontsevich, A.~Takeda, and Y.~Vlassopoulos, ``{Pre-Calabi-Yau algebras and
  topological quantum field theories},'' {\em arXiv preprint arXiv:2112.14667}
  (2021) .

\bibitem{Kontsevich:1997vb}
M.~Kontsevich, ``{Deformation quantization of Poisson manifolds. 1.},'' {\em
  Lett. Math. Phys.} {\bfseries 66} (2003) 157--216,
\href{http://arxiv.org/abs/q-alg/9709040}{{\ttfamily arXiv:q-alg/9709040
  [q-alg]}}.

\bibitem{Shoikhet:2000gw}
B.~Shoikhet, ``{A proof of the Tsygan formality conjecture for chains},'' {\em
  Advances in Mathematics} {\bfseries 179} no.~1, (2003) 7 -- 37.

\bibitem{Vasiliev:1986td}
M.~A. Vasiliev, ``Free massless fields of arbitrary spin in the de sitter space
  and initial data for a higher spin superalgebra,''
{\em Fortsch. Phys.} {\bfseries 35} (1987) 741--770.

\bibitem{Vasiliev:1988sa}
M.~A. Vasiliev, ``Consistent equations for interacting massless fields of all
  spins in the first order in curvatures,''
{\em Annals Phys.} {\bfseries 190} (1989) 59--106.

\bibitem{Ponomarev:2017nrr}
D.~Ponomarev, ``{Chiral Higher Spin Theories and Self-Duality},'' {\em JHEP}
  {\bfseries 12} (2017) 141,
\href{http://arxiv.org/abs/1710.00270}{{\ttfamily arXiv:1710.00270 [hep-th]}}.

\bibitem{Krasnov:2021nsq}
K.~Krasnov, E.~Skvortsov, and T.~Tran, ``{Actions for Self-dual Higher Spin
  Gravities},''
\href{http://arxiv.org/abs/2105.12782}{{\ttfamily arXiv:2105.12782 [hep-th]}}.

\bibitem{Hughston:1979tq}
L.~P. Hughston, R.~S. Ward, M.~G. Eastwood, M.~L. Ginsberg, A.~P. Hodges, S.~A.
  Huggett, T.~R. Hurd, R.~O. Jozsa, R.~Penrose, A.~Popovich, {\em et~al.},
  eds., {\em {Advances in twistor theory}}.
\newblock
1979.
\newblock

\bibitem{Eastwood:1981jy}
M.~G. Eastwood, R.~Penrose, and R.~O. Wells, ``{Cohomology and Massless
  Fields},''
\href{http://dx.doi.org/10.1007/BF01942327}{{\em Commun. Math. Phys.}
  {\bfseries 78} (1981) 305--351}.

\bibitem{Woodhouse:1985id}
N.~M.~J. Woodhouse, ``{Real methods in twistor theory},''
\href{http://dx.doi.org/10.1088/0264-9381/2/3/006}{{\em Class. Quant. Grav.}
  {\bfseries 2} (1985) 257--291}.

\bibitem{penroserindler}
R.~Penrose and W.~Rindler,
  \href{http://dx.doi.org/10.1017/CBO9780511564048}{{\em Spinors and
  Space-Time}}, vol.~1 of {\em Cambridge Monographs on Mathematical Physics}.
\newblock Cambridge University Press, 1984.

\bibitem{Alexandrov:1995kv}
M.~Alexandrov, M.~Kontsevich, A.~Schwarz, and O.~Zaboronsky, ``{The Geometry of
  the Master Equation and Topological Quantum Field Theory},'' {\em Int. J.
  Mod. Phys.} {\bfseries A12} (1997) 1405--1429,
\href{http://arxiv.org/abs/hep-th/9502010}{{\ttfamily arXiv:hep-th/9502010
  [hep-th]}}.

\bibitem{Vasiliev:1999ba}
M.~A. Vasiliev, ``{Higher spin gauge theories: Star-product and AdS space},''
\href{http://arxiv.org/abs/hep-th/9910096}{{\ttfamily hep-th/9910096}}.

\bibitem{Vasiliev:1990cm}
M.~A. Vasiliev, ``{Closed equations for interacting gauge fields of all
  spins},''
{\em JETP Lett.} {\bfseries 51} (1990) 503--507.

\bibitem{Didenko:2018fgx}
V.~E. Didenko, O.~A. Gelfond, A.~V. Korybut, and M.~A. Vasiliev, ``{Homotopy
  Properties and Lower-Order Vertices in Higher-Spin Equations},''
  \href{http://dx.doi.org/10.1088/1751-8121/aae5e1}{{\em J. Phys. A} {\bfseries
  51} no.~46, (2018) 465202}, \href{http://arxiv.org/abs/1807.00001}{{\ttfamily
  arXiv:1807.00001 [hep-th]}}.

\bibitem{Didenko:2019xzz}
V.~E. Didenko, O.~A. Gelfond, A.~V. Korybut, and M.~A. Vasiliev, ``{Limiting
  Shifted Homotopy in Higher-Spin Theory and Spin-Locality},''
  \href{http://dx.doi.org/10.1007/JHEP12(2019)086}{{\em JHEP} {\bfseries 12}
  (2019) 086}, \href{http://arxiv.org/abs/1909.04876}{{\ttfamily
  arXiv:1909.04876 [hep-th]}}.

\bibitem{Didenko:2020bxd}
V.~E. Didenko, O.~A. Gelfond, A.~V. Korybut, and M.~A. Vasiliev,
  ``{Spin-locality of $\eta^{2}$ and $ {\overline{\eta}}^2 $ quartic
  higher-spin vertices},''
  \href{http://dx.doi.org/10.1007/JHEP12(2020)184}{{\em JHEP} {\bfseries 12}
  (2020) 184}, \href{http://arxiv.org/abs/2009.02811}{{\ttfamily
  arXiv:2009.02811 [hep-th]}}.

\bibitem{Gelfond:2021two}
O.~A. Gelfond and A.~V. Korybut, ``{Manifest form of the spin-local higher-spin
  vertex $\varUpsilon ^{\eta \eta }_{\omega CCC}$},''
  \href{http://dx.doi.org/10.1140/epjc/s10052-021-09401-4}{{\em Eur. Phys. J.
  C} {\bfseries 81} no.~7, (2021) 605},
  \href{http://arxiv.org/abs/2101.01683}{{\ttfamily arXiv:2101.01683
  [hep-th]}}.

\bibitem{IYUDU202163}
N.~Iyudu, M.~Kontsevich, and Y.~Vlassopoulos, ``{Pre-Calabi-Yau algebras as
  noncommutative Poisson structures},''
  \href{http://dx.doi.org/https://doi.org/10.1016/j.jalgebra.2020.08.029}{{\em
  Journal of Algebra} {\bfseries 567} (2021) 63--90}.

\bibitem{PGr}
{Nima Arkani-Hamed, Jacob Bourjaily, Freddy Cachazo, Alexander Goncharov,
  Alexander Postnikov, Jaroslav Trnka}, {\em Grassmannian Geometry of
  Scattering Amplitudes}.
\newblock Cambridge University Press, 2016.

\bibitem{williams2021positive}
L.~K. Williams, ``{The positive Grassmannian, the amplituhedron, and cluster
  algebras},'' \href{http://arxiv.org/abs/2110.10856}{{\ttfamily
  arXiv:2110.10856}}.

\bibitem{Sharapov:2017yde}
A.~A. Sharapov and E.~D. Skvortsov, ``{Formal higher-spin theories and
  Kontsevich--Shoikhet--Tsygan formality},'' {\em Nucl. Phys.} {\bfseries B921}
  (2017) 538--584,
\href{http://arxiv.org/abs/1702.08218}{{\ttfamily arXiv:1702.08218 [hep-th]}}.

\bibitem{Sharapov:2022eiy}
A.~Sharapov and E.~Skvortsov, ``{Integrable Models From Non-Commutative
  Geometry With Applications to 3D Dualities},'' in {\em {21st Hellenic School
  and Workshops on Elementary Particle Physics and Gravity}}.
\newblock 4, 2022.
\newblock \href{http://arxiv.org/abs/2204.08903}{{\ttfamily arXiv:2204.08903
  [hep-th]}}.

\bibitem{FFS}
B.~Shoikhet, G.~Felder, and B.~Feigin, ``{Hochschild cohomology of the Weyl
  algebra and traces in deformation quantization},'' {\em Duke Mathematical
  Journal} {\bfseries 127} no.~3, (2005) 487--517.

\bibitem{Sharapov:2022phg}
A.~Sharapov, E.~Skvortsov, and A.~Sukhanov, ``{Deformation quantization of the
  simplest Poisson Orbifold},''
  \href{http://arxiv.org/abs/2207.08916}{{\ttfamily arXiv:2207.08916
  [math-ph]}}.

\bibitem{Feigin}
B.~Feigin, ``{The Lie algebras gl(l) and cohomologies of Lie algebras of
  differential operators},'' {\em Russ. Math. Surv.} {\bfseries 34} (1988) 169.

\bibitem{Sezgin:2005pv}
E.~Sezgin and P.~Sundell, ``{An Exact solution of 4-D higher-spin gauge
  theory},'' {\em Nucl.Phys.} {\bfseries B762} (2007) 1--37,
\href{http://arxiv.org/abs/hep-th/0508158}{{\ttfamily arXiv:hep-th/0508158
  [hep-th]}}.

\bibitem{Didenko:2009td}
V.~Didenko and M.~Vasiliev, ``{Static BPS black hole in 4d higher-spin gauge
  theory},'' {\em Phys.Lett.} {\bfseries B682} (2009) 305--315,
\href{http://arxiv.org/abs/0906.3898}{{\ttfamily arXiv:0906.3898 [hep-th]}}.

\bibitem{Aros:2017ror}
R.~Aros, C.~Iazeolla, J.~Nore\~na, E.~Sezgin, P.~Sundell, and Y.~Yin, ``{FRW
  and domain walls in higher spin gravity},''
  \href{http://dx.doi.org/10.1007/JHEP03(2018)153}{{\em JHEP} {\bfseries 03}
  (2018) 153}, \href{http://arxiv.org/abs/1712.02401}{{\ttfamily
  arXiv:1712.02401 [hep-th]}}.

\bibitem{Didenko:2021vdb}
V.~E. Didenko and A.~V. Korybut, ``{Planar solutions of higher-spin theory.
  Nonlinear corrections},''
  \href{http://dx.doi.org/10.1007/JHEP01(2022)125}{{\em JHEP} {\bfseries 01}
  (2022) 125}, \href{http://arxiv.org/abs/2110.02256}{{\ttfamily
  arXiv:2110.02256 [hep-th]}}.

\bibitem{Didenko:2021vui}
V.~E. Didenko and A.~V. Korybut, ``{Planar solutions of higher-spin theory.
  Part I. Free field level},''
  \href{http://dx.doi.org/10.1007/JHEP08(2021)144}{{\em JHEP} {\bfseries 08}
  (2021) 144}, \href{http://arxiv.org/abs/2105.09021}{{\ttfamily
  arXiv:2105.09021 [hep-th]}}.

\bibitem{Sharapov:2021drr}
A.~Sharapov and E.~Skvortsov, ``{Higher Spin Gravities and Presymplectic AKSZ
  Models},''
\href{http://arxiv.org/abs/2102.02253}{{\ttfamily arXiv:2102.02253 [hep-th]}}.

\bibitem{Sharapov:2020quq}
A.~Sharapov and E.~Skvortsov, ``{Characteristic Cohomology and Observables in
  Higher Spin Gravity},'' \href{http://dx.doi.org/10.1007/JHEP12(2020)190}{{\em
  JHEP} {\bfseries 12} (2020) 190},
\href{http://arxiv.org/abs/2006.13986}{{\ttfamily arXiv:2006.13986 [hep-th]}}.

\bibitem{Alkalaev:2019xuv}
K.~Alkalaev and X.~Bekaert, ``{Towards higher-spin AdS$_2$/CFT$_1$
  holography},'' \href{http://dx.doi.org/10.1007/JHEP04(2020)206}{{\em JHEP}
  {\bfseries 04} (2020) 206}, \href{http://arxiv.org/abs/1911.13212}{{\ttfamily
  arXiv:1911.13212 [hep-th]}}.

\bibitem{Sharapov:2019vyd}
A.~Sharapov and E.~Skvortsov, ``{Formal Higher Spin Gravities},'' {\em Nucl.
  Phys.} {\bfseries B941} (2019) 838--860,
\href{http://arxiv.org/abs/1901.01426}{{\ttfamily arXiv:1901.01426 [hep-th]}}.

\bibitem{Didenko:2022qga}
V.~E. Didenko, ``{On holomorphic sector of higher-spin theory},''
  \href{http://arxiv.org/abs/2209.01966}{{\ttfamily arXiv:2209.01966
  [hep-th]}}.

\bibitem{Giombi:2009wh}
S.~Giombi and X.~Yin, ``{Higher Spin Gauge Theory and Holography: The
  Three-Point Functions},'' {\em JHEP} {\bfseries 1009} (2010) 115,
\href{http://arxiv.org/abs/0912.3462}{{\ttfamily arXiv:0912.3462 [hep-th]}}.

\bibitem{Giombi:2010vg}
S.~Giombi and X.~Yin, ``{Higher Spins in AdS and Twistorial Holography},'' {\em
  JHEP} {\bfseries 1104} (2011) 086,
\href{http://arxiv.org/abs/1004.3736}{{\ttfamily arXiv:1004.3736 [hep-th]}}.

\bibitem{Skvortsov:2015lja}
E.~D. Skvortsov and M.~Taronna, ``{On Locality, Holography and Unfolding},''
  {\em JHEP} {\bfseries 11} (2015) 044,
\href{http://arxiv.org/abs/1508.04764}{{\ttfamily arXiv:1508.04764 [hep-th]}}.

\bibitem{Giombi:2011kc}
S.~Giombi, S.~Minwalla, S.~Prakash, S.~P. Trivedi, S.~R. Wadia, and X.~Yin,
  ``{Chern-Simons Theory with Vector Fermion Matter},'' {\em Eur. Phys. J.}
  {\bfseries C72} (2012) 2112,
\href{http://arxiv.org/abs/1110.4386}{{\ttfamily arXiv:1110.4386 [hep-th]}}.

\bibitem{stasheff2018linfty}
J.~Stasheff, ``{$L_\infty$ and $A_\infty$-structures: then and now},''
  \href{http://arxiv.org/abs/1809.02526}{{\ttfamily {arXiv}:1809.02526
  [{math.QA}]}}.

\bibitem{HK}
J.~Huebschmann and T.~Kadeishvili, ``Small models for chain algebras,'' {\em
  Mathematische Zeitschrift} {\bfseries 207} no.~2, (1991) 245--280.

\bibitem{GLS}
V.~K. A.~M. Gugenheim, L.~A. Lambe, and J.~D. Stasheff, ``{Perturbation theory
  in differential homological algebra II},'' {\em Illinois J. Math.} {\bfseries
  35} no.~4, (1991) 357--373. \url{http://dml.mathdoc.fr/item/1255987784}.

\bibitem{merkulov1999strongly}
S.~Merkulov, ``{Strong homotopy algebras of a K{\"a}hler manifold},'' {\em
  International Mathematics Research Notices} {\bfseries 1999} no.~3, (1999)
  153--164, \href{http://arxiv.org/abs/math/9809172}{{\ttfamily
  arXiv:math/9809172}}.

\bibitem{Kontsevuch:2006jb}
M.~Kontsevich and Y.~Soibelman, ``{Notes on $A_\infty$-Algebras,
  $A_\infty$-Categories and Non-Commutative Geometry},''
  \href{http://dx.doi.org/10.1007/978-3-540-68030-7\_6}{{\em Lect. Notes in
  Physics} {\bfseries 757} (2009) 153--220},
  \href{http://arxiv.org/abs/math/0606241}{{\ttfamily arXiv:math/0606241}}.

\end{thebibliography}
\end{document}